\documentclass[12pt,titlepage,twoside,a4paper]{report}
\usepackage[dvips]{epsfig}
\usepackage{textcomp}
\usepackage[german,english]{babel}

\newcommand{\dcauthorpre}{Dipl.-Phys. } 
\newcommand{\dcauthorsurname}{Gehrmann } 
\newcommand{\dcauthorname}{Bernd } 
\newcommand{\dcauthoradd}{geboren am 26.01.1972 in Gelsenkirchen } 
\newcommand{\dctitle}{The step scaling function of QCD at negative flavor number} 
\newcommand{\dcsubtitle}{\quad }
\newcommand{\dcapprovala}{Dr. Rainer Sommer} 
\newcommand{\dcapprovalb}{Prof. Dr. Istvan Montvay} 
\newcommand{\dcapprovalc}{Prof. Dr. Ulli Wolff}
\newcommand{\dcdegree}{doctor rerum naturalium\\ (Dr. rer. nat.)} 
\newcommand{\dcsubject}{Physik} 
\newcommand{\dcfaculty}{Mathematisch-Naturwissenschaftlichen Fakult\"at I}
\newcommand{\dcuniversity}{Humboldt-Universit\"at zu Berlin}
\newcommand{\dcdean}{Prof. Dr. Bernhard Ronacher}
\newcommand{\dcpresident}{Prof. Dr. J\"urgen Mlynek}
\newcommand{\dcdatesubmitted}{6. M\"arz 2002} 
\newcommand{\dcdateexam}{5. Juni 2002} 
\newcommand{\dckeydea}{Gitter-QCD}
\newcommand{\dckeydeb}{Schr\"odinger-Funktional}
\newcommand{\dckeydec}{Step-Scaling-Funktion}
\newcommand{\dckeyded}{Bermion-Modell}
\newcommand{\dckeywordsde}{\vspace*{2cm} \\{\bf{Schlagw\"orter:}}\\ \dckeydea, \dckeydeb, \dckeydec, \dckeyded \\}
\newcommand{\dckeyena}{Lattice QCD}
\newcommand{\dckeyenb}{Schr\"odinger functional}
\newcommand{\dckeyenc}{Step scaling function}
\newcommand{\dckeyend}{Bermion model}
\newcommand{\dckeywordsen}{\vspace*{2cm} \\{\bf{Keywords:}}\\ \dckeyena,
  \dckeyenb, \dckeyenc, \dckeyend \\}
\author{von \\ \dcauthorpre  \dcauthorname  \dcauthorsurname  \\ \dcauthoradd}

\title{ \vspace{-5cm}\dctitle \\ 
\vspace{0.5cm}
\large{\dcsubtitle} \\ 
\vspace{0.5cm} {\Large{D I S S E R T A T I O N }}\\ 
\vspace{0.5cm} \large{zur Erlangung des akademischen Grades \\ 
\dcdegree\\ im Fach \dcsubject \\ 
\vspace{0.5cm} eingereicht an der \\ 
\dcfaculty \\ 
\dcuniversity \\}}
\date{\vspace{0.5cm}
\raggedright{
Pr\"asident der Humboldt-Universit\"at zu Berlin:\\
\dcpresident \vspace{-0.3cm}
}\vspace{0.5cm}\\
\raggedright{
Dekan der \dcfaculty:\\
\dcdean \vspace{-0.3cm}
}\vspace{0.5cm}\\
\raggedright{
Gutachter:
\begin{enumerate} 
\item{\dcapprovala} \vspace{-0.3cm}
\item{\dcapprovalb} \vspace{-0.3cm}
\item{\dcapprovalc} \vspace{-0.3cm}
\end{enumerate}} \vspace{0.5cm}
\raggedright{
\begin{tabular}{lll}
Tag der m\"undlichen Pr\"ufung: & & \dcdateexam
\end{tabular}}\\ 
}

\newcommand{\beq}{\begin{equation}}
\newcommand{\eeq}{\end{equation}}
\newcommand{\beqn}{\begin{eqnarray}}
\newcommand{\eeqn}{\end{eqnarray}}
\newcommand{\bequ}{\begin{displaymath}}
\newcommand{\eequ}{\end{displaymath}}
\newcommand{\beqnu}{\begin{eqnarray*}}
\newcommand{\eeqnu}{\end{eqnarray*}}
\newcommand{\eqref}[1]{(\ref{#1})}
\newcommand{\xvec}{\mathbf{x}}
\newcommand{\yvec}{\mathbf{y}}
\newcommand{\zvec}{\mathbf{z}}
\newcommand{\muhat}{\hat\mu}
\newcommand{\nuhat}{\hat\nu}
\newcommand{\khat}{\hat k}
\newcommand{\zerohat}{\hat 0} 
\newcommand{\gbar}{\bar g} 
\newcommand{\mbar}{\kern1pt\overline{\kern-1pt m\kern-1pt}\kern1pt}

\newcommand{\calL}{{\cal L}}
\newcommand{\calO}{{\cal O}}
\newcommand{\calZ}{{\cal Z}}
\newcommand{\SUN}{{\rm SU}(N)}
\newcommand{\SUtwo}{{\rm SU}(2)}
\newcommand{\SUthree}{{\rm SU}(3)}
\newcommand{\sun}{{\rm su}(N)}
\newcommand{\suthree}{{\rm su}(3)}
\newcommand{\Othree}{{\rm O}(3)}
\newcommand{\Uone}{{\rm U}(1)}
\newcommand{\mq}{m_{\rm q}}
\newcommand{\mc}{m_{\rm c}}
\newcommand{\mR}{m_{\rm R}}
\newcommand{\gR}{g_{\rm R}}
\newcommand{\csw}{c_{\rm sw}}
\newcommand{\ct}{c_{\rm t}}
\newcommand{\cA}{c_{\rm A}}
\newcommand{\bA}{b_{\rm A}}
\newcommand{\bP}{b_{\rm P}}
\newcommand{\bg}{b_{\rm g}}
\newcommand{\bm}{b_{\rm m}}
\newcommand{\Zg}{Z_{\rm g}}
\newcommand{\Zm}{Z_{\rm m}}
\newcommand{\ZA}{Z_{\rm A}}
\newcommand{\ZP}{Z_{\rm P}}
\newcommand{\fA}{f_{\rm A}}
\newcommand{\fP}{f_{\rm P}}
\newcommand{\Sg}{S_{\rm g}}
\newcommand{\Sf}{S_{\rm f}}
\newcommand{\Sb}{S_{\rm b}}
\newcommand{\cttilde}{\tilde{c}_{\rm t}}
\newcommand{\alphas}{\alpha_{\rm s}}
\newcommand{\rmO}{{\rm O}}
\newcommand{\rmZ}{{\rm Z}}
\newcommand{\Nf}{N_{\rm f}}
\newcommand{\Nor}{N_{\rm or}}
\newcommand{\tauint}{\tau_{\rm{int}}}
\newcommand{\Lmax}{L_{\rm max}}
\newcommand{\Scost}{S_{\rm cost}}
\newcommand{\Mcost}{M_{\rm cost}}
\newcommand{\Pacc}{P_{\rm acc}}
\newcommand{\Nbin}{N_{\rm bin}}
\newcommand{\Munimpr}{M_{\rm unimproved}}
\newcommand{\MSbar}{{\rm \overline{MS\kern-0.05em}\kern0.05em}}
\newcommand{\MS}{{\rm MS}}
\newcommand{\MOM}{{\rm MOM}}
\newcommand{\Tr}{{\rm Tr}}
\newcommand{\diag}{{\rm diag}}
\renewcommand{\Re}{{\rm Re}}
\renewcommand{\Im}{{\rm Im}}

\begin{document}

\maketitle
\thispagestyle{empty}
\cleardoublepage

\selectlanguage{english}
\abstract 
As a computationally less costly test case for full QCD, we investigate an
$\SUthree$ Yang-Mills theory coupled to a bosonic spinor field. This theory
corresponds to QCD with minus two quark flavors and is known as the bermion
model. Our central object of interest is the step scaling function which
describes the scale evolution of the running coupling in the Schr\"odinger
functional scheme. With the help of a non-perturbative recursive finite size
technique, it can be used to determine the $\Lambda$ parameter, which
characterizes the coupling at high energy, from experimental input at low
energies.

We study in detail the lattice artefacts and the continuum extrapolation of
the step scaling function from lattice simulations when $\rmO(a)$ improvement
according to the Symanzik programme is used. Our results are compared to the
unimproved bermion and dynamical fermion cases, and to renormalized
perturbation theory in the continuum limit.

For the bermion model, we also examine the step scaling function with massive
quarks. According to the Appelquist-Carazzone theorem the contributions from
matter fields are expected to vanish for large masses, such that the step
scaling function converges to the pure gauge theory case. If one wants to
connect non-perturbatively different effective theories with different numbers
of active quarks over flavor thresholds, lattice artefacts should be
reasonably small.  In order to test the feasibility of such a method, we
investigate the step scaling function and its lattice artefacts for several
values of the mass.

For the Monte Carlo simulation of improved bermions, we develop a suitable
algorithm and compare its performance with unimproved bermions and full QCD.
As a preparative study, we compare the efficiency of algorithms in pure
gauge theory. \\
\dckeywordsen 
\clearpage
\thispagestyle{empty}
\quad
\clearpage

\selectlanguage{german} 
\abstract 
Wir untersuchen eine $\SUthree$ Yang-Mills-Theorie mit einer Kopplung an ein
bosonisches Spinorfeld. Diese als Bermion-Modell bekannte Theorie entspricht
formal QCD mit minus zwei Quark-Flavors. Gegen\"uber der vollen QCD erfordert
sie wesentlich weniger Computerzeit und ist deshalb als relativ
kosteng\"unstiges Testmodell geeignet. Im Mittelpunkt unseres Interesses steht
die Step-Scaling-Funktion, die die Skalenabh\"angigkeit der laufenden Kopplung
im Schr\"odinger-Funktional-Renormierungsschema beschreibt. Mit Hilfe einer
nicht-perturbativen Finite-Size-Technik kann sie benutzt werden, um den
$\Lambda$-Parameter, der die Kopplung bei hohen Energien charakterisiert, aus
experimentellen Daten bei niedrigen Energien zu bestimmen.

Wir studieren im Detail die Gitterartefakte und die Kontinuumsextrapolation
der aus Gittersimulationen bestimmten Step-Scaling-Funktion, wenn
$\rmO(a)$-Verbesserung nach Symanzik verwendet wird. Unsere Resultate stellen
wir dem Fall von unverbesserten Bermionen und dynamischen Fermionen
gegen\"uber, und vergleichen im Kontinuumslimes mit renormierter
St\"o\-rungs\-theo\-rie.

Weiterhin betrachten wir im Bermion-Modell die Step-Scaling-Funktion mit
massiven Quarks. Nach dem Appelquist-Carazzone-Theorem erwartet man, da{\ss}
Beitr\"age von Materiefeldern mit ansteigender Masse verschwinden, so da{\ss}
die Step-Scaling-Funktion gegen den Fall reiner Eichtheorie konvergieren
sollte. Wenn man nicht-perturbativ verschiedene effektive Theorien mit
verschiedener Anzahl von aktiven Quarks \"uber Massenschwellen hinweg
verbinden will, sollten Gitterartefakte klein sein. Um die Durchf\"uhrbarkeit
einer solchen Methode zu testen, untersuchen wir die Step-Scaling-Funktion und
ihre Gitterartefakte f\"ur verschiedene Massen.

F\"ur die Monte-Carlo-Simulation von verbesserten Bermionen entwickeln wir
einen geeigneten Algorithmus und vergleichen seine Effizienz mit unbesserten
Bermionen und mit voller QCD. Als vorbereitende Studie vergleichen wir die 
Effizienz verschiedener Algorithmen in reiner Eichtheorie.
\dckeywordsde
\clearpage
\thispagestyle{empty}
\quad
\clearpage

\selectlanguage{english}
\pagenumbering{roman}
\tableofcontents
\cleardoublepage

\pagenumbering{arabic}

\chapter{Introduction}

In the standard model of particle physics, Quantum Chromodynamics (QCD) is the
theory of the strong interaction. It covers the interaction between six
flavors of \emph{quarks} which are the constituents of hadronic matter. QCD is
a gauge theory based on the non-abelian $\SUthree$ gauge group. The fermionic
matter fields, the quarks, carry a quantum number called ``color'' and
transform according to the fundamental representation of the group. The gauge
fields transform according to the adjoint representation and describe an octet
of \emph{gluons}.

Classically, the structure of the strong interaction is relatively simple
compared to the electroweak theory. The gauge group is unbroken and the quark
states that participate in the strong interaction coincide with the mass
eigenstates. The only input parameters are the strong coupling constant and
the quark masses. As a quantum field theory, QCD nevertheless exposes a lot of
interesting phenomena, part of which are insuffiently understood even after
thirty years of research.

The quantization of continuum field theories leads to divergences that must be
removed by regularizing the theory. One can obtain finite results for physical
observables by a \emph{renormalization} of the parameters of the theory, which
become functions of a renormalization scale. The running coupling $\alphas$ of
QCD plays an essential role in the characterization of the theory.

An important property of non-abelian gauge theories is \emph{asymptotic
  freedom}. At high energies, the coupling vanishes asymptotically and quarks
behave like free particles. Before the discovery of QCD, the phenomenological
parton model used for deep inelastic scattering processes reflected this
behavior. The application of perturbation theory to QCD successfully describes
the corrections to the Bjorken scaling law that follows from the assumption of
free particles.

On the other hand, the behavior of the running coupling is such that the
coupling grows at low energies. As a consequence, the traditional perturbative
methods in Quantum Field Theory, which were developed in the framework of
Quantum Electrodynamics (QED), break down in this regime. A new feature
arising at large distances is the phenomenon of \emph{confinement}, which
expresses the observation that free quarks do not exist in nature. Hadronic
matter appears only in the form of color singlets, which are grouped into
mesons (quark-antiquark hadrons) and baryons (hadrons consisting of three
quarks).

A decisive step for the understanding of low energy QCD was done by Wilson in
1974, who formulated this theory as a lattice regularized euclidean Quantum
Field Theory. In this regularization scheme, matter fields are defined on the
sites of a hypercubic space-time lattice and gauge fields are parametrized as
parallel transporters between the sites. The continuum limit is obtained by
decreasing the lattice spacing to zero. This approach opened the possibility
of applying methods from the toolbox of statistical physics, like the strong
coupling (high temperature) expansion. Today, Monte Carlo simulations on the
lattice have become one of the most popular methods for the non-perturbative
investigation of QCD.

\begin{figure}[htb]
  \begin{center}
    \epsfig{file=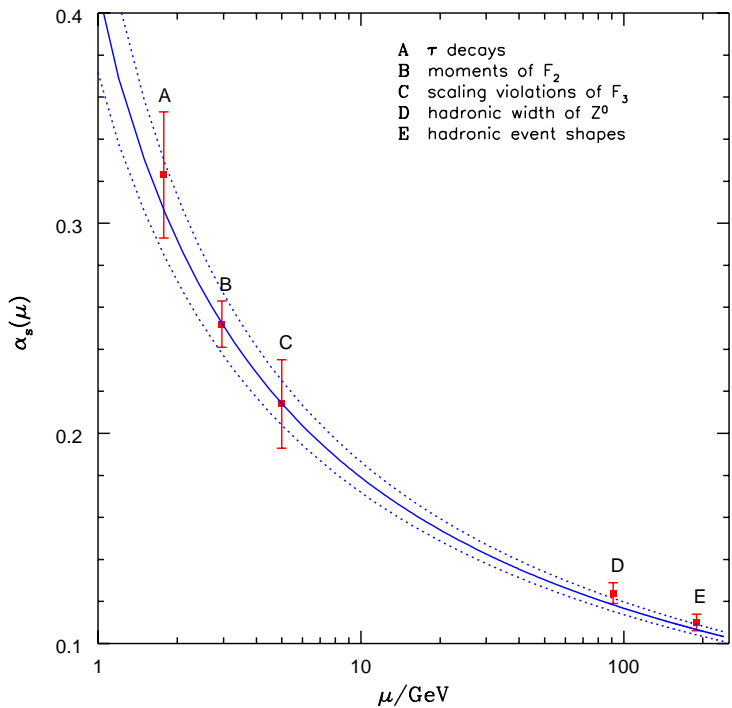,height=11cm,width=11cm}
    \caption{\sl Experimental values for $\alphas$, compared with the world average.}
    \label{fig:RunningCoupling}
  \end{center}
\end{figure}

The scale dependences of the running coupling and the running quark masses in
QCD are described by the renormalization group equations. As these are
differential equations of first order, in QCD with $\Nf$ quark flavors there
are $\Nf+1$ integration constants which have to be determined from experiment.
One way to express the free parameter of the running coupling is the $\Lambda$
parameter which is a measure for the asymptotic decay of the coupling at high
energies. Another way -- used for example in the particle data book \cite{PDB}
-- is to compute the coupling at a reference scale, conventionally taken as
the $\rmZ^0$ mass.

Figure~\ref{fig:RunningCoupling} shows an overview of some typical
experimental measurements of the QCD coupling $\alphas$ in the $\MSbar$
renormalization scheme at different scales \cite{hepex0004021}.  In order to
compare results obtained at different scales, one evolves the coupling to the
$\rmZ^0$ mass by employing the renormalization group equation in the $\MSbar$
scheme, computed in perturbation theory up to 4-loop.  In the above reference,
a world average of
\begin{eqnarray*}
  \bar \alphas(M_{\rmZ^0}) = 0.1184 \pm 0.0031
\end{eqnarray*}
is obtained. We have plotted the running $\MSbar$ coupling for this average
and its error as continuous respectively dashed lines\footnote{The running
  $\MSbar$ coupling has been computed using the {\rm RunDec} program from
  \cite{hepph0004189}.}.

It turns out that the experimental results for $\alphas(M_{\rmZ^0})$ obtained
from measurements at different scales agree well with each other. But it is
also clear that at some scale, $\alphas$ is not a small expansion parameter
anymore and perturbation theory must break down. At lower energies and larger
distances, confinement occurs and non-perturbative contributions like
instantons are known to play an important role. But not only does perturbation
theory fail in qualitatively describing these phenomena. Even at scales where
it apparently converges, it is impossible to specify the systematic error
caused by using perturbation theory up to a low finite order for the evolution
of the coupling to the reference scale. A renormalization scheme that is only
perturbatively defined is not suitable at low energies.

Still, one can investigate the low energy properties of QCD with the help of
computer simulations. In this context, QCD is regularized by a hypercubic
lattice. After the free parameters in the theory have been fixed by requiring
certain quantities - like hadron masses - to take their physical value,
further observables can be measured in Monte Carlo simulations and compared to
experimental results. While in principle, this method provides a
non-perturbatively defined scheme which is suitable in a certain low energy
range, the size of lattices which can be simulated in practice puts a tight
constraint on the energy scale which can be reached.

So while different renormalization approaches can be used to study different
ranges of scales, it is not a priori clear how the parameters in the different
schemes are connected to each other.  For a test of QCD at all scales, it is
indispensable to use a scheme which is defined non-perturbatively, practically
tractable and which does not use any uncontrolled approximations. A method for
this was proposed by L\"uscher, Weisz and Wolff \cite{LuescherWeiszWolff}.

The fundamental concept of this approach is that one does not need to
accommodate all relevant scales on a single lattice. Instead, one uses a
finite volume scheme where the coupling runs with the space-time volume. The
coupling can be tracked recursively along increasing scales using the
\emph{step scaling function}, which describes the change of the coupling under
discrete changes of the scale. The step scaling function at a single point can
be determined by simulating lattice pairs with decreasing lattice spacing and
extrapolating to the continuum. No large scale ratios occur within each
lattice pair, and the extrapolation to the continuum can be done with
relatively moderate lattice sizes.

This strategy can be used for any asymptotically free theory. First tests were
made with the nonlinear $\Othree$ model in two dimensions
\cite{LuescherWeiszWolff}. Later it has been generalized to pure $\SUtwo$ gauge
theory \cite{heplat9207009} and pure $\SUthree$ gauge theory \cite{heplat9309005}
within the framework of the ALPHA collaboration.  While in principle, there is
considerable freedom in the precise choice of boundary conditions and
definition of the coupling, for non-abelian gauge theories a special choice
has solidified where the theory is defined on a cylinder with Dirichlet
boundary conditions in the time direction.  The fields at the lower and upper
boundary induce a classical background field on the cylinder. A coupling can
then be defined as the response of the system towards changes of the
background field.  The partition function can be quantum mechanically
interpreted as the propagation kernel for going from the initial configuration
at the lower boundary to the final configuration at the upper boundary. For
this reason, this renormalization scheme is also dubbed the
\emph{Schr\"odinger functional scheme}.

In the context of QCD, the ALPHA programme has been fully implemented in the
$\SUthree$ pure gauge theory, which can also be understood as the quenched
approximation to QCD. This includes several aspects. At low energy, a
reference scale is set that allows to express quantities like the $\Lambda$
parameter in physical units. The Sommer scale $r_0$
\cite{heplat9310022,heplat9806005} is derived from the effective potential
between static quarks. It can be determined experimentally by measurements of
charmonium and bottomonium bound states.  At intermediate scales, pairs of
lattices with decreasing lattice spacing were simulated and the step scaling
function was then extrapolated to the continuum. Finally, when a scale is
reached where perturbative behavior can be shown to set in, one applies
perturbation theory to compute the $\Lambda$ parameter in the Schr\"odinger
functional scheme. A 2-loop calculation is required to confirm that no
uncontrolled error is introduced in this step \cite{heplat9911018}.

In the quenched approximation, the fermion determinant in the partition
function is formally set to a constant. Physically, this corresponds to
freezing the dynamics of quarks and neglecting their vacuum polarization effects.
The sole motivation for this approximation is the radically reduced cost of
computer simulations. While this model has turned out to yield surprisingly
good results in many areas like hadron spectroscopy and matrix elements, the
quenched approximation is of course not a substitute for full QCD. In
particular, certain phenomena like string breaking or the $\eta-\eta'$ mass
splitting are absent. Quantitatively, systematic deviations from
experiment up to 10\% are seen in the hadron spectrum \cite{heplat9904012}.

Meanwhile, the ground has been layed out for the extension of the ALPHA
programme to a pair of massless dynamical fermions. First results for the
running coupling have been published in \cite{heplat0105003}. The main
practical problem is the high cost of numerical simulations with dynamical
fermions.  Therefore, the range of simulated lattices does not yet reach very
close to the continuum limit, and simulations on larger lattices are
desirable. Also, an element still missing in the programme is the
determination of a physical scale as a substitute for $r_0$.

As the efficiency of Monte Carlo algorithms for dynamical fermions decreases
with a high inverse power of the lattice spacing, it is not possible to
perform simulations arbitrarily near to the continuum limit. In order to
extrapolate results to vanishing lattice spacing, one wants to accelerate the
convergence to the continuum as much as possible. Different solutions for this
problem have been invented. The aim of the \emph{perfect action} approach is
to apply theories that yield continuum results already at finite lattice
spacing \cite{heplat9308004}. The theoretical background for this method is
Wilson's renormalization group. In a computer implementation, approximations
have to be made to parametrize the action. Another problem is that the fixed
point actions used today are not ``quantum perfect'', i.e. they yield
continuum results only at vanishing value of the gauge coupling.

A complementary approach is the Symanzik programme. The idea here is to cancel
lattice artefacts order by order in the lattice spacing $a$.  This is done by
adding certain -- in the language of the renormalization group irrelevant --
terms to the action and to operators and adjusting their coefficients
appropriately.  In practice, the full non-perturbative determination of
improvement coefficients is a big task. Therefore currently only $\rmO(a)$
improvement is feasible, i.e.  observables converge to the continuum limit
with $\rmO(a^2)$ artefacts.

Symanzik's improvement programme is based on statements that are valid for
asymptotically small lattice spacings and are derived from perturbation
theory. In particular, one does not know when higher order terms in $a$ become
negligible and the asymptotic behavior sets in \cite{heplat0006021}.

\vspace{3mm}

In this thesis, we are going to study the approach to the continuum limit in
the Schr\"odinger functional and test whether perturbative expectations about
lattice artefacts hold. The aim of this investigation is to put the study of
the running coupling for dynamical fermions on a firmer ground and justify the
used continuum extrapolation and its error estimate. Since full QCD is so
notoriously costly to simulate, we study the step scaling function in a
different theory which also goes beyond the quenched approximation. By setting
the number of quark flavors to $\Nf=-2$, we obtain a Yang-Mills theory coupled
to a bosonic spinor field. This theory, also known as the \emph{bermion
  model}, has a local interaction in terms of Bose fields and is therefore
much cheaper to simulate.

Another topic which we treat with the bermion model is the massive step
scaling function. As soon as one wants to study QCD with quarks that have
masses, one has to cope with additional lattice artefacts which are especially
sizable when the mass becomes comparable to the inverse lattice spacing. An
attractive possibility for saving the evaluation of the running coupling from
this danger is to make use of the decoupling of heavy quarks. Since according
to the Appelquist-Carazzone theorem, quarks with masses very large compared to
some scale do not contribute to the physics at this scale, one may drop these
quarks from the theory in a certain energy regime.  Depending on the mass
cutoff where one matches the theories with and without a heavy quark, one
introduces a systematic error, which has previously been estimated in
perturbation theory \cite{heplat9508012}. Here we study the dependence of the
step scaling function in the bermion model on the mass, which should be an
indication whether the decoupling follows perturbative expectations.

\chapter{Theory}
\label{chap:Theory}

In this chapter, we are going to introduce the basic concepts used in this
thesis, and define the model used in later chapters. It goes without saying
that we cannot discuss in depth all issues involved. For an introduction into
gauge theories and perturbative renormalization we refer to \cite{BailinLove}.
The framework and terminology of the lattice regularization of quantum field
theories is layed out in \cite{MontvayMuenster}. Non-perturbative
renormalization and $\rmO(a)$ improvement are discussed in \cite{hepph9711243}
and \cite{heplat9802029}. These references also review the Schr\"odinger
functional approach.

\section{Perturbative renormalization: a brief reminder}

In perturbation theory, where transition amplitudes and other quantities are
expressed by Feynman graphs, the need for renormalization arises in loop
diagrams, which are divergent when evaluated in a naive way.  The first step
in the renormalization procedure is to construct a Lagrangian from the bare
one $\calL_B$ and additional counterterms,
\beq
  \calL_B \rightarrow \calL = \calL_B + \delta \calL.
\eeq
The bare Lagrangian develops the usual divergences, and $\delta \calL$ creates
additional diagrams. Now one has to \emph{regularize} the divergent diagrams.
A suitable method for gauge theories is the \emph{dimensional regularization}
\cite{DimensionalRegularization}, in which one continues the theory
analytically in $D=4-2\epsilon$ dimensions. Its advantage e.g. over momentum
cutoffs is that gauge invariance is manifestly retained (for parity conserving
theories). Divergences then emerge as poles in the limit $\epsilon=0$, while
convergent integrals are unaffected.

The poles can be canceled by choosing the coefficients of the counterterms
appropriately. This choice is not unambiguous, and possible choices differ by
finite amounts. The precise set of rules for fixing the coefficients is
called a \emph{renormalization scheme}.

In the minimal subtraction (\MS) scheme \cite{MS}, only the poles are
subtracted. It is a member of a larger class of schemes which are mass
independent, i.e. the renormalization condition does not depend on the
renormalized masses. A very popular member of this class is the $\MSbar$
scheme \cite{MSbar}, in which further terms are subtracted that frequently
appear in Feynman graphs. An example for a mass dependent scheme is the
momentum (\MOM) scheme \cite{MOM}, which is defined by imposing boundary
conditions on the Green's functions in momentum space.

In principle, it can happen for an arbitrary Lagrangian that different kinds
of divergences appear in every order of perturbation theory.  An infinite
number of coefficients would have to fixed then. Such a theory would have
little predictive power\footnote{It might still serve as a low-energy
  effective theory. The Fermi theory of weak interactions is an example for
  this \cite{Kaplan}.}. A theory is called \emph{renormalizable} if only a
finite number of counterterms is necessary. In such a theory, the renormalized
parameters can be adjusted to take their physical values. Once this has been
done, one can make predictions.

Fortunately, with gauge theories we are in a comfortable position. In his
famous articles \cite{tHooftProof1,tHooftProof2}, 't Hooft has proven the
renormalizability of unbroken and broken non-abelian gauge theories. The
renormalized theory is gauge invariant, and the counterterm structure is quite
simple in that all necessary terms are already present in the bare Lagrangian.

\section{Running coupling and masses}

In the course of dimensional regularization, one has to express dimensionful
quantities by some scale $\mu$ not present in the Lagrangian itself.
Similarly, other regularization schemes introduce some cutoff scale in order
to render integrals finite. Consequently, renormalized parameters unavoidably
acquire a dependency on a \emph{renormalization scale}.

In the QCD Lagrangian, the bare coupling $g_0$ and the bare quark masses $m_{0,i}$ for
the flavors $i=1\ldots \Nf$ are the bare parameters of the theory. These
parameters are fixed in a renormalization scheme such that a corresponding
number of physical observables take their prescribed values. After this
renormalization, there is no freedom any more, and other renormalized
parameters can be predicted.

A natural question to ask is ``How do the renormalized Green's functions
change with the renormalization scale when the bare ones are held fixed?''
This question is answered by the renormalization group equations. In the
following, we assume a mass-in\-de\-pen\-dent scheme, i.e. a scheme where the
definition of the renormalized parameters does not depend on the quark masses.
In that case, the renormalization group equations assume a simpler form.

For the coupling, one is led to a description of the scale dependence by the
Callan-Symanzik $\beta$-function,
\beq
  \mu \frac{\partial \gR}{\partial\mu} = \beta(\gR).
  \label{eq:RenormalizationGroupEquation}
\eeq
The $\beta$-function has an asymptotic expansion
\beq
  \beta(\gR) \stackrel{\gR\to 0}{=} -\gR^3 ( b_0 + b_1 \gR^2 + b_2 \gR^4 + \cdots ),
  \label{eq:BetaExpansion}
\eeq
where the first two coefficients are universal,
\beqn
  b_0 &=& \frac{1}{(4\pi)^2} \left(11 - \frac 2 3 \Nf \right) \nonumber\\
  b_1 &=& \frac{1}{(4\pi)^2} \left(102 - \frac{38}{3} \Nf \right),
\eeqn
and the higher order coefficients depend on the scheme. For $\Nf \le 16$, the
expansion of the $\beta$-function obviously begins with a negative term, i.e.
in the asymptotic high energy regime, the coupling decreases logarithmically
with increasing energy.  This property is known as \emph{asymptotic freedom}.
It reflects the observation that at high energies, quarks behave like free
particles.

In a way similar to the coupling, the scale dependence of the renormalized
masses is described by the equations
\beq
  \mu \frac{\partial m_{{\rm R},i}}{\partial \mu} 
    = \tau(\gR) m_{{\rm R},i}, \quad i=1\ldots \Nf,
\eeq
where the $\tau$-function has an expansion
\beq
  \tau(\gR) = -\gR^2 ( d_0 + d_1 \gR^2 + d_2 \gR^4 + \cdots)
\eeq
with a universal coefficient
\beq
  d_0 = \frac{8}{(4\pi)^2}.
\eeq
In the $\MSbar$ scheme, the $\beta$- and $\tau$-functions are known up to
4-loop in perturbation theory \cite{hepph9701390,hepph9703278}.

The asymptotic solutions of the renormalization group equations are
\beqn
  \gR^2(\mu) &\stackrel{\mu\to\infty}{=}&
  \frac{1}{2 b_0 \log(\mu/\Lambda)} \nonumber\\
  m_{{\rm R},i}(\mu) &\stackrel{\mu\to\infty}{=}&
  \frac{M_i}{\left[ \log(\mu/\Lambda) \right]^{d_0/2b_0}}.
\eeqn
The integration constants $\Lambda$ and $M_i$ can be regarded as the
fundamental parameters of QCD. This means, once these parameters are known,
they uniquely fix all running parameters at all scales.

The $\Lambda$ parameter depends on the renormalization scheme, but can be
exactly transformed between different schemes through the 1-loop coefficient
relating the couplings in those schemes. The $M_i$ are scheme independent, and
are therefore also called renormalization group invariant quark masses.

In order to obtain the fundamental parameters of QCD from the renormalized
parameters at a finite scale, one has to integrate the renormalization group
equations. This connection is given by the exact relations
\beqn
  \Lambda &=& \mu (b_0 \gR(\mu)^2)^{-b_1/2b_0^2} 
    \exp\left( - \frac{1}{2b_0 \gR(\mu)^2} \right) \times \nonumber\\
&&\quad\times
    \exp\left\{ -\int_{0}^{\gR(\mu)} dx \,
      \left[ \frac{1}{\beta(x)} + \frac{1}{b_0 x^3} 
        - \frac{b_1}{b_0^2 x} \right] \right\}  \nonumber\\
  M_i &=& m_{{\rm R},i} (2 b_0 \gR(\mu)^2)^{-d_0/2b_0} \times \nonumber\\
&&\quad\times
    \exp\left\{ -\int_{0}^{\gR(\mu)} dx \,
      \left[ \frac{\tau(x)}{\beta(x)} - \frac{d_0}{b_0 x}
        \right] \right\}.
  \label{eq:Lambda}
\eeqn
In practice, one inserts the $\beta$- and $\tau$-functions to a finite order
of perturbation theory here, provided that the perturbative behavior has
already set in at the scale $\mu$.

\section{Non-perturbative renormalization}

As described in the introduction, an important aim is to compute the running
coupling at all scales. A natural way to achieve this is to start with a
non-perturbative scheme that is based on a lattice regularized theory.  In
order to eliminate the bare parameters of the theory in favor of physical ones
at low energies, one uses this theory to compute hadronic observables like the
pion decay constant $F_\pi$ and hadron masses.  Then the
coupling is evolved to high energies and compared to experiments via jet cross
sections etc.  The connection between low-energy hadronic schemes and
perturbative energies however involves scales very different from each other,
thus imposing heavy demands on lattice simulations: on the one hand, one must
choose the lattice cutoff $a^{-1}$ away from the energy scale $\mu$, in order
to avoid large lattice artefacts hampering an extrapolation to the continuum.
The limiting quantity here is the energy scale at which the connection to
perturbation theory is made, which should be e.g. around 10~GeV. On
the other hand, the system size $L$ should be large enough to avoid finite
size effects.  The relevant energy scale for this is the confinement scale at
about 0.4~GeV in the quenched approximation, or even the pion mass $m_\pi$ at
about 0.14~GeV.  Together, these constraints imply simulations on lattices
with linear extent $L/a \gg 70$, which is difficult to achieve in practice.

\section{Strategy}

The Schr\"odinger functional scheme uses a trick to overcome the problem of
widely disparate scales: instead of regarding finite size effects as a
problem that is distorting physical states of the finite system compared to
the infinite system, one considers the finite volume behavior of the system as
origin of interesting observables.  This is analogous -- though not equivalent --
with the computation of critical exponents in statistical systems, which can
be extracted from the change of observables with the box size.

In a finite-volume renormalization scheme, the running of the coupling with
the energy scale is identified with the running of a coupling $\gbar(L)$ with
the system size $L=\mu^{-1}$. One starts at a low energy scale $\Lmax$ which is
fixed by requiring that the coupling takes some value,
\beq
  \gbar(\Lmax) = \mbox{prescribed value}.
\eeq
The physical value of $\Lmax$ has to be connected to a physical scale by
computing for instance $F_\pi$ in units of $\Lmax^{-1}$.

Then this coupling is traced non-perturbatively to energies high enough for
perturbation theory to apply. Finally, one can use the relations
\eqref{eq:Lambda} to calculate the $\Lambda$ parameter. This can be
transformed to the $\Lambda$ parameter in the $\MSbar$ scheme with a 1-loop
order calculation.

The last step in this technique actually corresponds to a transformation of a
small volume coupling to the infinite volume coupling $\alpha_\MSbar$. Since
these couplings are in one-to-one correspondence, this matching is entirely
justified if they are just small enough for perturbation theory to be applied.

The important point to notice here is that although finite-volume quantities
have been used in the scale evolution of the coupling, there is no reference
to the volume in the final result anymore. This is achieved by the standard
assumption that QCD physics is described by the same Lagrangian regardless of
the context in which it is used. In particular, the finite size effects of the
theory are predicted by the Lagrangian, as is for example the energy
dependence of scattering processes.

Now we can introduce an additional concept used to evolve the running
coupling. As introduced before, this evolution is given by the renormalization
group equation \eqref{eq:RenormalizationGroupEquation}. Thus, the coupling at
a scale $2L$ is related to the coupling at $L$ through a unique function,
called the \emph{step scaling function}\cite{LuescherWeiszWolff},
\beq
  \gbar^2(2L) = \sigma(\gbar^2(L)).
\eeq
With the help of this function, the coupling can be computed recursively at
scales $2^{-k} \Lmax$ beginning at a starting point $\Lmax$.

The step scaling function can be computed with the help of Monte Carlo
simulations. First, one simulates a lattice with $L/a$ lattice sites in each
direction and tunes the coupling to the desired value. Then one simulates a
lattice with twice the extent, $2L/a$, using the same bare parameters. The
coupling obtained from this simulation is an approximation $\Sigma(u,a/L)$ of
the step scaling function $\sigma(u)$. We expect this to have $\rmO(a)$
lattice artefacts. With the $\rmO(a)$ improvement programme discussed later,
we assume (apart from logarithmic corrections),
\beq
  \sigma(u) = \Sigma(u, a/L) + \rmO(a^2).
\eeq 
Therefore, by computing $\Sigma(u, a/L)$ for a number of lattice sizes $L/a$,
one can obtain $\sigma(u)$ by extrapolating to the continuum.

A notable property of this recursive scheme is that by identifying
$\mu=L^{-1}$, the constraints on the required lattice sizes are significantly
relaxed.  Instead of
\beq
  L \gg m_\pi^{-1} \gg \mu^{-1} \gg a,
\eeq
we only need
\beq
  L \gg a
\eeq
for an extrapolation to the continuum limit.

The step scaling function can be understood as an integrated version of the
$\beta$-function for finite changes of the scale. The couplings at $L$ and
$2L$ are related through an integral,
\beq
  \log 2 = \int_{L}^{2L} \frac{dL'}{L'}
         = -\int_{\gbar(L)}^{\gbar(2L)} \frac{dx}{\beta(x)}
  \label{eq:BetaIntegral}
\eeq
such that the perturbative expansion of $\sigma$ can be derived from the
expansion \eqref{eq:BetaExpansion} of the $\beta$-function,
\beq
  \sigma(u) = u + s_0 u^2 + s_1 u^3 + s_2 u^4 + \rmO(u^4),
  \label{eq:SigmaExpansion}
\eeq
where
\beqn
  s_0 &=& 2 \log 2 \, b_0 \nonumber\\
  s_1 &=& (2 \log 2)^2 b_0^2 + 2 \log 2 \, b_1 \nonumber\\
  s_2 &=& (2 \log 2)^3 b_0^3 + (2 \log 2)^2 \frac 5 2 b_0 b_1 
          + 2 \log 2 \, b_2.
\eeqn
In addition to the truncated $n$-loop expansion of the step scaling function
\beq
  \sigma^{n\rm -loop}(u) = u + \sum_{k=1}^{n} s_{k-1} u^{k+1},
  \label{eq:SigmaPerturbative}
\eeq
we define another perturbative step scaling function as the solution of
$\eqref{eq:BetaIntegral}$ with a truncated $\beta$-function,
\beq
  \log 2 = \int_{\sqrt{u}}^{\sqrt{\hat\sigma^{n\rm -loop}(u)}}
  \! dx \left( \sum_{k=1}^{n} b_{k-1} x^{k+2} \right)^{-1}.
  \label{eq:SigmahatPerturbative}
\eeq

The functions $\sigma^{n\rm -loop}(u)$ and $\hat \sigma^{n \rm -loop}(u)$
differ by terms of order $u^{n+2}$ from each other and from the exact function
$\sigma(u)$. In practice, $\hat\sigma^{n \rm -loop}$ seems to be a better
approximation for the exact function $\sigma$. The difference between both
variants may be used as an estimate of the neglected higher order terms.

\section{Improvement}

While a finite size technique nicely solves the problem of disparate scales,
it still inherits a problem from QCD that makes the precise measurement of
observables difficult.  In order to determine continuum quantities, one must
compute them for several values of the lattice spacing $a$ and then
extrapolate to the continuum limit $a\to 0$.  Unfortunately, this is also the
limit in which Monte Carlo algorithms suffer from critical slowing down, i.e.
the cost of conventional algorithms grows with a rate proportional to at least
$a^{-5}$ in the quenched approximation and typically more than $a^{-7}$ with
dynamical fermions.  As a consequence, one can not get very close to the
continuum limit with currently available hardware, and it is not obvious that
measurements are already in a range in which e.g. an asymptotic behavior
linear in $a$ can be assumed for an extrapolation.

Symanzik found that lattice theories are equivalent to effective continuum
theories that make the cutoff dependence explicit, order by order in $a$. This
means, the theory corresponds to an action
\beq
  S_{\rm eff} = \int d^4x \left\{ \calL_0(x) + a \calL_1(x) + \cdots \right\}
\eeq
and effective lattice fields are
\beq
  \phi_{\rm eff} = \phi_0 + a \phi_1 + \cdots.
\eeq
Here, $\calL_0$ stands for the naive continuum Lagrangian (in our
case, the QCD one) and the higher $\calL_k$ are linear combinations
of local operators also called \emph{counterterms}. The set of possible
operators in each order is restricted by the demand that they must
have dimension $4+k$ and be invariant under the symmetries of the
lattice theory. The \emph{improvement coefficients} of these terms are
functions of the bare couplings and are not known a priori.

Following this observation, Symanzik suggested to use improved lattice actions
and improved local fields in order to reduce the size of lattice artefacts and
accelerate the rate of convergence to the continuum
\cite{Symanzik1,Symanzik2}.  Meanwhile, it is common in the ALPHA
collaboration to use $\rmO(a)$ improvement in QCD, such that cutoff effects
linear in $a$ are removed in all on-shell quantities. In this context,
strategies have been developed to compute improvement coefficients
non-perturbatively \cite{heplat9512009}. Several improvement coefficients have
been determined perturbatively \cite{heplat9606016,heplat9704001} and
non-perturbatively
\cite{heplat9609035,heplat9710071,heplat9803017,heplat0009021}.

For lattices without boundaries, the only necessary counterterm is the
so-called \emph{clover}, or Sheikoleslami-Wohlert term
\cite{SheikholeslamiWohlert}. The boundary conditions used in the
Schr\"odinger functional approach are not translational invariant. Therefore,
additional counterterms have to be added at the boundary. A detailed analysis
can be found in \cite{heplat9605038}.

In the following sections, we shall assume degenerate quark masses and denote
the bare mass with $m_0$. In the lattice regularization, chiral symmetry is
explicitly broken.  As a consequence, the quark mass gets an additive
renormalization depending on the lattice spacing. One defines the critical
mass $\mc(g_0)$ as the bare mass for which the renormalized mass vanishes. A
subtracted mass is then defined as $\mq=m_0-\mc$. Furthermore, in the improved
theory, a rescaling of the bare parameters by factors $1+\rmO(a \mq)$ is
necessary \cite{heplat9605038}. The general connection between bare and
renormalized parameters is then given by

\parbox{1cm}{\hspace{1cm}}
\parbox{4.5cm}{\beqn
  \tilde g_0^2 &=& g_0^2 (1 + \bg a \mq) \nonumber\\
  \tilde \mq &=& \mq (1 + \bm a \mq) \nonumber
\eeqn}
\parbox{1cm}{\hspace{1cm}}
\parbox{4.5cm}{\beqn
  \gR^2 &=& \tilde g_0^2 \Zg(\tilde g_0^2, a\mu) \nonumber\\
  \mR &=& \tilde \mq \Zm(\tilde g_0^2, a\mu). \nonumber
\eeqn}
\hfill
\parbox{1cm}{\beqn
\eeqn}

\noindent The coefficients $\bg, \bm$ are again improvement coefficients which
are independent of the renormalization scheme.

\section{Model}

In the following, we will describe the model and the imposed boundary
conditions. For undefined notations, we refer to appendix \ref{app:Notation}.
We set up our theory on a four-dimensional hypercubic lattice with lattice
spacing $a$ and extent $T$ in the time direction and $L$ in the space
directions. Normally, we set $T=L$.  On the links between neighboring sites
$x$ and $x+a\muhat$ (where $\muhat$ denotes the unit vector in direction
$\mu=0,1,2,3$) lives a gauge field that is represented by $\SUthree$ link
variables $U(x,\mu)$. Furthermore, on the lattice sites reside $\Nf$ flavors
of mass degenerate fermionic quark fields $\psi_f(x)$ which also carry Dirac
and color indices. We do not specify $\Nf$ at the moment. Later we will
consider the theory in which $\Nf$ is continued to negative numbers. This has
to be done after the integration over the quark fields has been performed.

\begin{figure}[htb]
  \begin{center}
    \epsfig{file=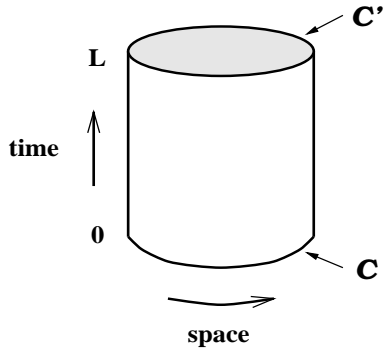,height=5cm,width=5cm}
    \caption{\sl Schr\"odinger functional boundary conditions.}
    \label{fig:cylinder}
  \end{center}
\end{figure}

We think of the lattice as being wrapped up on a cylinder, i.e.  for the gauge
fields we impose periodic boundary conditions in the space directions, while
the quark fields obey periodic boundary conditions up a to a phase factor
$\exp(i\theta)$ \cite{heplat9508012}.

The gauge field at the boundary takes the form 
\beqn
  U(x,k)|_{x_0=0} &=& \exp(aC) \nonumber\\
  U(x,k)|_{x_0=T} &=& \exp(aC').  
\eeqn
This still leaves open a wide range of possibilities. First, one imposes the
restriction that the matrices $C_k$ and $C_k'$ are diagonal and independent of
$k$,
\beq
  C_k = \frac i L \left( \begin{array}{ccc}
    \phi_1 & 0 & 0 \\
    0 & \phi_2 & 0 \\
    0 & 0 & \phi_3 \end{array} \right), \quad
  C_k' = \frac i L \left( \begin{array}{ccc}
    \phi_1' & 0 & 0 \\
    0 & \phi_2' & 0 \\
    0 & 0 & \phi_3' \end{array} \right),
  \label{eq:BoundaryField}
\eeq
with real $\phi_k$ and $\phi_1+\phi_2+\phi_3=\phi_1'+\phi_2'+\phi_3'=0$ such that the
corresponding link variables are in $\SUthree$. 

We choose the boundary fields as a line through the point ``A''
as discussed in \cite{heplat9309005}, parametrized by a
variable $\eta$,

\parbox{2cm}{\hspace{2cm}}
\parbox{3cm}{\beqn
  \phi_1 &=& \eta - \frac{\pi}{3} \nonumber\\
  \phi_2 &=& -\frac 1 2 \eta \nonumber\\
  \phi_3 &=& -\frac 1 2 \eta + \frac{\pi}{3} \nonumber
\eeqn}
\parbox{2cm}{\hspace{2cm}}
\parbox{3cm}{\beqn
  \phi'_1 &=& -\eta - \pi \nonumber\\
  \phi'_2 &=& \frac 1 2 \eta + \frac{\pi}{3} \nonumber\\
  \phi'_3 &=& \frac 1 2 \eta + \frac{2\pi}{3}. \nonumber
\eeqn}
\hfill
\parbox{1cm}{\beqn
\eeqn}

The boundary fields on the opposite sides of the cylinder are chosen such that
the partition function is invariant under a combination of a time reflection,
charge conjugation and a central conjugation.

A solution of the field equations for the link variables then is 
$V(x,\mu)=\exp(aB_\mu(x_0))$ with
\beq
  B_0(x_0) = 0,\quad B_k(x) = [x_0 C_k'+(T-x_0)C_k]/T.
  \label{eq:BackgroundField}
\eeq
In \cite{heplat9207009}, it has been shown that for $\SUthree$ and for the
lattice sizes of interest, this solution is the unique minimum of the action
introduced in the next section, in an environment of $\eta=0$. Hence, we use
the notion that the boundary field enforces a constant color-electric
(classical) background field.

The boundary conditions for the quark fields are discussed in detail
in \cite{heplat9606016}. The boundary quark fields serve as sources for
fermionic correlation functions. They are set to zero after differentiation,
\beqn
  P_+ \psi(x)|_{x_0=0}      = \rho{(\xvec)}, && \quad
  P_- \psi(x)|_{x_0=T}      = \rho'{(\xvec)} \nonumber\\
  \bar \psi(x) P_-|_{x_0=0} = \bar \rho{(\xvec)}, && \quad
  \bar \psi(x) P_+|_{x_0=T} = \bar \rho'{(\xvec)}.
\eeqn
Here we have used the projectors $P_\pm = \frac 1 2 (1 \pm \gamma_0)$.  For
notational reasons -- namely in order to avoid referencing undefined fields in
the Dirac operator --, we furthermore extend the time direction beyond the
boundaries and set
\beq
  \psi(x) = \bar\psi(x) = 0 \quad\mbox{for $x_0 < 0$ and $x_0 > T$}
\eeq
and
\beqn
  P_+ \psi(x)|_{x_0=T} = P_- \psi(x)|_{x_0=0} &=& 0 \nonumber\\
  \bar \psi(x) P_-|_{x_0=T} = \bar \psi(x) P_+|_{x_0=0} &=& 0.
\eeqn
Analogously, all link variables outside the cylinder are set to the
unity matrix.

The Schr\"odinger functional is the partition function of this system and is
defined as a path integral over all gauge and quark fields that fulfill the
given boundary conditions,
\beq
  \calZ[C',C] = e^{-\Gamma} 
  = \int D[U]D[\bar\psi]D[\psi] e^{-S[U,\bar\psi,\psi]}.
\eeq
$D[U]$ denotes the measure $\Pi_{x,\mu} dU(x,\mu)$, and $D[\psi]$ stands for
the product over sites, Dirac and color indices $\Pi_{xDc} d\psi_{Dc}(x)$.
The expectation value of any product of fields is now given by
\beq
  \langle \calO \rangle = \left\{ \frac{1}{\calZ}
    \int D[U]D[\bar\psi]D[\psi] \calO e^{-S[U,\bar\psi,\psi]}
    \right\}_{\rho=\rho'=\bar\rho=\bar\rho'=0}.
\eeq
Note that possible choices for $\calO$ include the variational
derivatives 
\beqn
  \zeta(\xvec) = \frac{\delta}{\delta\bar\rho(\xvec)}, && \quad
  \bar\zeta(\xvec) = -\frac{\delta}{\delta\rho(\xvec)} \nonumber\\
  \zeta'(\xvec) = \frac{\delta}{\delta\bar\rho'(\xvec)}, && \quad
  \bar\zeta'(\xvec) = -\frac{\delta}{\delta\rho'(\xvec)}.
\eeqn
These act on the Boltzmannian and have the effect of inserting
$\psi(x)$ terms near the boundary.

\section{Coupling}

We can interpret the effective action as a function of the background
field,
\beq
  \Gamma[B] = -\log\calZ[C',C]
\eeq
The background field can be varied by changing the parameter $\eta$.  We
define a derivative for $\Gamma$ as the response to a change of the background
field,
\beq
  \Gamma'[B] = \frac{\partial \Gamma[B]}{\partial \eta}.
\eeq
It has a perturbative expansion
\beq
  \Gamma'[B] = \frac{1}{g_0^2} \Gamma_0' + \Gamma_1 + g_0^2 \Gamma_2 + \cdots.
\eeq

The renormalization properties of the Schr\"odinger functional have been
studied in perturbation theory. Symanzik has proven the renormalizability of
the $\phi^4$ theory with Schr\"odinger functional boundary conditions to all
orders of perturbation theory \cite{SymanzikSchrodinger}. For QCD, the
renormalizability has been established up to 2-loop for $\Nf=0$
\cite{heplat9911018} and up to 1-loop for dynamical fermions
\cite{heplat9504005}.

The important result is that the effective action is finite after the bare
coupling has been eliminated in favor of a renormalized coupling, and the
fermionic boundary fields have been rescaled with a renormalization factor. We
infer that $\Gamma'$ is itself suitable as a renormalized coupling. It is
normalized such that its perturbative expansion begins with the bare coupling
at tree level. The Schr\"odinger functional coupling $\gbar$ is then defined as
\beq
  \gbar^2 = \left.\frac{\Gamma'_0[B]}{\Gamma'[B]}\right|_{\eta=0}.
  \label{eq:SchrodingerFunctionalCoupling}
\eeq
The normalization factor is calculated as
\beq
  \Gamma'_0[B] = 12 (L/a)^2 [\sin(\gamma)+\sin(2\gamma)],
  \quad \gamma = \frac 1 3 \pi (a/L)^2.
\eeq

It is clear that $\gbar^2$ is an inherently non-perturbative definition of the
coupling, as desired. Its only dependence on an external scale is on the
system size $L$. We can therefore speak of it as a coupling running with $L$.
The QCD coupling $\alphas(\mu)$ at the scale $\mu$ is related to it by
\beq
  \alphas(\mu) = \frac{\gbar^2(L=1/\mu)}{4\pi}.
\eeq

On the practical side, the coupling defined in this way can easily be computed
in Monte Carlo simulations as the expectation value
\beq
  \Gamma'[B] = \left\langle \frac{\partial S}{\partial \eta} \right\rangle.
\eeq
In order to completely define the scheme for $\Nf \neq 0$, one complements
this definition with the condition that the coupling is taken at vanishing
current mass $m_1$. The definition of the current mass will be introduced in
section~\ref{sec:Mass}. This condition can safely be imposed because the
Schr\"odinger functional is known to have a mass gap of order $1/T$ in
perturbation theory. The step scaling function is then defined as
\beq
  \Sigma(u,a/L) = \gbar^2(2L)\Big|_{u=\gbar^2(L),m_1(a/L)=0}.
\eeq

\section{Action}

The action is given as the sum $S[U,\bar\psi,\psi] = \Sg[U] +
\Sf[U,\bar\psi,\psi]$ of a pure gauge term and the fermionic action.  For the
pure gauge part, we use the Wilson plaquette action modified by $\rmO(a)$
improvement,
\beq
  \Sg[U] = \frac{1}{g_0^2} \sum_p w(p) \Tr(1-U(p))
\eeq
Here, $U(p)$ denotes the parallel transporter around a plaquette $p$,
\beq
  U(p) = U(x,\mu) U(x+a\muhat,\nu) U^\dagger(x+a\nuhat,\mu) U^\dagger(x,\nu)
\eeq
and the sum extends to all oriented (i.e. left-handed and right-handed)
plaquettes. In this thesis, we will alternatively express the bare coupling by
$\beta=6/g_0^2$.

On a system without boundaries, the Wilson action already reaches the
continuum limit with $\rmO(a^2)$ artifacts and the weights are $w(p)=1$ for
all plaquettes. In our Schr\"odinger functional setup however, $\rmO(a)$
improvement is achieved by adding a counterterm at the boundaries. The
addition of this term is equivalent to a modification of the weights such that
\beqn
  w(p) = \ct(g_0)
\eeqn
if $p$ is a time-like plaquette attached to a boundary plane. In all other
cases $w(p)=1$. The improvement coefficient $\ct$ is only known
perturbatively. Its 2-loop value depends quadratically on $\Nf$ and has the
form \cite{heplat9911018}
\begin{eqnarray}
&& \ct(g_0) = 1+\left( -0.08900(5) + 0.0191410(1) \Nf \right) 
              g_0^2 \nonumber\\
&&\quad + \left( -0.0294(3) + 0.002(1) \Nf + 0.0000(1)
           \Nf^2 \right) g_0^4 + \rmO(g_0^6).
  \label{eq:ct}
\end{eqnarray}

For the quark fields, we start with a fermionic action of the form
\beq
   \Sf[U,\bar\psi,\psi] = a^4 \sum_x \bar\psi(x) (D + m_0) \psi(x)
   \label{eq:FermionAction}
\eeq
with the Wilson-Dirac operator
\beq
  D = \frac 1 2 \sum_\mu \left[
  \gamma_\mu(\nabla_\mu^* + \nabla_\mu) - a \nabla_\mu^* \nabla_\mu
  \right].
\eeq
The derivative operators $\nabla_\mu$ are given by
\beqn
  \nabla_\mu \psi(x) &=& \frac 1 a 
    \Bigl[ \lambda_\mu U(x,\mu) \psi(x+a\hat\mu) - \psi(x) \Bigr], \nonumber\\
  \nabla_\mu^* \psi(x) &=& \frac 1 a
    \Bigl[ \psi(x) - \lambda_\mu^* U^\dagger(x-a\hat\mu,\mu) \psi(x-a\hat\mu) \Bigr].
\eeqn
As a difference to the conventionally used operators, they include
phase factors
\beq
  \lambda_\mu = e^{i\theta_\mu a/L}, \quad
  \theta_0 = 0, \,\, -\pi < \theta_k < \pi.
\eeq
As can be easily seen, these factors are equivalent to boundary
conditions
\beq
  \psi(x+L\hat k) = e^{i\theta_k} \psi(x), \quad
  \bar\psi(x+L\hat k) = \bar\psi(x) e^{-i\theta_k}
\eeq
in the space directions. However, in an implementation on the
computer, it is simpler to ``distribute'' this phase on the difference
operators and impose strict periodic boundary conditions on the
fields. 

In \cite{heplat9508012}, it is argued that the choice of $\theta_k$
should be guided by practical considerations. The lowest eigenvalue
of the Dirac operator on the classical background field varies with
$\theta_k$. In a Monte Carlo simulation, the square of the Dirac
operator is inverted frequently, so that a small condition number can
improve its performance. An optimal condition number has been found
around the value $\theta_k=\theta=\pi/5$ which we use here.
  
In the quark sector, $\rmO(a)$ improvement can be implemented by adding 
certain terms to the Dirac operator. One is a bulk term
\beq
  \delta D_{\rm v} \psi(x) = \csw \frac i 4 a \sigma_{\mu\nu}
  \hat F_{\mu\nu}(x) \psi(x),
\eeq
also known as Sheikoleslami-Wohlert term. In this term,
\beq
  \hat F_{\mu\nu} = \frac{1}{8a^2} ( Q_{\mu\nu} - Q_{\nu\mu} )
\eeq
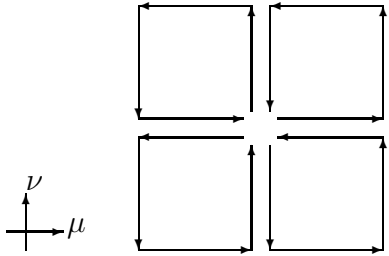
\begin{figure}[btp]
  \begin{center}
  \setlength{\unitlength}{0.5cm}
  \begin{picture}(10,7.5)
  \put(3.5,3.5){\vector(1,0){2.8}}
  \put(6.5,3.7){\vector(0,1){2.8}}
  \put(6.5,6.5){\vector(-1,0){3}}
  \put(3.5,6.5){\vector(0,-1){3.0}}
  \put(7.2,3.5){\vector(1,0){2.8}}
  \put(10,3.5){\vector(0,1){3}}
  \put(10,6.5){\vector(-1,0){3}}
  \put(7,6.5){\vector(0,-1){2.8}} 
  \put(3.5,0){\vector(1,0){3.0}}
  \put(6.5,0){\vector(0,1){2.8}}
  \put(6.3,3){\vector(-1,0){2.8}}
  \put(3.5,3){\vector(0,-1){3.0}}
  \put(7,0){\vector(1,0){3.0}}
  \put(10,0){\vector(0,1){3}}
  \put(10,3){\vector(-1,0){2.8}}
  \put(7,2.8){\vector(0,-1){2.8}}
  \put(0,0.5){\vector(1,0){1.5}}
  \put(1.6,0.5){$\mu$}
  \put(0.5,0){\vector(0,1){1.5}}
  \put(0.5,1.6){$\nu$}
  \end{picture}
  \caption{\sl Graphical representation of the products of links contributing
    to the clover term. The point in the middle is $x$.}
  \label{fig:clover}
  \end{center}
\end{figure}
is the lattice definition of the field strength tensor. The term
$Q_{\mu\nu}$ is visualized in figure \ref{fig:clover} and is explicitly
given by
\beqn
  \label{Clover definition}
&&  Q_{\mu\nu}(x) = \Bigl\{
    U(x,\mu) U(x+a\muhat,\nu)
    U^\dagger(x+a\nuhat,\mu) U^\dagger(x,\nu) \nonumber\\
&&\quad+{}
    U(x,\nu) U^\dagger(x+a\nuhat-a\muhat,\mu)
    U^\dagger(x-a\muhat,\nu) U(x-a\muhat,\mu) \nonumber\\
&&\quad+{}
    U^\dagger(x-a\muhat,\mu) U^\dagger(x-a\nuhat-a\muhat,\nu)
    U(x-a\nuhat-a\muhat,\mu) U(x-a\nuhat,\nu) \nonumber\\
&&\quad+{}
    U^\dagger(x-a\nuhat,\nu) U(x-a\nuhat,\mu)
    U(x-a\nuhat+a\muhat,\nu) U^\dagger(x,\mu) \Bigr\}.
\eeqn

As with Schr\"odinger functional boundary conditions, we do not have
translational invariance in the time direction, $\rmO(a)$ improvement here
requires an additional term
\beqn
  \delta D_{\rm b} \psi(x) = (\cttilde - 1) \frac 1 a \Bigl\{
  \delta_{x_0,a}    [ \psi(x) - U^\dagger(x-a\zerohat) P_+ \psi(x-a\zerohat)] \nonumber\\
 + \delta_{x_0,T-a} [ \psi(x) - U(x,0) P_- \psi(x+a\zerohat) ] \Bigr\}.
\eeqn
The coefficient $\cttilde$ is known perturbatively \cite{heplat9704001},
\beq
  \cttilde(g_0) = 1 - 0.01795(2) g_0^2 + \rmO(g_0^4).
  \label{eq:cttilde}
\eeq

At this point we want to introduce some further notations. Sometimes it is
useful to separate the integration of quark and gauge fields. We therefore
write the expectation value above as
\beq
  \langle \calO \rangle = \langle [ \calO ]_{\rm F} \rangle_{\rm G}.
\eeq
Here, $\langle \ldots \rangle_{\rm G}$ denotes the gauge field average with respect
to the distribution
\beq
  \det(D+\delta D+m_0) \exp(-\Sg[U]).
\eeq
The fermionic expectation value $[ \ldots ]_{\rm F}$ can be represented by a
generating functional. We refer to \cite{heplat9606016} for a detailed
discussion and only list the relevant results here. The quark propagator $S(x,y)$
on a given gauge field is defined as the solution of
\beq
  (D+\delta D+m_0) S(x,y) = a^{-4} \delta(x,y), \quad 0<x_0<T.
\eeq
with appropriate boundary conditions. It fulfills
\beq
  S^\dagger(x,y) = \gamma_5 S(y,x) \gamma_5.
  \label{eq:PropagatorWithGamma5}
\eeq
Furthermore, one can define a propagator $H(x)$ ``from the lower boundary to
point $x$'' through
\beq
  (D+\delta D+m_0) H(x) = a^{-1} \delta(x_0,a) 
    \cttilde U^\dagger(x-a\zerohat,0) P_+.
  \label{eq:BoundaryPropagator}
\eeq
Fermionic expectation values $[ \ldots ]_{\rm F}$ necessary for the computation of
correlation functions can be expressed in terms of these propagators. The
basic boundary-bulk 2-point functions are
\beqn
  a^3 \sum_\yvec [ \psi(x) \bar\zeta(\yvec) ]_{\rm F} &=& \gamma_5 H(x) \gamma_5 \nonumber\\
  a^3 \sum_\yvec [ \zeta(\yvec) \bar\psi(x)  ]_{\rm F} &=& H^\dagger(x).
  \label{eq:BasicTwoPoint}
\eeqn
One proceeds similarly for the upper boundary.

\section{Mass}
\label{sec:Mass}

For the definition of a quark mass, we use chiral symmetry to derive a
relation between correlation functions, following \cite{heplat9605038}. In the
continuum, the PCAC (partially conserved axial current) relation $\partial_\mu
A_\mu^a = 2m P_\mu^a$ is a special case of the chiral Ward identity and
connects the isovector axial current
\beq
  A^a_\mu(x) = \bar \psi(x) \gamma_\mu \gamma_5 \frac{\tau^a}{2} \psi(x)
\eeq  
and its associated density
\beq
  P^a(x) = \bar \psi(x) \gamma_5 \frac{\tau^a}{2} \psi(x).
\eeq
The matrices $\tau_a, a=1 \ldots 3$ are the Pauli matrices and act on the
flavor indices.  On the lattice, we demand that the PCAC relation holds for
renormalized quantities. This means, we use it to derive relations between
correlation functions and require these to converge to the proper continuum
limit. Since we are going to implement improvement in the action, we also have
to use improved operators in order to get a mass definition that has only
$\rmO(a^2)$ artefacts. An analysis \cite{heplat9605038} shows that this
amounts to an addition of a term
\beq
  \delta A_\mu^a = \cA \tilde\partial_\mu P_\mu^a.
\eeq 
to the axial current operator.  Here, we use the notation
$\tilde\partial_\mu=1/2(\partial_\mu+\partial_\mu^*)$ for the average of
forward and backward derivative on the lattice.  $\cA$ is a further
improvement coefficient. It has been computed to 1-loop order in perturbation
theory \cite{heplat9606016},
\beq
  \cA(g_0) = -0.00756(1) g_0^2.
\eeq
Non-perturbative data in the quenched approximation is also
available \cite{heplat9609035}, but not used in this work.
With this improvement term, the renormalized axial current and its
associated pseudo-scalar density are given by the expressions
\beqn
  (A_{\rm R})_\mu^a &=& \ZA(1+\bA a \mq) \left[
    A_\mu^a + \cA a \tilde\partial_\mu P_\mu^a \right] \nonumber\\
  (P_{\rm R})_\mu^a &=& \ZP(1+\bP a \mq) P_\mu^a.
\eeqn
Here, $\ZA$ and $\ZP$ are renormalization factors. While the former
depends only on the bare coupling, the latter is scale dependent.

We now define a renormalized mass through the relationship
\beq
  \langle \tilde\partial_\mu (A_{\rm R})_\mu^a(x) \calO \rangle = 2 \mbar
    \langle (P_{\rm R})^a(x) \calO \rangle + \rmO(a^2),
  \label{eq:RenormalizedMass}
\eeq
where $\calO$ may be any product of improved renormalized fields
located at non-zero distance from $x$.
Furthermore, we define a current quark mass by the relation
\beq
  \left\langle \{ \tilde\partial_\mu A_\mu^a 
    + \cA a \partial_\mu^* \partial_\mu P^a
    \} \calO^a \right\rangle = 2 m 
    \langle P^a \calO^a \rangle,
  \label{eq:UnrenormalizedMass}
\eeq
where $\calO^a$ is the operator
\beq
  \calO^a = a^6 \sum_{\yvec, \zvec}
    \bar\zeta(\yvec) \gamma_5 \frac{\tau^a}{2} \zeta(\zvec).
\eeq
We sum over all space-like $\xvec$ and define bare correlation functions $\fA$
and $\fP$ as
\beqn
  \fA(x_0) 
&=& -\frac{a^9}{L^3}\sum_{\xvec,\yvec,\zvec} \frac 1 3
  \left\langle A_0^a(x) \bar{\zeta}(\yvec) \gamma_5
  \frac{\tau^a}{2} \zeta(\zvec) \right\rangle \nonumber\\
  \fP(x_0) 
&=& -\frac{a^9}{L^3}\sum_{\xvec,\yvec,\zvec} \frac 1 3
  \left\langle P^a(x) \bar{\zeta}(\yvec) \gamma_5
  \frac{\tau^a}{2} \zeta(\zvec) \right\rangle.
\eeqn
Similar correlations functions can be defined at the
upper boundary,
\beqn
  \fA(T-x_0) 
&=& \frac{a^6}{L^3}\sum_{\xvec,\yvec,\zvec} \frac 1 3
  \left\langle A_0^a(x) \bar{\zeta'}(\yvec) \gamma_5
  \frac{\tau^a}{2} \zeta'(\zvec) \right\rangle \nonumber\\
  \fP(T-x_0) 
&=& \frac{a^6}{L^3}\sum_{\xvec,\yvec,\zvec} \frac 1 3
  \left\langle P^a(x) \bar{\zeta'}(\yvec) \gamma_5
  \frac{\tau^a}{2} \zeta'(\zvec) \right\rangle.
\eeqn
A time-dependent mass is then obtained as
\beq
  m(x_0) = \frac{\tilde\partial_0 \fA(x_0)
           + a \cA \partial_0^* \partial_0 \fP(x_0)}
           {2 \fP(x_0)}.
  \label{eq:CurrentMass}
\eeq
In practice, $m(x_0)$ turns out to have large lattice artifacts at the
boundaries, with a plateau in the middle. Thus, choosing $x_0$ to be
in the middle of the lattice is a good idea,
\beq
  m_1 = \left\{ \begin{array}{ll}
      m\left(\frac T 2\right) 
      & \mbox{for even $T/a$} \\
      \frac 1 2 \left( m\!\left(\frac{T-a}{2}\right)
                      +m\!\left(\frac{T+a}{2}\right)\right)
      & \mbox{for odd $T/a$.}
      \end{array} \right.
  \label{eq:m1}
\eeq
This is a quantity that can be actually measured in a simulation and
plays the role of an unrenormalized mass. We also use the notion of a
PCAC or \emph{current mass}. In contrast, a computation of the
renormalized mass requires knowledge about the renormalization
factors.  By combining \eqref{eq:RenormalizedMass} and
\eqref{eq:UnrenormalizedMass}, we find the relation
\beq
  \mbar = \frac{\ZA(1+\bA a \mq)}{\ZP(1+\bP a \mq)} m_1 + \rmO(a^2).
  \label{eq:RenormalizedAndUnrenormalizedMass}
\eeq
This relation also reflects the knowledge that the current quark mass has no
additive renormalization.
The definitions of the correlation functions still contain fermionic
expectation values. For a practical measurement in a simulation, these have to
be integrated out.  From an application of Wick's theorem,
\eqref{eq:BasicTwoPoint} and \eqref{eq:PropagatorWithGamma5}, one gets
\beqn
  \fA(x_0) 
&=& - \frac{a^9}{L^3} \sum_{\xvec,\yvec,\zvec} \frac 1 2
    \left\langle \Tr \left\{ [\zeta(\zvec) \bar\psi(x)]_{\rm F} \gamma_0 \gamma_5
    [\psi(x) \bar\zeta(\yvec)]_{\rm F} \gamma_5 \right\} \right\rangle_{\rm G} \nonumber\\
&=& -\frac 1 2 \frac{a^3}{L^3} \sum_\xvec
    \left\langle \Tr \left\{ H^\dagger(x) \gamma_0 H(x) 
    \right\} \right\rangle_{\rm G},
  \label{eq:fa}
\eeqn
where the trace is over Dirac and color indices and not over flavor.
Analogously, the correlation function $\fP$ can be written as
\beq
  \fP(x_0) = \frac 1 2 \frac{a^3}{L^3} \sum_\xvec
    \left\langle \Tr \left\{ H^\dagger(x) H(x) 
    \right\} \right\rangle_{\rm G} \nonumber\\
  \label{eq:fp}.
\eeq
With the help of the equations \eqref{eq:fa}, \eqref{eq:fp} together with
\eqref{eq:BoundaryPropagator} one can compute $\fA$ and $\fP$ in a Monte Carlo
simulation.

\chapter{Performance of algorithms in pure gauge theory}
\label{chap:Performance}

\section{Motivation}

From an algorithmic perspective, the simulation of gauge theories with
fermions is fundamentally different from the simulation of pure gauge
theories. A pure $\SUN$ theory has a \emph{local} action. Therefore, it is
natural to apply local algorithms for its simulation, unless such algorithms
fail in effectively decorrelating successive configurations. The Hybrid
Overrelaxation algorithm (HOR) has become a classical method for this purpose.

On the other hand, an important property of QCD with fermions is the Pauli
principle, which in the path integral formalism leads to a formulation with
Grassmann variables. In practice, Grassmann variables cannot easily be handled
numerically, and thus the common way of treating them is to integrate the path
integral analytically and represent the resulting \emph{fermion determinant}
as a path integral over boson fields also called \emph{pseudofermions}.

In terms of these boson fields, the action is highly non-local. This means
that any algorithm consisting of a sweep over the lattice and updating links
locally needs an amount of work at least quadratic in the lattice volume for a
complete update of all link variables. In general, two groups of algorithms
have been proposed to solve this problem. One is based on using molecular
dynamics and includes the Hybrid Monte Carlo (HMC) method and variants
thereof. The other is the MultiBoson (MB) method.

The MultiBoson algorithm has been proposed by L\"uscher \cite{heplat9311007}.
The basic idea is to transform the theory into a local bosonic theory. To this
end, the fermion determinant -- here for two degenerate flavors -- is
approximated by a polynomial,
\beq
  \det Q^2 = \lim_{n\to\infty} [\det P_n(Q^2)]^{-1},\quad Q=\gamma_5 (D+m_0),
\eeq
in an interval that covers the spectrum of the matrix $Q^2$.
Using the hermiticity property of $Q$, one can express the
approximate fermionic action by
\beqn
  \label{eq:MultiBosonAction}
  \det P_n(Q^2)^{-1} = \int D[\phi] D[\phi^\dagger] D[\phi] 
  \exp\Bigl\{ -a^4 \sum_x \sum_{k=1}^n \nonumber\\
  \bigl( | (Q-\mu_k) \phi_k(x)|^2 
  + \nu_k^2 |\phi_k(x)|^2 \bigr) \Bigr\},
\eeqn
where the constants $\mu_k$ and $\nu_k$ are given by the roots of the
polynomial. There are $n$ boson fields $\phi_k$ which carry color indices.
When $Q$ is local, these fields may be updated with algorithms known from pure
gauge theory.  The deviation of the polynomial approximation from the exact
measure may be corrected for instance by reweighting.

It has turned out that in practice, a sufficiently precise approximation
requires large $n$, i.e. a large number of boson fields. This seems to
compensate the advantage of having to simulate only a local action. As an
additional disadvantage, it is difficult to simulate the action
\eqref{eq:MultiBosonAction} when the fermion matrix includes a clover term,
because in that case the matrix $(Q-\mu_k)^2$ contains terms beyond the
hypercube around the lattice site. A review of the MultiBoson technique can be
found in \cite{heplat9903035}.

The standard algorithm for the simulation of a pair of dynamical fermions is
the Hybrid Monte Carlo algorithm \cite{DuaneKennedy}.  It generates new
configurations by computing discretized molecular dynamics trajectories. An
acceptance step makes this algorithm exact. Hybrid Monte Carlo is not
applicable to either odd flavor simulations with Wilson fermions or
simulations with ($\Nf=4$) staggered fermions.  In those cases, it is possible
to apply e.g. the R algorithm \cite{GottliebToussaint}, which does not include
an acceptance step. It therefore has systematic errors dependent on the step
size, and one has to extrapolate to zero step size in order to get reliable
results.

It is clear from the preceding discussion that the development of the Hybrid
Monte Carlo and other modern algorithms is mainly motivated by the wish to
simulate dynamical fermions efficiently. Since the characteristics of these
theories are very different from pure gauge theory, one cannot transfer any
experience about the relative merits of different algorithms from dynamical
fermions to pure gauge theory, and vice versa.  In this part of our work, we
first compare different algorithms in pure gauge theory. This gives results
about the overhead of HMC compared to HOR before dynamical quarks are
included.

\section{Hybrid Overrelaxation}
\label{sec:HybridOverrelaxation}

A standard algorithm for the simulation of pure gauge theory is a combination
of overrelaxation and heatbath steps. Both are local updates. The
overrelaxation algorithm is a microcanonical update which is known to
decorrelate configurations very efficiently.  It is not ergodic though. The
heatbath step is necessary in order to guarantee a simulation of the correct
(non-microcanonical) statistical ensemble.

\subsection{Heatbath}

In general, the notion of a heatbath algorithm is used for Markov chains in
which each new configuration is chosen from the canonical distribution, i.e.
independent from the previous configuration,

\beq
  P(U' \leftarrow U) \propto \exp(-S[U']).
\eeq

For arbitrary actions and arbitrary fields it is a non-trivial task to find an
efficient way for generating such an ensemble.  We begin here with an $\SUtwo$
theory. The $\SUthree$ case can be connected to $\SUtwo$ by updating subgroups and
will be discussed later.

In the following, we drop the arguments $x$ and $\mu$ so that it is obvious
that we have to cope with the distribution of a mere $\SUN$ matrix.

\subsubsection*{SU($2$)}

We can formulate the goal of finding a heatbath update for $\SUtwo$ as finding
an algorithm which produces the distribution
\beq
  dP(U') \propto e^{\frac{1}{2} \Tr(U' W^\dagger)} dU',
  \label{eq:HeatbathSU2}
\eeq 
where $W$ is a real multiple of an $\SUtwo$ matrix $\hat W$. This is e.g.  the
case for an $\SUtwo$ theory with a Wilson action. In that case, $\det W>0$, so
that we can write $W=\sqrt{\det W} \, \hat W$.  This distribution factorizes
when the $\SUtwo$ variables are expressed in the quaternionic representation,
\beqn
  U' &=& a_0 + i a_j\sigma_j \nonumber\\
  W  &=& w_0 + i w_j\sigma_j.
\eeqn
From the invariance of the Haar measure under the multiplication with group
elements and \eqref{eq:HaarMeasureQuaternionic} follows
\beqn
  dP(U'\hat W) 
&\propto& e^{\frac{1}{2} \sqrt{\det W} \, \Tr(U')} dU' \nonumber\\
&\propto& \frac{1}{\pi^2} \delta(a^2-1) 
          e^{\sqrt{W_\mu W_\mu} \, a_0} \, da_0 d^3 ¸\mathbf{a} \nonumber\\
&\propto& \frac{1}{\pi^2} \delta(\mathbf{n}^2-1) 
          e^{\rho a_0}  \sqrt{1-a_0^2} \,\, da_0 d^3 \mathbf{n}.
\label{eq:Distributiona0}
\eeqn
In the last step, we have set $a_j=n_j\sqrt{1-a_0^2}$ and $\rho=\sqrt{W_0^2 +
  W_j W_j}$.

This means, we have to generate a flat distribution in $a_0$ and a uniform
distribution on the surface of a three dimensional sphere for $\mathbf{n}$.
These produce a matrix $U'$, which is according to \eqref{eq:Distributiona0}
to be multiplied with $\hat W$, resulting in a new value for the link
variable.

Since the angular element can be parametrized as $d\Omega=d(\cos\theta)
d\phi$, the $\mathbf{n}$ vector can simply be generated by drawing a pair
$(u_1,u_2)$ of numbers with a uniform distribution in [0,1) and setting
\beqn
  n_1 &=& 1 - 2 u_1 \nonumber\\
  n_2 &=& \sqrt{1-n_1^2} \, \cos(2\pi u_2) \nonumber\\
  n_3 &=& \sqrt{1-n_1^2} \, \sin(2\pi u_2).
\eeqn

A method to generate the distribution in $a_0$ has been described by Fabricius
and Haan in \cite{FabriciusHaan} and shortly later by Kennedy and Pendleton
\cite{KennedyPendleton}. The basic idea is to generate a variable $y$
according to
\beq
  P(y) \propto \exp(-y) \sqrt y \, \theta(y)
\eeq
and to set $a_0=1-y\rho$. The distribution of the generated $a_0$ is given
by
\beqn
  \bar P(a_0) da_0 &=& P(y) dy \nonumber\\
  \Rightarrow \bar P(a_0) &=& P(y(a_0)) \frac{dy}{da_0} \nonumber\\
 &\propto& e^{\rho a_0} \sqrt{1-a_0} \, \theta(1-a_0).
  \label{eq:Distributiona0prime}
\eeqn
The difference to \eqref{eq:Distributiona0} is a factor 
$\sqrt{1+a_0} \, \theta(1+a_0)$, which can be accommodated by an
acceptance step: the proposal from \eqref{eq:Distributiona0prime}
is accepted if a flat random number $c$ fulfills $2c^2 \le 1+a_0$.
This corresponds to a $P_1(c)=\theta(2c^2-1-a_0)$ and therefore
\beqn
  \int dc \bar P(a_0) P_1(c) 
  &=& \int_{0}^{\sqrt{(1-a_0)/2}} dc \, e^{\rho a_0} \sqrt{1-a_0} \,
      \theta(1-a_0) \theta(1+a_0) \nonumber\\ 
  &=& e^{\rho a_0} \sqrt{1-a_0} \, \theta(1-a_0) \nonumber\\
  &\propto& P(a_0).
\eeqn

In the form proposed by Fabricius and Haan, one maps a uniform 
distribution to the needed distribution in $a_0$ by a pretty 
complicated function involving logarithms of logarithms,
which is difficult to calculate in a sufficiently precise way.

Here we use a slightly different method discussed in
\cite{WeiszFabHaan}, which combines two distributions in order
to obtain the desired one, and therefore requires twice the
number of random numbers per update. In order to create a value
for $a_0$, one generates a number $a$ with a Gaussian distribution
\beq
  P_1(a) \propto \exp(-a^2) \theta(a)
\eeq
and a number $b$ distributed like
\beq
  P_2(b) \propto \exp(-b) \theta(b).
\eeq
Then one makes a variable transformation $y=a^2+b$ and $z=a$. The
numbers $z$ are finally dropped. As for the Jacobian of this 
transformation,
\beq
  \det \frac{\partial(a,b)}{\partial(y,z)} = 1,
\eeq
the combined distribution of $y$ and $z$ is
\beqn
  P(y,z) &=& P_1(a(y,z),b(y,z)) \times P_2(a(y,z),b(y,z)) \nonumber\\
         &=& \exp(-z^2) \theta(z) \times \exp(-y+z^2) \theta(y-z^2) \nonumber\\
         &=& \exp(-y) \theta(z) \theta(\sqrt{y}-z).
\eeqn
Since the generated numbers $z$ are dropped, the resulting
distribution of $y$ is obtained by integrating over $z$,
i.e. one really gets the desired
\beq
  P(y) = \int dz P(y,z) = \sqrt{y} \exp(-y) \theta(y).
\eeq
The distributions $P_1$ and $P_2$ are well known: the former is the
Gaussian distribution explained in appendix~\ref{app:RandomNumbers}. The
latter can be generated from uniformly distributed random numbers $u$
by taking $b=-\log(1-u)$.

\subsubsection*{SU($N$)}

We assume that the action has the form
\beq
  \label{eq:ActionLinearInU}
  S[U] = -\Re \Tr \left\{ U(x,\mu) V^\dagger(x,\mu) \right\} + \cdots,
\eeq
with a part that depends linearly on the link variable $U(x,\mu) \in \SUN$
($N\ge 3$) that is updated, and a part that is independent. This form is
fulfilled by the Wilson action, including the Schr\"odinger functional
improvement term in pure gauge theory. In this case, the term $V(x,\mu)$ is
proportional to the sum of the six ``staples'' of the link,
\beqn
  V(x,\mu) &=& \frac{\beta}{N} S(x,\mu) + \cdots \nonumber\\
  S(x,\mu) &=& \sum_{\nu\neq\mu} \Bigl\{
    U(x,\nu) U(x+a\hat\nu,\mu) 
    U^\dagger(x+a\hat\mu,\nu) + \nonumber\\
  &&\quad U^\dagger(x-a\hat\nu,\nu) U(x-a\hat\nu,\mu) 
    U(x-a\hat\nu+a\hat\mu,\nu) \Bigr\}.
  \label{eq:staples}
\eeqn
If one of the link variables in the definition of $S(x,\mu)$ lies in a
boundary plane it is to be multiplied with a factor $\ct$.

The essential difference to the $\SUtwo$ case is that the sum of staples is
not in general a real multiple of an element of the gauge group.  This makes
it non-trivial to factorize the distribution for $U$ into several flat
distributions.

For the $\SUthree$ case -- and more generally $\SUN$ --, the method described in
the previous section was generalized by Cabibbo and Marinari
\cite{CabibboMarinari} in terms of projecting to subgroups successively. In
this approach, one chooses a set $\cal F$ of $\SUtwo$ subgroups of $\SUN$ with
the property that no subset of $\SUN$ is left invariant under left
multiplication by $\cal F$, except for the whole group,
\beq
  {\cal F} = \{ \SUtwo_1, \ldots, \SUtwo_m \}, \quad m\ge N-1.
\eeq
This guarantees that by successively multiplying a given $\SUN$ matrix with
random matrices from these subgroups, one can reach any other $\SUN$ matrix.
So we are left with the task of giving the random matrices the correct
distribution.  Again, in order to simplify the notation, we do not explicitly
display the indices $x,\mu$.

An update $U' \leftarrow U$ starts with $U^{(0)}=U$ and moves in $m$ steps to
$U^{(m)}=U'$. The current link variable in each step is obtained by applying
to the one from the previous step a member $A^{(k)} \in \SUtwo_k$ of a
subgroup,
\beq
  U^{(k)} = A^{(k)} U^{(k-1)}.
\eeq
The matrices $A^{(k)}$ are chosen according to the distribution
\beq
  dP(A_k) = dA^{(k)} \frac{\exp\{ -S[A^{(k)} U^{(k-1)}] \}}
            {\int_{\SUtwo_k} dA \exp\{ -S[A U^{(k-1)}] \}}.
  \label{eq:DistributionCabibboMarinari}
\eeq
In this work, we select as subgroups of $\SUthree$ the three most natural
possibilities,
\beq
  A_1 = \left( \begin{array}{ccc} 
          a_{11} & a_{12} & 0 \\
          a_{21} & a_{22} & 0 \\
          0      & 0      & 1 
        \end{array} \right)\!,\,
  A_2 = \left( \begin{array}{ccc} 
          1      & 0      & 0      \\
          0      & a_{11} & a_{12} \\
          0      & a_{21} & a_{22} 
        \end{array} \right)\!,\,
  A_3 = \left( \begin{array}{ccc} 
          a_{11} & 0      & a_{12} \\
          0      & 1      & 0      \\ 
          a_{21} & 0      & a_{22}
        \end{array} \right),
  \label{eq:subgroup_matrices}
\eeq
where $a \in \SUtwo$. This in fact implies one subgroup more than necessary
for the condition explained above. It is more symmetric and may lead to better
autocorrelation times than the minimum of two subgroups. To our knowledge, a
detailed study on this topic has however not been done.

In our choice of subgroups, it is obvious that
\beq
  S[A^{(k)} U^{(k-1)}] = -\Re \Tr(a v),
\eeq
where $v$ is the $(i,j) \in \{ (1,2), (2,3), (1,3)\}$ submatrix 
of $X=U^\dagger V$,
\beq
   v = \left( \begin{array}{ccc}
       X_{ii} & X_{ij} \\
       X_{ji} & X_{jj}
       \end{array} \right).
\eeq
By using the quaternionic representation of $a_k$ and $v$, we can bring this
into a slightly different form,
\beqn
  S[A^{(k)} U^{(k-1)}]
&=& -\Re \Tr(a v) \nonumber\\
&=& -\Re (a_0 v_0 - a_j w_j) \nonumber\\
&=& -\frac 1 2 \Tr ( a w^\dagger ),
  \label{eq:DistributionSubgroups}
\eeqn
where $w$ is a real multiple of an $\SUtwo$ matrix. Its quaternionic components
are given by
\beqn
  w_0 &=& 2 \, \Re \, v_0 \nonumber\\
  w_j &=& -2 \, \Re \, w_j,
\eeqn
and explicitly by
\beqn
  w_0 &=& \Re X_{ii} + \Re X_{jj} \nonumber\\
  w_1 &=& -\Im X_{ij} - \Im X_{ji} \nonumber\\
  w_2 &=& -\Re X_{ii} + \Re X_{ji} \nonumber\\
  w_3 &=& -\Im X_{ii} + \Im X_{jj}.
\eeqn
Therefore, from \eqref{eq:DistributionSubgroups} we can see that
\eqref{eq:DistributionCabibboMarinari} can be obtained by applying the
Fabricius-Haan algorithm for $\SUtwo$, as described above.

\subsection{Overrelaxation}
\label{sec:Overrelaxation}

The idea of overrelaxation is to choose the new variable as far as possible
from the original one, in an intuitive sense. We again assume that the action
has the form \eqref{eq:ActionLinearInU}.

For a pure $\SUtwo$ gauge theory with a Wilson-like action, where $W$ is
the sum of elements of the gauge group, it is easy to write down an update
$U'\leftarrow U$ with the property
\beq
  \Tr(U' W^\dagger) = \Tr(U W^\dagger),
\eeq
namely
\beq
  U' = \frac{2}{\Tr (W^\dagger W)} W U^\dagger W.
\eeq
This mechanism is not directly transferable to $\SUthree$, because the matrix $W
U^\dagger W$ is in general not a real multiple of an $\SUthree$ matrix. But we
can make microcanonical updates in some $\SUtwo$ subgroups of the variable in
question. In analogy to the heatbath case, we create a new U' by applying
matrices $A_k$ as in \eqref{eq:subgroup_matrices},
\beq
  U' = A_3 A_2 A_1 U.
  \label{eq:OverrelaxationSU3}
\eeq
In order for the action to stay constant, the $\SUtwo$ matrices $a$
corresponding to the $A_k$ must fulfill the condition
\beq
  \Tr (a w^\dagger) = \Tr w^\dagger.
\eeq
This is achieved by setting them to
\beq
  a = -1 + 2 \frac{2 w_0}{\Tr (w w^\dagger)} w.
\eeq
A potentially dangerous operation in this computation is the division by the
value $\Tr(w w^\dagger)$, which may become very small and cause an overflow.
In our implementation, when it becomes smaller then $10^{-10}$, we leave the
link variable at its old value instead of flipping it. This does not
invalidate the algorithm.

Finally, the Hybrid Overrelaxation is a combination of the two parts described
previously \cite{heplat9205001}. In each update, one heatbath step is
alternated with $\Nor$ overrelaxation steps. $\Nor$ is a parameter which we
may tune to reduce autocorrelation times on larger lattices.

In \cite{heplat9411017}, it has been found that the efficiency of the HOR
algorithm depends on the order in which a sweep over the lattice processes the
links. It is advantageous to perform an outer loop over the Lorentz index
$\mu$ and to embed a loop over $x$ \emph{within} this one. We have only used
this variant.

\section{Hybrid Monte Carlo}

The basic idea of the Hybrid Monte Carlo (HMC) is to employ molecular dynamics
in order to collectively move the configuration through the configuration
space.  This means, in each step all field variables are updated by computing
their trajectory through a coupled set of equations of motions. It is obvious
that such a step goes at least linear with the volume. But of course, it also
has to be considered that trajectories can only be computed by discretizing
them, so there are further parameters like the trajectory length and the step
size. With increasing volume, it may be necessary to adjust these parameters,
making the cost of a trajectory higher. As a consequence, the dependence of
the cost on the system size can not be trivially predicted.

As an alternative to molecular dynamics, one may also consider moving fields
through the configuration space by other small-step-size mechanisms, such as
random walks. However, a molecular dynamics trajectory is assumed to move more
rapidly away from the original configuration, as the single steps of the
discretized trajectory move into the same ``direction''
\cite{GottliebToussaint}.

For the derivation of the equations of motions, we note that expectation
values with respect to the action $S[U]$ are equal to the expectation values
of the respective observables with respect to the Hamiltonian
\beq
  H[P,U] = \frac 1 2 \sum_{x\mu j} P(x,\mu,j)^2 + S[U]
  \label{eq:HMCHamilton}
\eeq
with variables $P(x,\mu,j)$ from a canonical ensemble.  The real-valued
variables $P(x,\mu,j)$ are the expansion coefficients of variables $P(x,\mu)$
which come from the Lie algebra $\suthree$ of $\SUthree$,
\beq
  P(x,\mu) := \sum_{j=1}^8 i\lambda_j P(x,\mu,j)
\eeq
and which are to be interpreted as conjugate momenta of the link variables
$U(x,\mu) \in \SUthree$. Expectation values are defined by
\beq
  \langle F \rangle = Z^{-1} \int D[P]D[U] 
    \exp(-H[P,U]) F[U],
\eeq
with the normalization factor
\beq
   Z = \int D[P]D[U] \exp(-H[P,U]),
\eeq
and obviously factorize into integrals over link variables $U$ and the new
fictitious momenta $P$. The expectation value defined in this way corresponds
to a canonical ensemble.  For sufficiently large lattices, one should be able
to approximate this by a microcanonical ensemble, which is characterized by a
constant energy of all configurations summed over. Provided ergodicity holds,
a chain of configurations in the microcanonical ensemble can be generated by
solving the canonical equations of motion for the Hamiltonian $H[P,U]$.

Naturally, for a complex interacting theory, we cannot solve the equations of
motions exactly, but have to resort to a discretization to compute a
trajectory from a given starting point to an end point giving a new
configuration. This not only produces errors in the configurations, but also
changes the total energy, so that we cannot expect to produce the correct
ensemble in this way.  In the Hybrid Monte Carlo, this problem is overcome by
an acceptance step, which makes the algorithm exact.

The classical equations of motion for the Hamiltonian
and for the variables $U(x,\mu)$ and $P(x,\mu,j)$ are

\beqn
  \dot P(x,\mu,j) &=& - D_{x\mu j} S[U] \nonumber\\
  \dot U(x,\mu) &=& i P(x,\mu) U(x,\mu)
  \label{eq:EquationsOfMotion}
\eeqn
with a derivative $D_{x\mu j}$ for the gauge fields.
This derivative is defined as

\beq
  D_{x\mu j} f(U(x,\mu)) = 
  \left.\frac{d}{d\alpha}\right|_{\alpha=0}
  f\left( e^{i\alpha \lambda_j} U(x,\mu) \right),
\eeq
where $\lambda_j, j=1\ldots 8$ are the Gell-Mann matrices.  The form of the
second equation in \eqref{eq:EquationsOfMotion} guarantees that $U(x,\mu)$
remains in the $\SUthree$ group.  For an action linear in the $\suthree$ components
of $U(x,\mu)$ it is easy to prove that with these equations, the energy is
conserved.  Then $D_{x\mu j} S[U(x,\mu)] = S[i\lambda_j U(x,\mu)]$ and $\dot
S[U(x,\mu)]=S[\dot U(x,\mu)]$ and therefore\footnote{We only display the
  dependence of $S$ on a single link variable here.}
\beq
  \dot H[P,U] = \sum_{x\mu} P(x,\mu) \dot P(x,\mu)
  + S[\dot U] = 0.
\eeq

When applying the derivative operator $D_{x\mu j}$ to the pure gauge action
$\Sg[U]$ only those plaquettes contribute which contain $U(x,\mu)$. Using the
definition of the staples \eqref{eq:staples} we can write

\beqn
  D_{x\mu j} \Sg[U] = -\frac{2}{g_0^2} 
  \Re \Tr \Bigl[ i\lambda_j U(x,\mu) S^\dagger(x,\mu) \Bigr].
\eeqn

\subsection{Detailed balance}

In the following, we prove that the Hybrid Monte Carlo algorithm
fulfills detailed balance. While the previous sections motivated
the algorithm by the physics of the system represented by the
Hamilton operator $H$, this proof is based on weaker prerequisites.
We assume that the probability of a transition $U' \leftarrow U$
is based on three probability distributions: at the beginning of
a trajectory, momenta are drawn from a Gaussian distribution
\beq
  P_{\rm M}(P) = \exp\left(-\frac 1 2 \sum_{x\mu j} P(x,\mu,j)^2 \right).
\eeq
Following that, the combined set of variables $(P,U)$ is used
as initial condition for a deterministic process resulting in
a new set $(P',U')$ according to $(P',U') = T(P,U)$. The corresponding
probability matrix is
\beq
  P_{\rm T}((P',U')\leftarrow (P,U)) = \delta((P',U'), T(P,U)).
\eeq 
Finally, the new set of link variables is accepted with the
condition
\beq
  P_{\rm A}((P',U')\leftarrow (P,U)) 
  = \min\Bigl\{1, \exp(-H[P',U']+H[P,U]) \Bigr\}.
\eeq
with the Hamiltonian from equation \eqref{eq:HMCHamilton}.
As the momentum variables are discarded, we have to study the 
combined transition probability
\beqn
  P(U'\leftarrow U) = \int D[P']D[P]
   P_{\rm A}((P',U')\leftarrow (P,U)) \times \nonumber\\
   P_{\rm T}((P',U')\leftarrow (P,U)) P_{\rm M}(P).
\eeqn
As further input, we demand that the computed trajectory is
reversible,
\beq
  P_{\rm T}((P',U')\leftarrow(P,U)) = P_{\rm T}((-P,U)\leftarrow (-P',U')).
  \label{eq:TrajectoryReversible}
\eeq
Now we can use $P_{\rm M}(P) e^{-S[U]} = e^{-H[P,U]}$ to rearrange the left side of
the detailed balance condition,
\beqn
  \lefteqn{P(U'\leftarrow U) e^{-S[U]}} \nonumber\\
    &=& \int D[P']D[P] \min\left\{1, \exp(-H[P',U']+H[P,U])\right\} 
    \times\nonumber\\
    &&\quad\times P_{\rm T}((P',U')\leftarrow (P,U)) \exp(-H[P,U]) \nonumber\\
    &=& \int  D[P']D[P] \min\left\{1, \exp(-H[P,U]+H[P',U'])\right\} 
    \times\nonumber\\
    &&\quad\times P_{\rm T}((P',U')\leftarrow (P,U)) \exp(-H[P',U']) \nonumber\\
    &=& \int D[P']D[P] \min\left\{1, \exp(-H[P,U]+H[P',U'])\right\}
    \times\nonumber\\
    &&\quad\times P_{\rm T}((P,U)\leftarrow (P',U')) \exp(-H[P',U']) \nonumber\\
    &=& P(U\leftarrow U') e^{-S[U']}.
\eeqn

Thus, we have proven detailed balance for the described algorithm. Note that
we have not required that the Hamiltonian is used in the computation of the
trajectory. The trajectory may be computed with different Hamiltonians or
different discretizations of it, as long as these mechanisms are reversible.
While the validity of the algorithm is not disturbed by such a choice, it may
be important for achieving reasonable acceptance rates.

\subsection{Computation of the trajectory}

For the computation of a trajectory, we need an integration scheme for the
equations of motion fulfilling \eqref{eq:TrajectoryReversible}.  Here we use
the \emph{leapfrog} algorithm. As parameters, it has a step size $\delta\tau$
and the trajectory length $\tau=n\delta \tau$.

The integration begins with an update step for the conjugate momenta with a
step size $\Delta \tau = \delta\tau/2$. It is followed by $n-1$ update steps
with $\Delta \tau = \delta\tau/2$ for the link variables, alternating with the
momentum variables. Finally, the link variables are updated with $\Delta \tau
= \delta\tau$, and the momentum variables with $\Delta \tau = \delta\tau/2$.

The explicit formulae for these steps are
\beqn
  P'(x,\mu,j) &=&
    P(x,\mu,j) - D_{x\mu j} S[U] \Delta\tau \nonumber\\
  U'(x,\mu) &=&
    \exp\left\{ \sum_{j=1}^8 i\lambda_j P(x,\mu,j) \Delta\tau \right\}
    U(x,\mu).
\eeqn

The single steps cause a discretization error of the order $\delta\tau^3$
each. Therefore, the action for the final configuration is expected to differ
from the initial configuration by an error of order $\delta\tau^2$.

\subsection{Computation of the exponential}

In the Hybrid Monte Carlo algorithm, each step on a trajectory involves for
each link the evaluation of the exponential of an element of the gauge group
algebra. On the one hand, it is desirable to minimize the cost of this
evaluation. On the other hand, the HMC algorithm is only valid if the
computation of trajectories fulfills the requirement of reversibility. If we
approximate the exponential $\exp(A) \approx E(A)$, reversibility demands that
$E(-A) E(A) = 1$.

It is remarkable that this requirement can be easily fulfilled in the case of
a $\Uone$ group, where $A \in i {\bf R}$.  Here, $E(A) = (1+A/2)/(1-A/2)$ is a
valid approximation (considerations about acceptance rates notwithstanding).
It is straightforward to generalize this to a $\SUtwo$ group, whereas the
necessary inversion of a matrix becomes an obstacle in $\SUthree$.

\subsubsection{Functions of SU(3) matrices}

Following \cite{KennedyCommunication}, we can transform any polynomial in an
$\SUthree$ matrix into a second order polynomial.  We note that for any $3\times
3$ matrix $A$, the determinant can be written as
\beq
  \det A = \frac 1 6 \left[ (\Tr A)^3 - 3 (\Tr A)(\Tr A^2) + 2(\Tr A^3) \right].
\eeq
In the $\SUthree$ case, the argument of the exponential is an element of the Lie
algebra $\suthree$ with $\Tr A = 0$, so we have
\beq
  \det A = \frac 1 3 \Tr A^3.
\eeq
According to the Cayley-Hamilton theorem, $A$ is a solution of its
characteristic equation\footnote{By $I$, we denote the unity matrix.}
\beqn
  \lefteqn{ \det (A - \lambda I) = } \nonumber\\
&=& \frac 1 6 \left\{ [\Tr(A-\lambda I)]^3
                      - 3 \Tr(A-\lambda I) \Tr(A-\lambda I)^2
                      + 2 \Tr(A-\lambda I)^3 \right\} \nonumber\\
&=& -\lambda^3 + \frac 1 2 \lambda \Tr A^2 + \frac 1 3 \Tr A^3.
\eeqn
Therefore,
\beq
  A^3 = \left( \frac 1 2 \Tr A^2 \right) A +
        \left( \frac 1 3 \Tr A^3 \right) I.
  \label{eq:A3relation}
\eeq
Thus, any power series in $A$ is equivalent with a second order polynomial in
$A$ with complex coefficients. These coefficients are in turn power series in
$\Tr A^2$ and $\Tr A^3$. From $A=-A^\dagger$ we can conclude that $\Tr A^2 =
-\Tr(A A^\dagger)$ is real, and $\Tr A^3 = -\Tr({A^\dagger}^3)$ is imaginary.
More explicitly, we set $a=-\frac 1 2 \Tr A^2 \in {\bf R}$ and $b = \frac i 3
\Tr A^3 \in {\bf R}$. Then an analytic function $f(A)$ can be written as
\beq
  f(A) = a_2 A^2 + a_1 A + a_0 I.
\eeq
The three coefficients $a_2$, $a_1$ and $a_0$ are basis-independent and can in
principle be computed exactly. Practically, computing $f(A)=\exp(A)$ in this
manner is complicated. When eigenvalues of $A$ are approximately degenerate,
different formulae have to be used to avoid numerical problems. In particular,
on a SIMD (single instruction, multiple data) computer such case distinctions
are inefficient and difficult to implement.

\subsubsection{Taylor expansion}

One will in general opt for an approximation of the exponential in the form of
a Taylor expansion. In this case, reversibility is not obvious, and one has to
make sure that the approximation is accurate to machine precision in the range
of arguments which are reasonably expected to occur.

When $f(A)$ is a Taylor expansion to order $N$, the coefficients $a_2$, $a_1$
and $a_0$ are polynomials of degree $N-2$ in $a$ and $b$. It is easy to see
that the number of floating point operations is significantly reduced compared
to the naive computation of the matrix polynomial, with one of the standard
techniques which are discussed e.g. in \cite{GolubVanLoan}.

One way to evaluate the scalar polynomials is to compute them explicitly (e.g.
with Maple V) to the given order for general arguments and implement Horner
schemes for them in the simulation program. It is however advantageous for the
implementation of the above algorithm to take into account its recursive
nature.  To see this, we define a series of coefficient sets by
\beqn
 \sum_{k=0}^N \frac{A^k}{k!}
&=& a_2^{(N)} A^N + a_1^{(N)} A^{N-1} + a_0^{(N)} A^{N-2}
    + \sum_{k=0}^{N-3} \frac{A^k}{k!} \nonumber\\
&=& a_2^{(N-1)} A^{N-1} + a_1^{(N-1)} A^{N-2} + a_0^{(N-1)} A^{N-3}
    + \sum_{k=0}^{N-4} \frac{A^k}{k!} \nonumber\\
&=& \ldots
\eeqn
Obviously we have to start with the triplet
\beq
  a_2^{(N)} = \frac{1}{N!}, \;
  a_1^{(N)} = \frac{1}{(N-1)!}, \;
  a_0^{(N)} = \frac{1}{(N-2)!},
  \label{eq:TaylorInitial}
\eeq
and we are finally interested in
\beq
  a_2 = a_2^{(2)}, \;
  a_1 = a_1^{(2)}, \;
  a_0 = a_0^{(2)}.
\eeq
With the help of 
\beq
  A^3 = -a A - i b I,
\eeq
the recursion relation for the $a_i^{(k)}$ comes out as
\beqn
  a_2^{(k)} &=& a_1^{(k+1)} \nonumber\\
  a_1^{(k)} &=& a_0^{(k+1)} - a \, a_2^{(k+1)} \nonumber\\
  a_0^{(k)} &=& \frac{1}{(k-2)!} - ib \, a_2^{(k+1)}.
  \label{eq:TaylorRecursion}
\eeqn
This recursion scheme can be implemented as a loop $k=(N-1) \ldots 2$.  In such
an implementation, one can conveniently adjust the order $N$ to which the
Taylor expansion is used.

\subsubsection{Accuracy}

The error by approximating the exponential by a truncated Taylor expansion of
a $3\times 3$ matrix can be estimated by

\beqn
  \lefteqn{ \left\| \exp{A} - \sum_{k=0}^N \frac{A^k}{k!} \right\|_2 } \nonumber\\
&\le& \frac{3}{(N+1)!} \, \max_{0 \le s \le 1}
      \left\| A^{N+1} \exp^{(N+1)}(As) \right\|_2.
\eeqn
Using the Cauchy-Schwarz inequality and the fact that for a unitary matrix A,
$\|A\|_2=1$, one gets an estimate for the error

\beqn
  \lefteqn{ \left\| \exp{A} - \sum_{k=0}^N \frac{A^k}{k!} \right\|_2 } \nonumber\\
&\le& \frac{3}{(N+1)!} \| A^{N+1} \|_2 \,
      \max_{0 \le s \le 1} \| \exp(As) \|_2 \nonumber\\
&\le& \frac{3}{(N+1)!} \|A\|_2^{N+1}.
\eeqn
In the leapfrog algorithm, the relevant values of $A$ come from a statistical
ensemble. We have

\beq
  A = i \delta\tau \sum_{i=j}^8 \lambda_j P_j,
\eeq
where $\delta\tau$ is the step size, $\{\lambda_j,j=1\ldots 8\}$ are the
Gell-Mann matrices, and $P_j$ are the normally distributed momenta.
Obviously, the norm of $A$ is linear in $\delta\tau$. Numerically, one can see
that the norm very rarely ($\ll 10^{-5}$) exceeds $7\delta\tau$.

If we take this as a realistic bound on $\| A \|_2$ and demand
that the error in the result be smaller than $10^{-8}$, we obtain
an upper bound on $\delta\tau$, depending on $N$,
\beqn
  \frac{3}{(N+1)!} (7\delta\tau)^{N+1} = 10^{-8} \nonumber\\
\Rightarrow 
  \delta\tau = \frac 1 7 
    \left( \frac{10^{-8} (N+1)!}{3} \right)^\frac{1}{N+1}.
\eeqn
In table~\ref{tab:DeltaTauBound}, values for the order $N$ necessary for some
example values of $\delta\tau$ are listed.


\begin{table}[t]
  \begin{center}
    \begin{tabular}{l|llllllll}
      \hline\hline
      $N$ & 4 & 6 & 8 & 10 & 12 & 14 & 16 \\
      $\delta\tau$ & 0.0075 & 0.0297 & 0.0677 & 0.1189 & 0.1804 & 0.2498 &
      0.3252 \\
      \hline\hline
    \end{tabular}
    \caption{\sl Upper bound for $\delta\tau$ depending on the order
      $N$ of the Taylor expansion for the exponential.}
    \label{tab:DeltaTauBound}
  \end{center}
\end{table}

\subsection{Tests}

We have tested our implementation of the Hybrid Monte Carlo algorithm in
several ways.

\subsubsection{Reversibility}

For guaranteeing the correctness of the molecular dynamics evolution, one
should check whether trajectories are truly reversible. It has been seen
previously in \cite{heplat9708017} that the classical equations of motions in
the Hybrid Monte Carlo are chaotic in nature. This means, roundoff errors
introduced at any place in the computation are amplified exponentially with
the Monte Carlo time, even for $\delta\tau\to 0$. It is not evident however
whether this phenomenon plays a role for our choice of parameters.

A remarkable example in \cite{heplat9606004} demonstrates that reversibility
violations do not necessarily cause the acceptance rate to be low. As the
leapfrog integration scheme has a preference for conserving energy even when
the computed trajectory deviates much from the true trajectory, the energy
difference $\delta H$ may be small although there are reversibility violations
in the configuration itself and in observables.

To check reversibility explicitly, we compute the trajectory once, change the
sign of all momentum variables, and then compute a new trajectory with the
same length and step size. The resulting configuration should be identical
with the starting configuration, apart from unavoidable roundoff errors with a
magnitude comparable to the machine precision . A measure for the error is the
quantity
\beq
  \sqrt{ \sum_{x\mu} \left( \| \tilde U(x,\mu) - U(x,\mu) \|_2 \right)},
\eeq
where $\tilde U$ is the endpoint configuration. Obviously, this observable is
a more direct measure than typical observables only involving a subset of all
degrees of freedom. With several initial configurations and parameters $\tau$
and $\delta\tau$ used in our study, we could see that this error measure is of
the order of some $10^{-14}$ on a PC with 64 bit arithmetics.  On the APE100
machine with 32 bit arithmetics, it turns out to be of the order of some
$10^{-7}$.

First, this indicates that the trajectory computation is indeed correct apart
from roundoff errors. Second, it shows that at least in the range of
our parameters $\tau$ and $\delta\tau$, there is no sign that errors grow
exponentially with the number of steps on the trajectory. For larger lattices,
we expect that the number of steps per trajectory has to be increased.
Therefore, at some point the integration should become unstable. A detailed
analysis can be found in \cite{heplat0005023}.

As an additional check, we have also investigated which effect rounding errors
have on the measured coupling. To this end, we have implemented a version of
the program that artificially adds noise to the force in the update of the
momentum variables.  Precisely, in each computation of a $P(x,\mu,j)$
proposal, we substitute
\beq 
  D_{x\mu j} S[U] \rightarrow (1 + n) D_{x\mu j} S[U].
\eeq
Here, $n$ is drawn from a Gaussian distribution with width $\sigma$. The
parameter $\sigma$ can be chosen to model different magnitudes of errors.

\begin{table}[htb]
  \begin{center}
    \begin{tabular}{lll}
      \hline\hline
      $\sigma$ & $\gbar^{-2}$ \\ 
      \hline
      0.100000 & 1.0251(147) \\
      0.010000 & 1.0203(11) \\
      0.001000 & 1.0203(10) \\
      1.0e-7   & 1.0218(4) \\
      1.0e-15  & 1.0218(7)   \\
      \hline\hline
    \end{tabular}
    \caption{\sl Results for the coupling for different values of 
      artificial error magnitudes $\sigma$.}
    \label{tab:roundeff}
  \end{center}
\end{table}

On an APE100 computer with single precision, we have performed simulations at
$\beta=9.596$ on a $L/a=4$ lattice with $\sigma=10^{-7}, 10^{-3}, 10^{-2},
10^{-1}$. Table~\ref{tab:roundeff} shows the obtained results. In this table,
we have also inserted a value $\sigma=10^{-7}$ for the undisturbed simulation,
assuming that this is the typical magnitude of errors caused by single
precision arithmetics. This simulation result is averaged over $270000$
iterations, while the other runs consisted of only $50000-60000$ iterations.
As comparison, there is also a value of the coupling for $\sigma=10^{-15}$
obtained from a simulation on a PC with double precision. A simulation with
the HOR algorithms yields the result $\gbar^2=1.02175(44)$.

\begin{figure}[htb]
  \begin{center}
    \epsfig{file=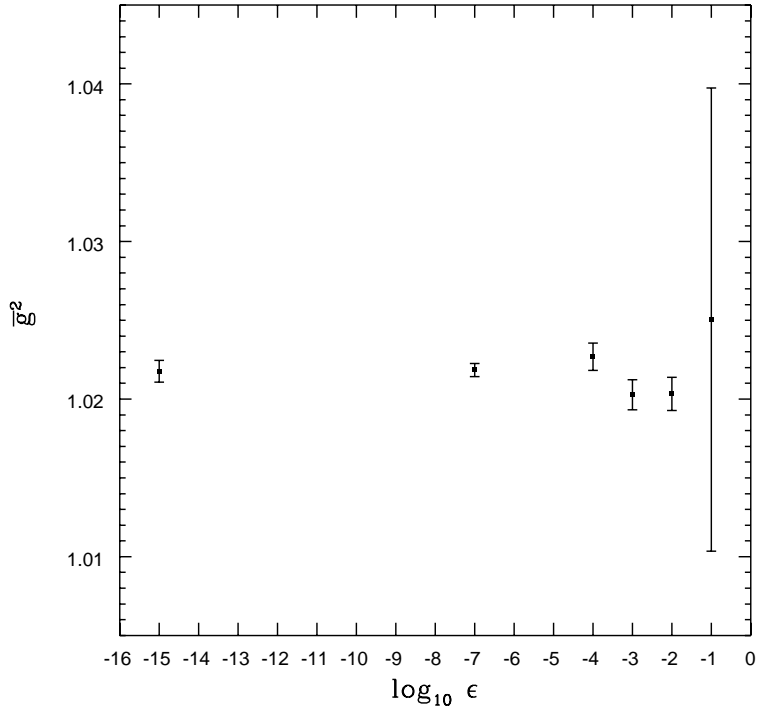,width=11cm,height=11cm}
    \vspace{-0.6cm}
    \caption{\sl Coupling in dependence of the artificial error $\sigma$.}
    \label{fig:roundeff}
  \end{center}
\end{figure}

Figure \ref{fig:roundeff} shows the results for $\gbar^{-2}$ in a plot where
the ordinate is logarithmic in $\sigma$. This illustrates that the coupling is
very insensitive against the Gaussian distributed errors in the force. Even at
$\sigma=0.01$, the coupling drifts only on a permille level. At even larger
$\sigma$, the acceptance rate collapses and the error grows, so that
systematic effects can not be measured.

This investigation increases our trust in the feasibility of the Hybrid Monte
Carlo on a single precision computer. Nevertheless, we note that it is not
clear that our model of Gaussian rounding errors is realistic; different
asymmetric models may introduce larger biases.  We have also restricted
ourselves to a specific $L/a$ and $\beta$.

In addition, we emphasize that in global sums in the APE100 code, a double
precision data type implemented in software is used \cite{AccurateSummation,
  Kahan}. Thus, on large lattices with many terms the loss of precision due to
cancellations is significantly reduced.

Rounding errors do not only play a role for the reversibility
of the HMC. For the implementation of any algorithm, the validity
of the simulation requires that all link variables remain in the
$\SUthree$ group. Due to rounding errors, matrices move away from
this group in the course of typically thousands of iterations
in a simulation. To avoid this, from time to time it is necessary
to ``project'' them back. Assuming that a variable has the value 
$v=u(1+\epsilon)$, where $\epsilon$ is the deviation from the
true $\SUthree$ matrix $u$, one can construct a corrected variable
$w$ with
\beqn
  x &=& \frac 1 2 v ( 3 - v^\dagger v) \nonumber\\
  w &=& x \left(1 - \frac i 3 \Im \det x \right).
\eeqn
This formulae are based on the assumption that when $\epsilon$
is in the order of typical roundoff errors of the machine,
$\epsilon^2$ is negligible. Then
\beqn
  v^\dagger v &\approx& 1 + \epsilon + \epsilon^\dagger \nonumber\\
  \det v &\approx& 1 + \Tr \, \epsilon,
\eeqn
further
\beqn
  x &\approx& u \left (1+ \frac{\epsilon-\epsilon^\dagger}{2} \right) \nonumber\\
  x^\dagger x &\approx& 1 \nonumber\\
  \det x &\approx& 1 + \Tr \, \frac{\epsilon-\epsilon^\dagger}{2},
\eeqn
and finally
\beqn
  w &\approx& u \left(1+\frac{\epsilon-\epsilon^\dagger}{2} \right)
        \left(1-\Tr\,\frac{\epsilon-\epsilon^\dagger}{6} \right) \nonumber\\
  w^\dagger w &\approx& 1 \nonumber\\
  \det w &\approx& 1.
\eeqn
All the preceding equations are meant to be valid up to order
$\epsilon^2$.

\subsubsection{Control variable}

Each step in the leapfrog is area preserving
\cite{MontvayMuenster}. This means, if we denote with $(P',U')$
the configuration going into the acceptance step,
\beq
  D[P'] D[U'] = D[P] D[U].
\eeq
Consequently, 
\beqn
 &&  \int D[P'] [DU'] \exp(-H[P',U']) \nonumber\\
 &&  \quad\quad= \int D[P] D[U] \exp(-H[P,U]-(H[P',U']-H[P,U])).
\eeqn
If we measure $E:=\exp(-H[P',U']+H[P,U])$ as an observable,
we expect exactly
\beq
  \langle E \rangle = 1.
\eeq
This can be used as a test for the correctness of the
leapfrog implementation and its reversibility.

\section{Local Hybrid Monte Carlo}
\label{sec:LocalHybridMonteCarlo}

We also study here a local version of the Hybrid Monte Carlo which was
apparently first discussed in \cite{heplat9306013}. As this algorithm does not
collectively move all degrees of freedom through the phase space, one
naturally does not expect it to be suitable for non-local actions. For local
theories, it is unlikely to compete with finite step size algorithms, like
Hybrid Overrelaxation. But in contrast to HOR, it is suitable for actions
which are not linear in the link variables. It is therefore a candidate for
the update of improved bermions, which we will investigate in
chapter~\ref{chap:Bermions}.

In the global Hybrid Monte Carlo, in each step on a trajectory the global
candidate configuration of new momentum and link variables are updated.  In
the Local Hybrid Monte Carlo (LHMC), a sweep over the lattice is performed.
For each link variable, a trajectory is computed with all other link variables
kept fixed.

When applied to a pure gauge theory, there are some advantages in this local
approach:

\begin{itemize}
\item Since for each trajectory only a single link variable
is changed, the discretization error is much smaller than for
the global case. Therefore, larger step sizes should be possible than
for the HMC. 

\item
Computing the ``force'' $D_{x\mu j} S[U]$ in each
step is quite cheap because the sum of staples stays constant over
a whole trajectory and has to be computed only once per trajectory.
Therefore, additional steps are quite cheap. In this study, this
advantage has turned out to be of minor importance, as the optimal
parameter set has $\tau/\delta \tau \approx 2 - 3$.
\end{itemize}

It is evident that the LHMC is not a suitable algorithm for the simulation of
pseudofermions. With a fermionic action, the computation of the force goes
with the volume of the lattice, which means that the cost of the algorithm is
quadratic in the volume.

We also mention here that for a pure Wilson action, the leapfrog algorithm
for calculating a trajectory can be replaced by an exact integration of the
equations of motion \cite{heplat9311017}. The method in this reference makes
use of the fact that in certain subgroups of $\SUthree$ the action takes the form
of the energy of a pendulum.

\section{Results}

Our aim is to represent an efficiency measure in a way that allows a
machine-independent comparison of different algorithms. Furthermore, we want
to investigate the cost of computing the observable $\gbar$, which is directly
relevant for the computation of the running coupling. Therefore, we define a
measure $\Scost$ such that in order to compute $\gbar^{-2}$ at $1\%$ accuracy,
the equivalent in complexity of $\Scost$ computations of all staples is
required. Explicitly,
\beqn
  \Scost 
&=&  (\mbox{number of computations of all staples}) \nonumber\\
&&\quad \times \left[ 
  \frac{100 \cdot (\mbox{error of $\gbar^{-2}$})}{\gbar^{-2}} \right]^2.
  \label{eq:Scost}
\eeqn
This measure already excludes a trivial volume factor. For the error of the
cost, we estimate
\beq
  \delta \Scost = \Scost \frac{\delta\tauint}{\tauint},
\eeq
which implies that we neglect the error of the autocorrelation function at
zero, $\Gamma(0)$, as explained in appendix~\ref{app:ErrorAnalysis}.

For the HMC algorithm, the number of staples computations is the number of
trajectories times $n+1$, when $n\delta\tau=\tau$ is the trajectory length.

For the LHMC algorithm, we have already noted that further additional steps on
a trajectory are cheaper than the initial one. In our experience, an
appropriate estimate for an equivalent of staples computations on our
computers is $1.6+0.4n$. For example, a trajectory with $n=6$ is only twice as
expensive as one with $n=1$. For the Hybrid Overrelaxation algorithm, we use
$1+\Nor$ as an equivalent for the staples computations.

We have performed this study at constant physics in the sense of the
Schr\"odinger functional. This means, for each lattice size $L/a$, the bare
parameter $\beta$ is tuned such that the measured coupling of the system
matches $\gbar^2=2.100$. We have considered lattice sizes $L/a=4,6,8,10$.
Except for the $L/a=4$ case, the tuned $\beta$ values were taken from
\cite{heplat9309005}. Table~\ref{tab:tuned_beta} shows the values we have
used.

\begin{table}[htbp]
  \begin{center}
    \begin{tabular}{rl}
      \hline\hline
      $L/a$ & $\beta$ \\
      \hline
      4  & 6.7796 \\
      6  & 7.1214 \\
      8  & 7.3632 \\
      10 & 7.5525 \\
      \hline\hline
    \end{tabular}
    \caption{\sl Tuned $\beta$ values for various $L/a$ belonging
      to $\gbar^2=2.100$.}
    \label{tab:tuned_beta}
  \end{center}
\end{table}

Most of the simulations were done on APE100 machines with 8 nodes, while the
largest lattices were simulated on machines with 128 nodes. For the Hybrid
Monte Carlo algorithms, we have used trajectory lengths of 0.5, 1.0 and 2.0
and varied the step size $\delta \tau$ in a range with reasonable acceptance
rates $\Pacc$.  In addition to the coupling, we have also measured the
acceptance rates. This was motivated by the common folklore that optimal
efficiency can be achieved with an acceptance of about $70\%$. It thus makes
sense to parametrize plots of the cost also by the acceptance rate. For the
Hybrid Overrelaxation algorithm, we have varied $\Nor$ from 0 (corresponding
to a pure heatbath) to 3.

\begin{table}[p]
  \begin{center}
    \begin{tabular}{rrrrrrrr}
      \hline\hline
      $L/a$ & $\Nor=0$ & $\Nor=1$ & $\Nor=2$ & $\Nor=3$ & 
              $\Nor=4$ & $\Nor=5$ & $\Nor=6$ \\
      \hline
      4  & 5.03(1)  & 3.41(5)  & 4.07(6) & 4.98(7)  \\
      5  & 9.20(26) & 5.68(10) & 6.42(9) & 7.61(11) \\
      6  & 13.8(4)  & 8.71(16) & 9.31(2) & 11.0(2)
         & 12.4(2)  & 14.4(2) & 16.3(3) \\
      8  & 28(1)    & 16.5(3)  & 17.9(3) & 19.2(3) \\
      10 & 46.3(2)  & 25.4(7)  & 26.4(6) & 30.1(7) \\
      \hline\hline
    \end{tabular}
    \caption{\sl $\Scost/1000$ for the HOR algorithm for different $L/a$ and $\Nor$.}
    \label{tab:scost_hor}
  \end{center}
\end{table}

\begin{table}[htbp]
  \begin{center}
    \begin{tabular}{rrrrrrrr}
      \hline\hline
      $L/a$ & $\Nor=0$ & $\Nor=1$ & $\Nor=2$ & $\Nor=3$ 
            & $\Nor=4$ & $\Nor=5$ & $\Nor=6$ \\
      \hline
      4  & 2.10(5) & 0.72(1) & 0.57(1) & 0.52(1) \\
      5  & 2.74(8) & 0.86(1) & 0.65(1) & 0.58(1) \\
      6  & 3.21(8) & 1.02(2) & 0.72(1) & 0.64(1) & 0.57(1) & 0.55(1) & 0.53(1)\\
      8  & 4.5(2)  & 1.27(2) & 0.92(2) & 0.75(1) \\
      10 & 5.4(2)  & 1.47(4) & 1.02(2) & 0.86(2) \\
      \hline\hline
    \end{tabular}
    \caption{\sl $\tauint$ for the HOR algorithm for different $L/a$ and $\Nor$.}
    \label{tab:tauint_hor}
  \end{center}
\end{table}

\begin{figure}[p]
  \begin{center}
    \epsfig{file=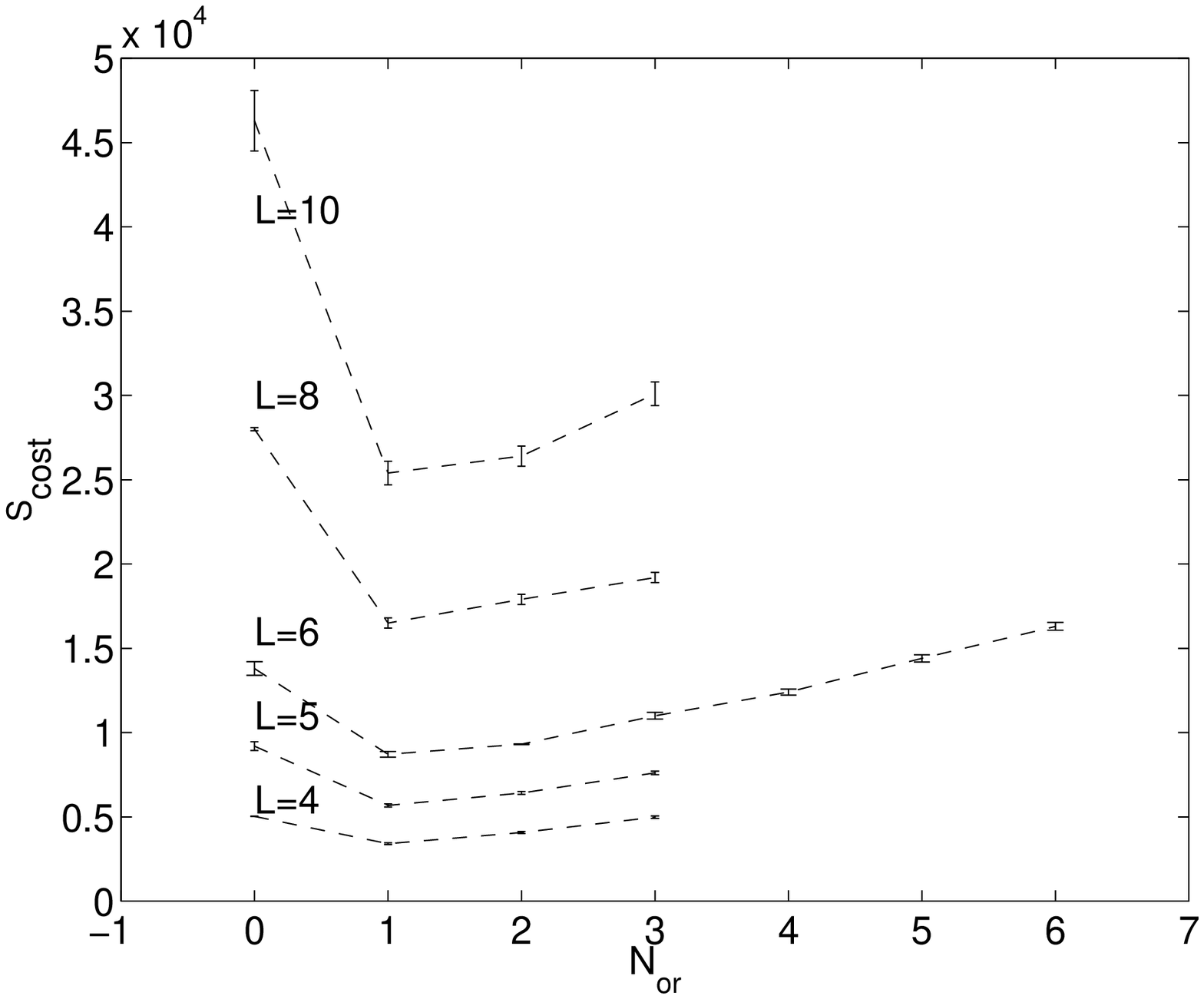,width=8cm,height=8cm}
    \caption{\sl $\Scost$ for the HOR algorithm.}
    \label{fig:hor}
  \end{center}
\end{figure}

Table~\ref{tab:scost_hor} shows the cost of the HOR algorithm. The
corresponding plot can be found in figure \ref{fig:hor}. For all lattice sizes
$L/a$, the optimal choice of $\Nor$ turns out to be 1.
Table~\ref{tab:tauint_hor}, in which we list the autocorrelation times for the
same runs, indicates the reason for this: for $\Nor=1$, there are almost no
autocorrelations. If the number of overrelaxation sweeps per measurement is
increased, the computational cost increases, while $\tauint$ in units of
updates cannot decrease further.

\begin{figure}[p]
  \begin{center}
    \epsfig{file=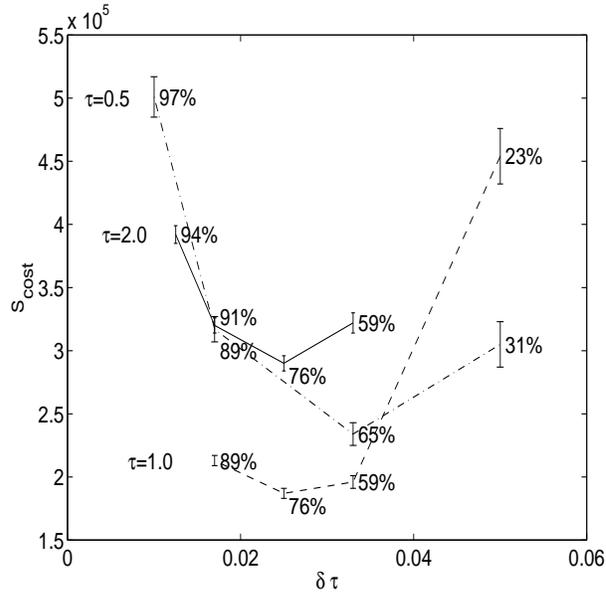,width=8cm,height=8cm} 
    \caption{\sl $\Scost$ for global Hybrid Monte Carlo at $L/a=6$.
      For each data point, the corresponding acceptance is shown.}
    \label{fig:hmc6}
  \end{center}
\end{figure}

\begin{figure}[p]
  \begin{center}
    \epsfig{file=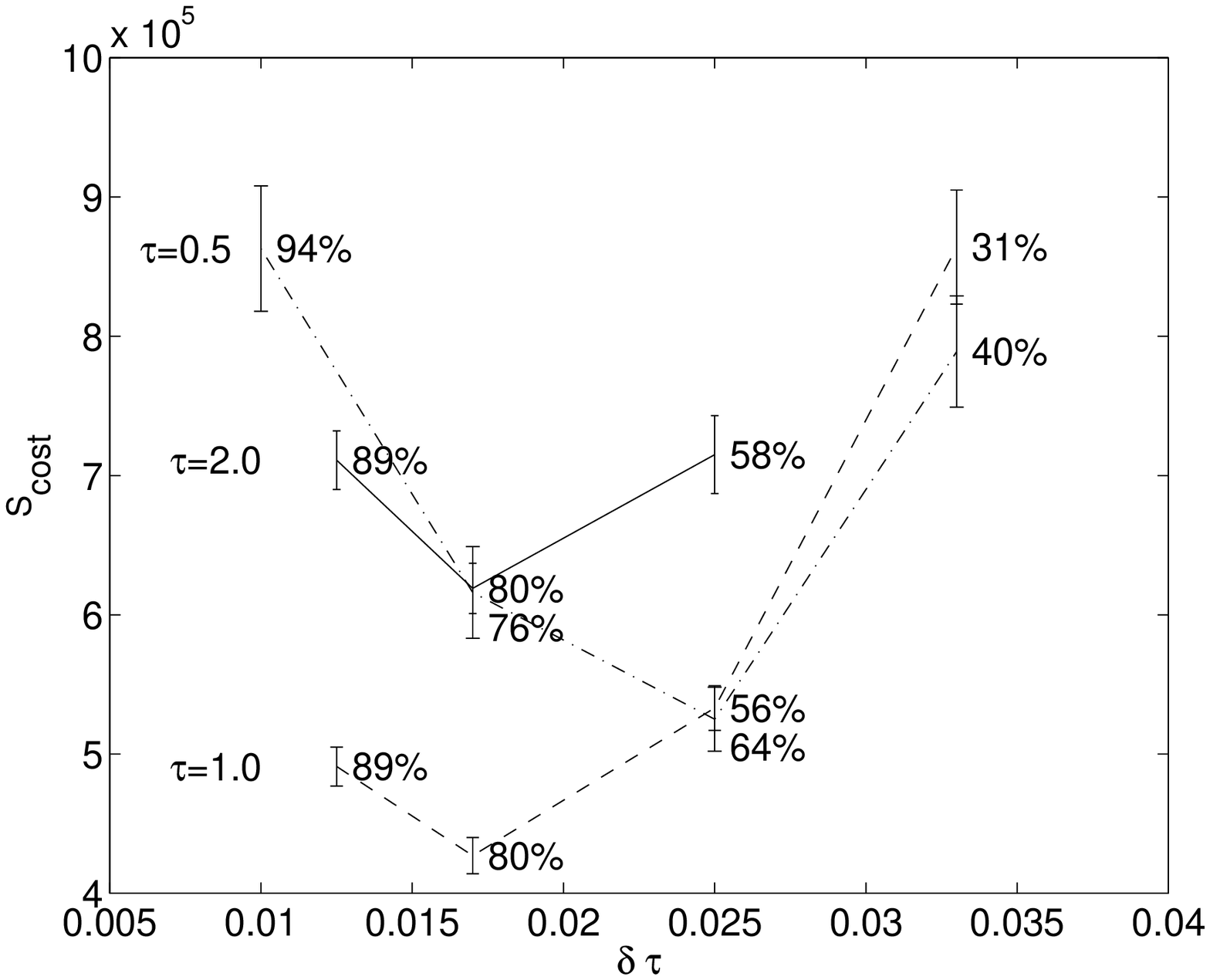,width=8cm,height=8cm} \\
    \caption{\sl $\Scost$ for global Hybrid Monte Carlo at $L/a=8$.}
    \label{fig:hmc8}
  \end{center}
\end{figure}

In table~\ref{tab:scost_hmc}, the results for the HMC algorithm are listed.
The figures \ref{fig:hmc6} and \ref{fig:hmc8} contain plots of these data for
$L/a=6$, $L/a=8$ respectively. Both plots show the data with constant
trajectory length connected with dashed lines.

In all cases, the optimal trajectory length is 1. One can also see that for
each trajectory length, the minimal cost is achieved at an acceptance rate of
roughly 70 \%. This is consistent with conventional wisdom. When $L/a$ is
increased, the step size has to be scaled down in order to gain optimal
performance. In \cite{GuptaKilcup} it is argued that for the leapfrog discretization,
the average acceptance scales as
\beq
  \langle \Pacc \rangle = \exp\left\{-(\delta\tau L/a)^4\right\},
\eeq
and therefore the step size must be changed with $\delta\tau\propto a/L$ in
order to keep the acceptance rate fixed.  This seems to be roughly the case in
our simulations.

\begin{table}[p]
  \begin{center}
    \begin{tabular}{rrrcr}
      \hline\hline
      $L/a$ & \multicolumn{1}{l}{$\tau$} & \multicolumn{1}{l}{$\tau/\delta\tau$} 
      & \multicolumn{1}{l}{acceptance in \%} 
      & \multicolumn{1}{l}{$\Scost$} \\
      \hline
      4  &  0.5  &   8  &  50.5(2) & $ 6.71(18) \cdot 10^4 $ \\
      4  &  0.5  &  12  &  77.1(1) & $ 5.97(12) \cdot 10^4 $ \\
      4  &  0.5  &  15  &  85.1(1) & $ 6.66(16) \cdot 10^4 $ \\
      4  &  0.5  &  20  &  91.5(1) & $ 7.72(14) \cdot 10^4 $ \\
      4  &  1.0  &  20  &  62.8(2) & $ 5.55(11) \cdot 10^4 $ \\
      4  &  1.0  &  30  &  83.2(1) & $ 5.07(8)  \cdot 10^4 $ \\
      4  &  1.0  &  40  &  90.2(1) & $ 5.77(8)  \cdot 10^4 $ \\
      4  &  1.0  &  60  &  95.7(1) & $ 7.84(14) \cdot 10^4 $ \\
      4  &  2.0  &  30  &  38.3(2) & $ 1.85(4)  \cdot 10^5 $ \\
      4  &  2.0  &  60  &  83.0(1) & $ 1.09(1)  \cdot 10^5 $ \\ 
      4  &  2.0  & 100  &  93.9(1) & $ 1.46(2)  \cdot 10^5 $ \\
      \hline
      6  &  0.5  &  10  &  30.5(3) & $ 3.05(18) \cdot 10^5 $ \\
      6  &  0.5  &  15  &  64.8(2) & $ 2.34(9)  \cdot 10^5 $ \\   
      6  &  0.5  &  30  &  91.0(1) & $ 3.17(10) \cdot 10^5 $ \\   
      6  &  0.5  &  50  &  96.8(1) & $ 5.01(16) \cdot 10^5 $ \\   
      6  &  1.0  &  20  &  22.8(9) & $ 4.54(22) \cdot 10^5 $ \\    
      6  &  1.0  &  30  &  59.1(3) & $ 1.96(5)  \cdot 10^5 $ \\   
      6  &  1.0  &  40  &  76.3(2) & $ 1.87(4)  \cdot 10^5 $ \\   
      6  &  1.0  &  60  &  89.3(2) & $ 2.13(4)  \cdot 10^5 $ \\   
      6  &  2.0  &  60  &  59.4(2) & $ 3.22(8)  \cdot 10^5 $ \\   
      6  &  2.0  &  80  &  76.4(2) & $ 2.90(6)  \cdot 10^5 $ \\   
      6  &  2.0  & 120  &  89.2(1) & $ 3.20(6)  \cdot 10^5 $ \\   
      6  &  2.0  & 160  &  94.1(1) & $ 3.92(7)  \cdot 10^5 $ \\  
      \hline
      8  &  0.5  &  15  &  39.6(3) & $ 7.89(40) \cdot 10^5 $ \\
      8  &  0.5  &  20  &  63.6(2) & $ 5.25(23) \cdot 10^5 $ \\
      8  &  0.5  &  30  &  83.0(2) & $ 6.16(35) \cdot 10^5 $ \\
      8  &  0.5  &  50  &  94.1(2) & $ 8.63(45) \cdot 10^5 $ \\
      8  &  1.0  &  30  &  30.7(3) & $ 8.64(41) \cdot 10^5 $ \\
      8  &  1.0  &  40  &  56.4(2) & $ 5.33(16) \cdot 10^5 $ \\
      8  &  1.0  &  60  &  79.9(3) & $ 4.27(13) \cdot 10^5 $ \\
      8  &  1.0  &  80  &  89.3(4) & $ 4.91(14) \cdot 10^5 $ \\
      8  &  2.0  &  80  &  57.5(4) & $ 7.15(28) \cdot 10^5 $ \\
      8  &  2.0  & 120  &  80.4(2) & $ 6.19(18) \cdot 10^5 $ \\  
      8  &  2.0  & 160  &  88.5(2) & $ 7.11(21) \cdot 10^5 $ \\
      \hline\hline
    \end{tabular}
    \caption{\sl $\Scost$ for the HMC algorithm for different $L/a$, $\tau$ and $\delta\tau$.}
    \label{tab:scost_hmc}
  \end{center}
\end{table}

\begin{table}[p]
  \begin{center}
    \begin{tabular}{rrrcr}
      \hline\hline
      $L/a$ & \multicolumn{1}{l}{$\tau$} & \multicolumn{1}{l}{$\tau/\delta\tau$} 
      & \multicolumn{1}{l}{acceptance in \%} 
      & \multicolumn{1}{l}{$\Scost$} \\
      \hline
      4  &  0.5  &   2  &  85.3  &  $ 7.39(16) \cdot 10^3  $ \\ 
      4  &  0.5  &   3  &  93.7  &  $ 9.08(18) \cdot 10^3  $ \\   
      4  &  0.5  &   5  &  97.8  &  $ 1.37(3)  \cdot 10^4  $ \\    
      4  &  1.0  &   3  &  67.6  &  $ 6.11(30) \cdot 10^4  $ \\   
      4  &  1.0  &   4  &  82.6  &  $ 3.41(11) \cdot 10^4  $ \\   
      4  &  1.0  &   5  &  89.5  &  $ 3.19(9)  \cdot 10^4  $ \\ 
      4  &  1.0  &  10  &  97.6  &  $ 4.33(10) \cdot 10^4  $ \\   
      4  &  2.0  &   6  &  73.9  &  $ 4.00(11) \cdot 10^4  $ \\ 
      4  &  2.0  &  10  &  93.5  &  $ 2.92(6)  \cdot 10^4  $ \\  
      4  &  2.0  &  20  &  98.2  &  $ 5.70(11) \cdot 10^4  $ \\  
      4  &  4.0  &  20  &  92.0  &  $ 2.36(9)  \cdot 10^5  $ \\   
      \hline
      6  &  0.5  &   2  &  85.1  &  $ 2.46(7)  \cdot 10^4  $ \\ 
      6  &  0.5  &   3  &  93.7  &  $ 2.22(5)  \cdot 10^4  $ \\
      6  &  0.5  &   5  &  97.8  &  $ 2.74(6)  \cdot 10^4  $ \\
      6  &  0.5  &  10  &  99.5  &  $ 4.14(10) \cdot 10^4  $ \\
      6  &  1.0  &   4  &  80.7  &  $ 1.15(6)  \cdot 10^5  $ \\
      6  &  1.0  &   5  &  88.2  &  $ 8.60(34) \cdot 10^4  $ \\
      6  &  1.0  &  10  &  97.3  &  $ 8.74(28) \cdot 10^4  $ \\
      6  &  2.0  &   6  &  71.7  &  $ 1.04(4)  \cdot 10^5  $ \\
      6  &  2.0  &  10  &  94.7  &  $ 4.62(11) \cdot 10^4  $ \\
      6  &  2.0  &  20  &  98.6  &  $ 7.32(17) \cdot 10^4  $ \\
      \hline
      8  &  0.5  &   2  &  85.1  &  $ 4.94(18) \cdot 10^4  $ \\
      8  &  0.5  &   5  &  97.7  &  $ 4.96(15) \cdot 10^4  $ \\
      8  &  0.5  &  10  &  99.4  &  $ 7.71(23) \cdot 10^4  $ \\
      8  &  1.0  &   4  &  79.7  &  $ 2.78(21) \cdot 10^5  $ \\
      8  &  1.0  &   5  &  87.5  &  $ 2.16(13) \cdot 10^5  $ \\
      8  &  1.0  &  10  &  97.1  &  $ 2.09(11) \cdot 10^5  $ \\
      8  &  1.0  &  20  &  99.3  &  $ 3.30(16) \cdot 10^5  $ \\
      8  &  2.0  &   8  &  89.5  &  $ 9.82(37) \cdot 10^4  $ \\
      8  &  2.0  &  10  &  94.6  &  $ 9.40(31) \cdot 10^4  $ \\
      8  &  2.0  &  20  &  98.8  &  $ 1.44(5)  \cdot 10^5  $ \\
      \hline\hline
    \end{tabular}
    \caption{\sl $\Scost$ for the LHMC algorithm for different $L/a$, $\tau$ and $\delta\tau$.}
    \label{tab:scost_lhmc}
  \end{center}
\end{table}

\begin{figure}[p]
  \begin{center}
    \epsfig{file=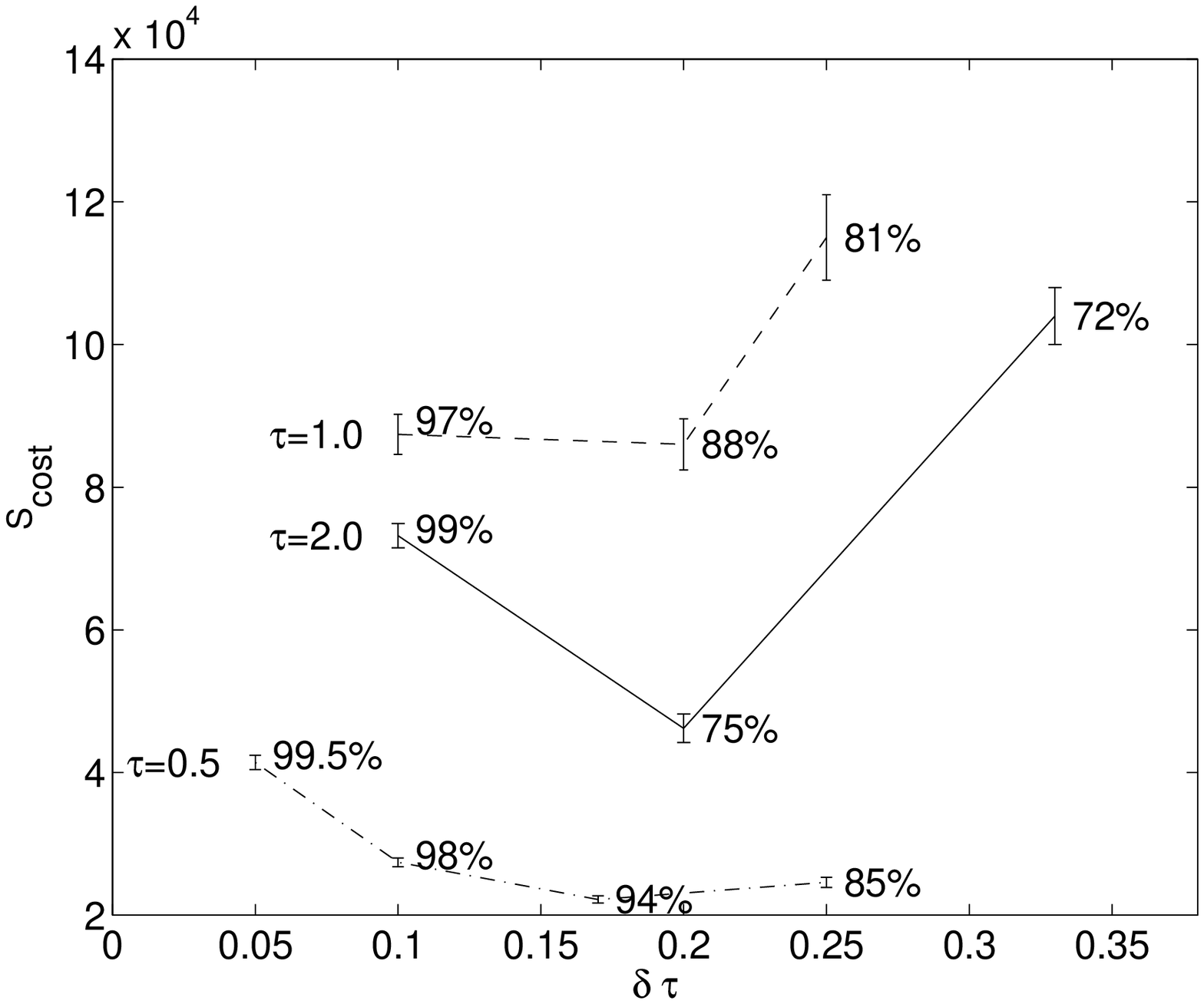,width=8cm,height=8cm}
    \caption{\sl $\Scost$ for Local Hybrid Monte Carlo at $L/a=6$.}
    \label{fig:lhmc6}
  \end{center}
\end{figure}

\begin{figure}[p]
  \begin{center}
    \epsfig{file=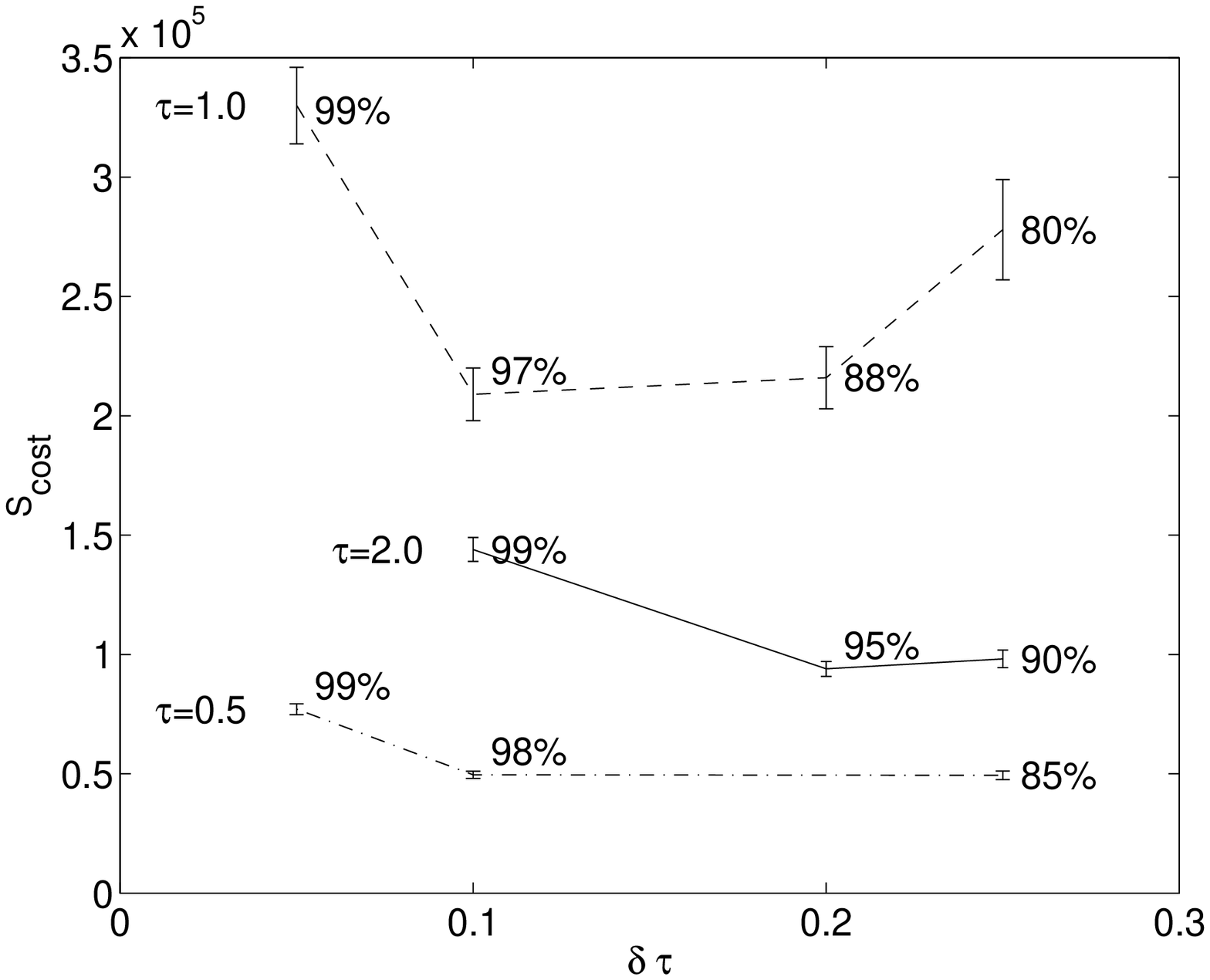,width=8cm,height=8cm} \\
    \caption{\sl $\Scost$ for Local Hybrid Monte Carlo at $L/a=8$.}
    \label{fig:lhmc8}
\end{center}
\end{figure}

\begin{figure}[p]
  \begin{center}
    \epsfig{file=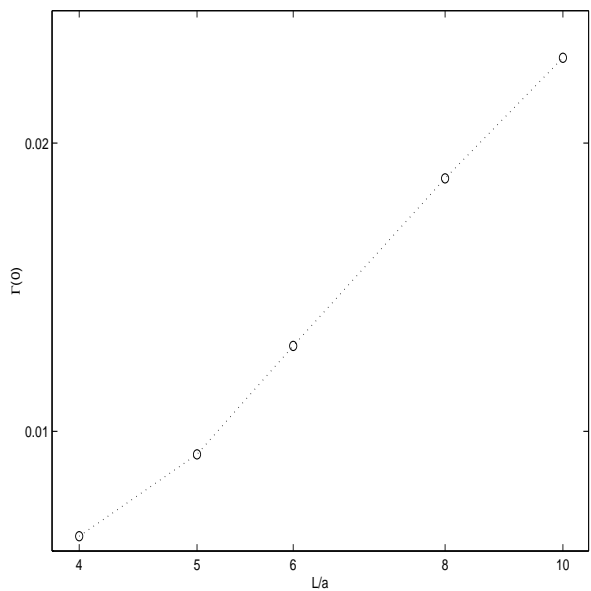,width=7.8cm,height=7.8cm}
    \caption{\sl Scaling behavior of $\Gamma(0)$ from runs with the HOR
      algorithm, in a double-logarithmic plot.}
    \label{fig:var}
  \end{center}
\end{figure}

\begin{figure}[p]
  \begin{center}
    \epsfig{file=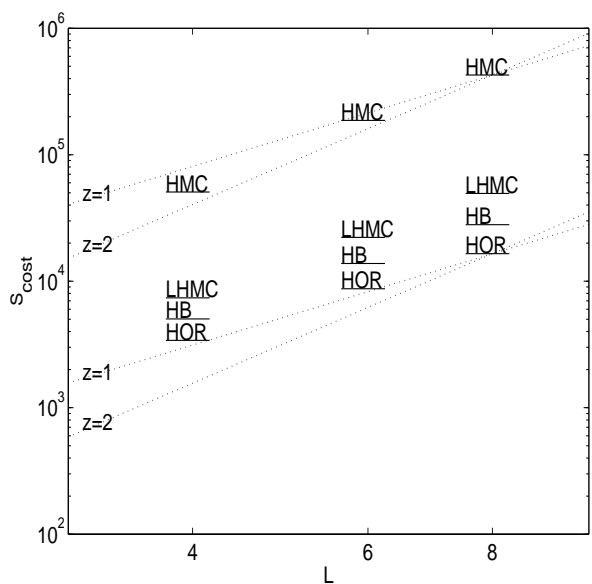,width=7.8cm,height=7.8cm}
    \caption{\sl Optimal $\Scost$ for the different algorithms in a
      double-logarithmic representation. For comparison, dotted lines which
      correspond to exponents $z=1,2$ are displayed.  The points denoted with
      HB correspond to a pure heatbath.}
    \label{fig:scale}
\end{center}
\end{figure}

Table~\ref{tab:scost_lhmc} contains our results for the cost and acceptance
rates of the LHMC algorithm. These data are illustrated in figure
\ref{fig:lhmc6} for $L/a=6$ and \ref{fig:lhmc8} for $L/a=8$.  For all $L/a$,
the optimal trajectory length is $\tau=0.5$.  It is remarkable that the cost
for $\tau=2.0$ is smaller than for $\tau=1.0$. In order to check the behavior
for even longer trajectories, we have performed runs with lower statistics and
found that the cost increases again. This may be an indication that
performance oscillates in the trajectory length. The reason for this could be
that the phase space that can be exhausted by a local trajectory is quite
small, because a link variable is restricted to the $\SUthree$ manifold. A
``good'' trajectory is one that is equivalent with an overrelaxation update.
When the trajectory length is further increased, it moves on cycles on the
sphere.

Compared to global Hybrid Monte Carlo, step sizes can be made much larger, as
expected. Optimal acceptance rates are also much larger.

In the following, we investigate how the performance of these algorithms
scales with $L/a$. In lattice field theories, the continuum limit is defined
near second order phase transitions in the bare parameter space. Simulations
in this area suffer from \emph{critical slowing down}, i.e.  the
autocorrelation time increases rapidly as the continuum is approached. One
expects that the autocorrelation time diverges as \cite{Sokal}
\beq
  \tauint \sim \min(L,\xi)^z,
\eeq
where $\xi$ is the correlation length and $z$ is an
exponent specific to the algorithm used. It is not clear that
the definitions of $z$ in the regimes $L>\xi$ and $L<\xi$
are equivalent \cite{heplat9212017}.
For simple local
algorithms like the heatbath, an intuitive understanding is that
modifications of variables are transported around by diffusive
dynamics, which suggests a critical dynamical exponent $z=2$
near the critical point. As the main cost of QCD simulations
arises from the largest lattices, $z$ is an important parameter
to characterize algorithms. 

\begin{table}[hbtp]
  \begin{center}
    \begin{tabular}{rrrrrrrr}
      \hline\hline
      $L/a$ & $\Gamma(0)$ \\
      \hline
      4 & 0.0311 \\
      5 & 0.0378 \\
      6 & 0.0491 \\
      8 & 0.0735 \\
      10 & 0.0982   \\
      \hline\hline
    \end{tabular}
    \caption{\sl $\Gamma(0)$ for runs with the HOR algorithm with
      different $L/a$. We have not determined errors for these data.}
    \label{tab:var_hor}
  \end{center}
\end{table}

According to \eqref{eq:SigmaOfGammaAndTauint}, the error of an
observable $A$ is the product of $\Gamma(0)$ and the autocorrelation
time. The former is a property of the statistical ensemble whereas the
latter is a property of the simulation algorithm. Both have their
separate scale dependence. In table~\ref{tab:var_hor} we have listed
$\Gamma(0)$ for the HOR algorithm, for which our data goes to
$L/a=10$. Figure \ref{fig:var} shows a logarithmic plot.  For a fit of
$\log(\Gamma(0))$ against $\log(L/a)$, we drop the value for $L/a=4$
because the scaling behavior is only reached asymptotically. From the
fit, we find approximately
\beq
  \Gamma(0) \propto \left( L/a \right)^{1.38}.
\eeq
Therefore, we expect our cost measure to scale like
\beq
  \Scost \propto \left( L/a \right)^{1.38+z}.
\eeq
Figure \ref{fig:scale} shows the cost of the studied algorithms
at different lattice sizes in comparison. For each $L/a$, we
have plotted the data with optimally tuned parameters. The
figure also indicates exponents $z=1,2$ by dashed lines. The
Hybrid Overrelaxation algorithm almost reaches $z=1$, while
the other algorithms scale slightly worse.

At $L/a=8$, the observed ratios in $\Scost$ for tuned parameters
are 1 : 3 : 26 for HOR : LHMC : HMC. This illustrates the overhead 
of HMC even before dynamical quarks are included.

\chapter{Bermions}
\label{chap:Bermions}

In this chapter, we are going to investigate the Schr\"odinger functional
coupling in a Yang-Mills theory coupled to a bosonic spinor field. Formally,
it corresponds to setting the number of flavors in the fermionic partition
function to $\Nf=-2$. In comparison to the full QCD case, this theory has a
local interaction and is therefore much cheaper to simulate.

In the literature, the term \emph{bermion model} has been coined for this
theory. It was introduced for the purpose of studying unquenching effects
without immediately moving to the full QCD theory
\cite{Bermions,heplat9507020,heplat9605044}. The emphasis in these articles
lies on the extrapolation of results from negative to positive flavor numbers.
In perturbation theory, this can surely be justified. There, the flavor number
is just a parameter in the Feynman graphs, and increasing powers of $\Nf$
appear with increasing loop order. Therefore, quantities calculated to finite
order are polynomials in $\Nf$.

Non-perturbatively, the $\Nf$ dependence is not straight-forward. An
extrapolation cannot be done in a reliable way and with errors under control.
Here, we are not trying to obtain quantitative information for the full theory
from the bermion theory. The model serves as tool for studying improvement and
the approach to the continuum limit.

\section{Model}

The Schr\"odinger functional, as the partition function of the system, is an
integral over all gauge and quark fields which fulfill the given boundary
conditions. After integrating out the quark fields, it is given by
\begin{eqnarray}
  \calZ = e^{-\Gamma} &=& \int D[U] D[\bar\psi] D[\psi]
                          e^{-S[U,\bar\psi,\psi]} \nonumber\\
                      &=& \int D[U] e^{-\Sg} \det(M^\dagger M)^{\Nf/2}
\end{eqnarray}
with the fermion matrix
\beq
  M = 2\kappa(D + \delta D + m_0),\quad \kappa = (8+2am_0)^{-1}.
\eeq
For $\Nf=-2$, the determinant can be integrated in,
\begin{equation}
  \calZ = \int D[U] D[\phi^\dagger] D[\phi] e^{-\Sg-\Sb}.
\end{equation}
Here, the fields $\phi(x)$ are located on the lattice sites and carry Dirac
and color indices. However, the components are complex numbers and can
therefore be regarded as boson fields. The bermion action $\Sb$ is given by
\beq
  \Sb[\phi,\phi^\dagger,U] = \sum_x |M\phi(x)|^2.
\eeq

In order to completely define the action, we have to fix the improvement
coefficients $\ct$, $\cttilde$ and $\csw$. For the first two, the choice is
easy: these coefficients are only known in perturbation theory, and
perturbative results are computed with general $\Nf$. Therefore, we can take
the formulae \eqref{eq:ct} and \eqref{eq:cttilde} and insert $\Nf=-2$ and the
value $g_0$ used in the respective simulation.

For the Sheikoleslami-Wohlert coefficient $\csw$, the situation is
more complicated. It has been determined non-perturbatively 
for the range of bare couplings $0\le g_0<1$ interesting for the ALPHA programme
for $\Nf=0$ \cite{heplat9609035} and $\Nf=+2$ \cite{heplat9803017}.
The results can be represented as smooth functions of the bare
coupling,
\beqn
  \csw(g_0)\big|_{\Nf=0} &=& 
    \frac{1-0.656 g_0^2-0.152 g_0^4-0.054 g_0^6}{1-0.922 g_0^2} \nonumber\\
  \csw(g_0)\big|_{\Nf=2} &=&    
    \frac{1-0.454 g_0^2-0.175 g_0^4+0.012 g_0^6+0.045 g_0^8}{1-0.720 g_0^2}. 
  \label{eq:cswnonpert}
\eeqn
These interpolation formulae are arranged such that they go over to the
$\Nf$-independent 1-loop result for $g_0\to 0$ \cite{heplat9606016},
\beq
  \csw(g_0) = 1 + 0.2659(1) g_0^2 + \rmO(g_0^4).
\eeq

For the purpose of this work, performing the method proposed in these papers
also for $\Nf=-2$ would be a large undertaking.  We have therefore chosen to
extrapolate the non-perturbative data from $\Nf=0$ and $\Nf=+2$ linearly in
$\Nf$. Only for the most critical parameters used in our simulations, we have
explicitly calculated $\csw$ with the method utilized in the papers cited
above and checked that our extrapolation matches the obtained values with good
accuracy.  This will be explained in more detail in section~\ref{sec:csw}.

\section{Simulation algorithm}

In \cite{heplat9907007}, the bermion model was investigated without
improvement. There, it was straight-forward to find a suitable algorithm:

\begin{itemize}
\item The gauge fields are updated with a Hybrid Overrelaxation
  algorithm as described in section~\ref{sec:HybridOverrelaxation}.
\item The boson fields are updated with an overrelaxation algorithm.
\end{itemize}

The important property of the unimproved action that made this possible is
that it is linear in each link variable. We found it practically impossible to
generalize these finite step size methods to the improved case, where the
action depends quadratically on the individual link variables.

From our experience with the Local Hybrid Monte Carlo algorithm, as discussed in
section~\ref{sec:LocalHybridMonteCarlo}, this was an alternative candidate for
the update of the gauge field.  However, the inclusion of the clover term
defeats its main advantage. In that case, the force $D_{x\mu j} S[U]$ depends
on the link $U(x,\mu)$ in a non-trivial way and has to be recomputed in each
step on a trajectory.  We therefore expect this algorithm to be relatively
expensive for improved bermions.

After some experimentation, we decided to update the gauge field by applying
local overrelaxation with respect to the action without clover term.  An
acceptance step with the local action difference with respect to the full
action then corrects for the error made. The assumption behind this method is
that the clover term only represents a small correction to the unimproved
action. Therefore, the acceptance rate should be reasonably high. Our results
confirm this assumption.

As the proposals for the gauge field update are fully deterministic and set up
in a special way, ergodicity may be severely hampered when the bermion fields
are also updated with an overrelaxation method. Therefore, we have changed the
boson field update to a global heatbath.

\subsection{Gauge fields}
\label{sec:GaugeUpdate}

In order to apply the overrelaxation method described in
section~\ref{sec:Overrelaxation}, we have to isolate a part of the full action
that is of the form \eqref{eq:ActionLinearInU}.  For the derivation of such a
contribution, we utilize the operator $M_\mu$ defined by
\beqn
  M_\mu\phi(x) = -\frac{\kappa}{a} \Bigl\{
    \lambda_\mu (1-\gamma_\mu) U(x,\mu) \phi(x+a\hat\mu) + \nonumber\\
    \lambda_\mu^* (1+\gamma_\mu) U(x-a\muhat,\mu) \phi(x-a\muhat)
  \Bigr\}.
\eeqn
Furthermore, we introduce $\overline M_\mu = \Munimpr-M_\mu$, which is the
contribution to the bermion matrix from the directions orthogonal to $\mu$.
Then the boson action can be written as
\beqn
  \Sb 
&=& |\Munimpr\phi(x)|^2 + |\Munimpr\phi(x+a\hat\mu)|^2 + \cdots \nonumber\\
&=& 2\, \Re (M_\mu\phi(x), \overline M_\mu \phi(x) ) \nonumber\\
&&\quad +
    2\, \Re (M_\mu\phi(x+a\hat\mu), \overline M_\mu \phi(x+a\hat\mu) ) 
    + \cdots \nonumber\\
&=& -\frac{2\kappa}{a}\, \Re \Bigl\{
    (\overline M_\mu \phi)^\dagger(x) \lambda_\mu (1-\gamma_\mu) 
      U(x,\mu) \phi(x+a\hat\mu) \nonumber\\
&&\quad + 
    (\overline M_\mu \phi)^\dagger(x+a\hat\mu) \lambda_\mu^* (1+\gamma_\mu)
      U^\dagger(x,\mu) \phi(x) \Bigr\} + \cdots \nonumber\\
&=& -\frac{2\kappa}{a}\, \Re \sum_{A\alpha\beta} \lambda_\mu \Bigl\{
    \Bigl[ (\overline M_\mu \phi)^\dagger(x) \Bigr]_{A\alpha} 
    U_{\alpha\beta} \, \phi(x+a\hat\mu)_{A\beta} \nonumber\\
&&\quad +
    \phi(x)_{A\alpha}^* U_{\alpha\beta} 
    \Bigl[ (\overline M_\mu \phi)^\dagger(x+a\hat\mu) 
    (1+\gamma_\mu) \Bigr]^*_{A\beta}
    \Bigr\} + \cdots \nonumber\\
&=& -2\kappa \, \Re \Tr \left\{ U(x,\mu) \xi^\dagger(x,\mu) \right\} + \cdots
\eeqn
Here we have used $A$ to denote the Dirac index and $\alpha,\beta$ to denote
color indices. The $\SUthree$ matrix $\xi$ can be computed as the dyadic
product of matrices $v$ and $w$, $\xi_{\alpha\beta}= \lambda_\mu^* (w^\dagger
v)_{\beta\alpha}$. These matrices in turn are given by
\beqn
  v &=& (1-\gamma_\mu) \overline M_\mu \phi(x) 
        + \frac{1}{a} (1+\gamma_\mu) \phi(x) \nonumber\\
  w &=& (1+\gamma_\mu) \overline M_\mu \phi(x+a\muhat)
        + \frac{1}{a} (1-\gamma_\mu) \phi(x+a\muhat).
\eeqn
The action we use to generate a proposal for a link variable now is
\beq
  S_{\rm prop} = -\Re \Tr \left\{ U(x,\mu) V^\dagger(x,\mu) \right\}
  \label{eq:BermionProposalAction}
\eeq
with
\beq
  V(x,\mu) = \frac{\beta}{3} S(x,\mu) + 2 \kappa \, \xi(x,\mu),
  \label{eq:BermionActionLinearInU}
\eeq
where $S(x,\mu)$ is the staples matrix as defined in \eqref{eq:staples}.

Note that in the action $S_{\rm prop}$ we have only considered terms from the
unimproved bermion action. In principle, we could have included the
contribution to $\Sb$ which stems from the improvement coefficient $\cttilde$
and which is linear in $U(x,\mu)$. But this term is very small, and it turns
out that varying $\cttilde$ does not change the acceptance rate of this
algorithm significantly. On the other hand, including it in the proposal
action would complicate the implementation of the algorithm such that
optimization blocks are broken and the program is slowed down. It is therefore
probably more efficient to include the tiny $\cttilde$ contribution into the
acceptance step.

The $\SUthree$ overrelaxation mechanism described in
section~\ref{sec:HybridOverrelaxation}, in which three $\SUtwo$ subgroups are
``flipped'' subsequently, is not set up symmetrically. This means, the
probability of going from one value to another is not the same as the
probability of the reverse transition.

This problem can be cured by choosing the order of subgroups randomly, i.e.
with a probability of $1/2$ for the sequences $1,2,3$ and $3,2,1$. Thus,
starting from a variable $U$ we generate a proposal $U'$ with the probability
\beq
  P_{\rm P}(U'\leftarrow U) = \frac 1 2 \Bigl\{
    \delta(U' - A_3 A_2 A_1 U) + \delta(U' - A_1 A_2 A_3 U) \Bigr\}.
\eeq
The application of these matrices is discussed after
\eqref{eq:OverrelaxationSU3}. The proposal is then accepted with
the probability
\beq
  P_{\rm A}(U'\leftarrow U) = \min\Bigl\{ 1, \exp(-S[\phi,\phi^\dagger,U']
   +S[\phi,\phi^\dagger,U]) \Bigr\}.
  \label{eq:BermionAcceptanceProbability}
\eeq
We can show that the combined probability 
\beq
  P(U'\leftarrow U) = P_{\rm A}(U'\leftarrow U) P_{\rm P}(U'\leftarrow U)
\eeq
fulfills detailed balance. Since $P_{\rm P}(U'\leftarrow U) = P_{\rm P}(U\leftarrow U')$,
we transform
\beqn
  \lefteqn{ P(U'\leftarrow U) \exp(-S[U]) } \nonumber\\
  &=& P_{\rm P}(U'\leftarrow U) \min\Bigl\{ \exp(-S[U]), \exp(-S[U']) \Bigr\} \nonumber\\
  &=& P_{\rm P}(U'\leftarrow U) P_{\rm A}(U\leftarrow U') \exp(-S[U']) \nonumber\\
  &=& P(U\leftarrow U') \exp(-S[U]).
\eeqn

We continue with a discussion of the implementation of this method.  In the
acceptance step, the change of the action for the proposed change of a link
variable is needed. For the gauge part of the action, the difference
\beq 
  \Sg[U'] - \Sg[U]
    = - \frac \beta 3 \Re\Tr \left\{ \Bigl[ U'(x,\mu)-U(x,\mu) \Bigr]
      S^\dagger(x,\mu) \right\}
  \label{eq:SgDifference}
\eeq
can easily be obtained. The analogous difference for the bermion contribution
is more complicated. A change of the variable $U(x,\mu)$ affects the clover
leafs centered around fourteen lattice sites,
\beq
  Y = \{ x, x+a\muhat, x \pm a\nuhat, x+a\muhat \pm a\nuhat \},\quad \nu\neq\mu.
\eeq
Hence, a naive computation of $\Sb[U']-\Sb[U]$ implies fourteen applications
of the Dirac operator for each local step, which would make the acceptance
step very expensive.

Instead, we introduce an auxiliary field $\psi=M[U]\phi$ which is always kept
up to date during a sweep over the lattice.  At the beginning of a sweep, the
Dirac operator is applied globally.  Then, for each proposed local change
$U'(x,\mu)\leftarrow U(x,\mu)$, proposals $\psi'(z)\leftarrow \psi(z)$ are
computed for the fourteen affected sites $z$. The bermion action difference
used in the acceptance step is
\beq
  \Sb[\phi,\phi^\dagger,U'] - \Sb[\phi,\phi^\dagger,U]
   = \sum_{z\in Y} \Bigl\{ |\psi'(z)|^2 - |\psi(z)|^2 \Bigr\}.
  \label{eq:SbDifference}
\eeq 
The proposals for the auxiliary field are accepted or rejected together with
the proposal of the link variable. With this method, in each local step only
the difference $M[U']\phi(z)-M[U]\phi(z)$ has to be calculated, which involves
fewer floating point operations than the computation of $M\phi(z)$.

In order to make this explicit, we note that the action depends on a link
variable $U(x,\mu)$ through either the hopping or the clover term. We split $M
= M_1 + M_2$ into a term $M_1$ which is diagonal in coordinate space and a
term $M_2$ which contains nearest-neighbor contributions. Then the proposals
are given by
\beq
  \psi'(z) = \psi(z) + \Delta^{(1)}_{x\mu} \phi(z) + \Delta^{(2)}_{x\mu} \phi(z),
  \label{eq:PsiUpdate}
\eeq
where we have to compute
\beqn
  \Delta^{(1)}_{x\mu} \phi(z) &=& M_1[U']\phi(z) - M_1[U]\phi(z) \nonumber\\
  \Delta^{(2)}_{x\mu} \phi(z) &=& M_2[U']\phi(z) - M_2[U]\phi(z).
\eeqn
The Hopping term affects the sites $x$ and $x+a\muhat$,
\beqn
  \Delta^{(2)}_{x\mu}\phi(x) &=& 
  - \frac{\kappa}{a} \lambda_\mu\ (1-\gamma_\mu)\
    \Bigl[ U'(x,\mu)-U(x,\mu) \Bigr] \, \phi(x+a\muhat) \nonumber\\
  \Delta^{(2)}_{x\mu}\phi(x+a\muhat) &=&
  - \frac{\kappa}{a} \lambda_\mu^*\ (1+\gamma_\mu)\
    \Bigl[ U'^\dagger(x,\mu)-U^\dagger(x,\mu) \Bigr] \, \phi(x).
\eeqn
The clover term affects all $z \in Y$.  At $z=x+a\muhat+a\nuhat$ we get for
example
\beqn
&&  \Delta^{(1)}_{x\mu}\phi(x+a\muhat+a\nuhat) = 
    \frac{i}{8a} \kappa \csw \sigma_{\mu\nu}
    U^\dagger(x+a\nuhat,\mu) U^\dagger(x,\nu) \times \nonumber\\
&&\quad \times \Bigl[ U'(x,\mu)-U(x,\mu)\Bigr] 
    U(x+a\muhat,\nu) \phi(x+a\muhat+a\nuhat).
\eeqn
Similar four-link products with $\phi(z)$ insertions are obtained at the other
sites.

Let us summarize our method. The update of the gauge fields begins with a
global application of $M$ in order to compute the field $\psi$. Then, a sweep
over the whole lattice is performed. In each local step, we do the following:
\begin{itemize}
\item Compute the $\SUthree$ matrix $V(x,\mu)$ given by
  \eqref{eq:BermionActionLinearInU}. 
\item Compute a proposal $U'(x,\mu)$ for the local link variable
  by an overrelaxation step,
  which leaves the action \eqref{eq:BermionProposalAction} constant.
\item Compute $\psi'(z)$ for $z\in Y$, following \eqref{eq:PsiUpdate}.
\item Compute $S[\phi,\phi^\dagger,U']-S[\phi,\phi^\dagger,U]$ using
  \eqref{eq:SgDifference} and \eqref{eq:SbDifference}.
\item Accept or reject the $U$ and $\psi$ proposals with the
  probability \eqref{eq:BermionAcceptanceProbability}.
\end{itemize}


\subsection{Boson fields}
\label{sec:BosonUpdate}

In a given gauge field background, one can easily find an update algorithm for
the boson fields. The action is quadratic in $\phi$, so the needed
distribution is
\beq
  P(\phi) \propto \exp\{-\sum_x(M\phi(x),M\phi(x))\}.
  \label{eq:BosonHeatbath}
\eeq
This can be achieved by generating a field $\zeta$ with the distribution
\beq
  P(\zeta) \propto \exp\{-\sum_x(\zeta(x), \zeta(x))\}
  \label{eq:DistributionZeta}
\eeq
and then computing $\phi' = M^{-1} \zeta$.
Equation~\eqref{eq:DistributionZeta} is a Gaussian distribution with
variance~2 for the real and imaginary components of $\phi(x)_{A\alpha}$.  An
algorithm for generating Gaussian numbers on the APE100 is described in
appendix~\ref{app:RandomNumbers}.

The direct usage of this heatbath method implies that one has to solve the
equation $\zeta=M\phi'$ with full precision, which can be quite costly. If one
reduces the precision of the solver, the correct distribution
\eqref{eq:BosonHeatbath} gets disturbed in an uncontrolled way. By introducing
additional degrees of freedom, one can get around this problem. Following
\cite{condmat9811025}, we introduce a field $\chi$ with the same indices as
the boson field $\phi$, distributed like
\beq
  P(\chi) = \exp\left\{-\sum_x | \chi(x) - M\phi(x) |^2 \right\}.
\eeq
As $\int D\chi P(\chi) = \mbox{const}$ is independent of $\phi$, one can then
simulate the combined distribution
\beq
  P(\phi,\chi) = \exp\left\{-\sum_x \Bigl[
   | M\phi(x)|^2 - |\chi(x) - M\phi(x)|^2 \Bigr] \right\}
\eeq
without affecting the distribution of the $\phi$ field. This can be done as
follows. One first updates $\chi$ by a global heatbath step. To this end, one
generates a normal random vector $\eta$ and computes
\beq
  \chi'(x) = M\phi(x) + \eta(x).
\eeq
In a second step, one updates $\phi$ with an overrelaxation method that leaves
the action constant,
\beq
  \phi'(x) = \zeta(x) - \phi(x),\quad \zeta(x) = M^{-1} \chi(x).
\eeq
We notice that in each update, the new $\chi$ field is constructed from $\phi$
and a random vector, and it is not used after the update is finished.
Therefore, it can be dropped and does not need to be stored in memory.

Obviously, if the equation $\chi = M\zeta$ is solved exactly, this algorithm
is completely identical with the one described above. However, we notice that
this version can be turned into a Metropolis algorithm. By inverting the
fermion matrix only up to a residue vector $r$, one can generate a proposal
$\phi'$ for an acceptance step,
\beqn
  M\zeta(x) &=& \chi(x) + r(x) \nonumber\\
  \phi'(x)  &=& \zeta(x) - \phi(x).
  \label{eq:PhiSymmetric}
\eeqn

For a Metropolis algorithm with a deterministic proposal, detailed balance is
equivalent with the requirement is that the proposal transition must be
symmetric in $\phi$ and $\phi'$. As in \eqref{eq:PhiSymmetric}, $\zeta$ is
independent of the old $\phi$ field, this is the case here.  The action
difference in the acceptance step can be computed as
\beqn
  \Delta S 
&=& \sum_x \left\{ |M\phi'(x)|^2 + |\chi(x)-M\phi'(x)|^2
                 - |M\phi(x)|^2 - |\chi(x)-M\phi(x)|^2 \right\} \nonumber\\
&=& 2\,\Re \sum_x r^\dagger(x) \Bigl( M\phi'(x) - M\phi(x) \Bigr).
\eeqn

For the inversion of the bermion matrix, we use a stabilized biconjugate
gradient (BiCGStab) solver with SSOR preconditioning. A study of an
implementation of this solver for the APE100 machine and with Schr\"odinger
functional boundary conditions can be found in \cite{heplat9910024}.

\section{The size of $\csw$}
\label{sec:csw}

In all our simulations, we have chosen the improvement coefficient $\csw$ by
extrapolating the non-perturbative formulae in $\Nf$.  This ansatz is
consistent with 1-loop perturbation theory, and also supported by the fact
that the 2-loop contribution is linear in $\Nf$. Nevertheless, we have
computed $\csw$ non-perturbatively for $\Nf=-2$ for the bare coupling
$\beta=8.99$ to check our assumptions.

By definition, improvement coefficients for $\rmO(a)$ improvement have an
ambiguity.  Different choices for these coefficients correspond to different
$\rmO(a^2)$ artefacts. Therefore, it is important to use the same improvement
condition as was used in the computation of \eqref{eq:cswnonpert}.
Consequently, we follow the method described in
\cite{heplat9609035,heplat9803017}. The idea there is to use an observable
that vanishes in the continuum limit and has only $\rmO(a^2)$ corrections if
the improvement coefficients are chosen properly.  Such an observable can be
deduced from the PCAC relation.

One first observes that the computation of the current quark mass $m(x_0)$ in
\eqref{eq:CurrentMass} includes the coefficient $\cA$ which is not known a
priori. In order to eliminate it, we define
\beqn
  r(x_0) &=& \frac 1 2 \tilde\partial_0 \fA(x_0) \nonumber\\
  s(x_0) &=& \frac 1 2 a \partial_0^* \partial_0 \fP(x_0)
\eeqn
and write
\beq
  m(x_0) = r(x_0) + \cA s(x_0).
\eeq
From the correlation function $\fA', \fP'$ at the upper boundary, one gets
another current mass $m'$. Then the mass defined as
\beqn
  M(x_0,y_0) &=& m(x_0) - s(x_0)
    \frac{m(y_0) - m'(y_0)}{s(y_0) - s'(y_0)} \nonumber\\
&=& r(x_0) - s(x_0)
    \frac{r(y_0) - r'(y_0)}{s(y_0) - s'(y_0)}
\eeqn
is independent of $\cA$. Again, a similar quantity $M'$ is obtained by
exchanging variables with and without a prime.  For this section, we define
the current mass $M$ as $M(T/2,T/4)$. Finally, the difference
\beq
  \Delta M = M\Biggl(\frac{3T}{4}, \frac{T}{4} \Biggr) -
            M'\Biggl(\frac{3T}{4}, \frac{T}{4} \Biggr)
\eeq
is zero in the continuum and inherits lattice artefacts from the correlation
functions used in its definition.  In order to ensure the perturbative
asymptotics $\csw(g_0) \to 1$ for $g_0 \to 0$, we precisely demand that
$\Delta M$ takes its tree-level value at $M=0$ \cite{heplat9609035},
\beq
  \Delta aM\Big|_{M=0} = 0.000277.
  \label{eq:cswImprovement}
\eeq
This improvement condition is imposed at $L/a=8$ and $T=2L$, furthermore
$\theta_k=0$ and with boundary fields
\beqn
  (\phi_1,\phi_2,\phi_3) &=& \frac 1 6 (-\pi,0,\pi) \nonumber\\
  (\phi_1',\phi_2',\phi_3') &=& \frac 1 6 (-5\pi,2\pi,3\pi).
\eeqn

Using the setup described above, we have measured the current mass $aM$ and
$a\Delta M$ for some values of $\kappa$ at three values of $\csw$. As shown in
table~\ref{tab:csw1}, these results are in line with the observations in
\cite{heplat9609035,heplat9709022}, which indicate that $\Delta M$ depends
only weakly on the mass $M$. Since it is not critical whether the condition
\eqref{eq:cswImprovement} is precisely imposed at zero current mass, we have
not extrapolated the mass to zero, but instead taken the results where it
roughly vanishes. The error introduced hereby is negligible. Our results for
the lattice artefact $a\Delta M$ are summarized in table~\ref{tab:csw2}.

\begin{table}[htbp]
  \begin{center}
    \begin{tabular}{llrr}
      \hline\hline
      $\kappa$ & $\csw$ & \multicolumn{1}{l}{$aM$} & \multicolumn{1}{l}{$a\Delta M$} \\
      \hline
      0.13209 & 1.171815 &  0.01899(12) &  0.00219(18)  \\
      0.13213 & 1.171815 &  0.01825(14) &  0.00237(21)  \\  
      0.13241 & 1.171815 &  0.00946(13) &  0.00218(19)  \\
      0.13209 & 1.271815 & -0.00030(19) &  0.00066(24)  \\
      0.13105 & 1.371815 &  0.01173(13) & -0.00124(18)  \\ 
      0.13139 & 1.371815 &  0.00061(17) & -0.00127(21) \\
      \hline\hline
    \end{tabular}
    \caption{\sl $a\Delta M$ at three trial values of $\csw$ at
      $\beta=8.99$ and some values of $\kappa$}
    \label{tab:csw1}
  \end{center}
\end{table}

\begin{table}[htbp]
  \begin{center}
    \begin{tabular}{lrr}
      \hline\hline
      $\csw$ & \multicolumn{1}{l}{$aM$} & \multicolumn{1}{l}{$a\Delta M$} \\
      \hline
      1.171815 &  0.0095(1) &  0.00218(19) \\
      1.271815 & -0.0003(2) &  0.00066(24) \\
      1.371815 &   0.0006(2) & -0.00127(21) \\
      \hline\hline
    \end{tabular}
    \caption{\sl $a\Delta M$ for different values of $\csw$ at $\beta=8.99$.}
    \label{tab:csw2}
  \end{center}
\end{table}

A linear interpolation of these three points to \eqref{eq:cswImprovement}
gives $\csw=1.285(7)$, which is very near to the value $\csw=1.271815$ from
our extrapolation in $\Nf$. By 1-loop perturbation theory, one can see that
this difference causes an error in the step scaling function smaller than
$4\cdot 10^{-4}$ \cite{KurthCommunication} which is negligible compared to the
statistical errors as listed in table~\ref{tab:res_improved_2L}.

\section{Measurement of the coupling}

The Schr\"odinger functional coupling is defined by
\eqref{eq:SchrodingerFunctionalCoupling}. In a Monte Carlo simulation, we have
to measure the average value of $\partial S/\partial \eta$, which gets
contributions from the pure gauge action $\Sg$ and the bermion action $\Sb$.
In both terms, one needs in the end the derivative of a boundary link with
respect to $\eta$. With the given boundary conditions, we get
\beq
  \left.\frac{\partial C_k}{\partial \eta}\right|_{\eta=0} 
    = \frac i L \Omega_0, \quad
  \left.\frac{\partial C_k'}{\partial \eta}\right|_{\eta=0}
    = -\frac i L \Omega_0,
\eeq
where we have defined $\Omega_0 = \diag(1, -1/2, -1/2)$. For the gauge action,
we evaluate
\beqn
  \label{eq:Observable}
&&  \left.\frac{\partial \Sg}{\partial \eta}\right|_{\eta=0}
  = -\frac{\beta \ct}{3}\frac{a}{L} \sum_{\xvec} 
  \sum_k \Re\Tr \; i \Omega_0 \Bigl\{ \nonumber\\
&&\quad
    \left[ U(x,k) U(x+a\khat,0)
    U^\dagger(x+a\zerohat,k) U^\dagger(x,0) \right]_{x_0=0} \nonumber\\
&&\quad - {}
    \left[ U(x,k) U^\dagger(x+a\khat-a\zerohat,0)
    U^\dagger(x-a\zerohat,k) U(x-a\zerohat,0) \right]_{x_0=T}   
  \Bigr\}. \quad\quad
\eeqn
The computation of the bermion contribution to the coupling is slightly more
complicated. It suffices to compute $\partial M\phi(x)/\partial\eta$, because
\begin{eqnarray}
  \left.\frac{\partial \Sb}{\partial \eta}\right|_{\eta=0}
&=& \frac{\partial}{\partial\eta} \sum_x 
    \left\{ (M\phi)^\dagger(x) (M\phi)(x) \right\} \nonumber\\
&=& 2 \, \Re \sum_x \psi^\dagger(x) \frac{\partial}{\partial\eta} (M\phi(x)).
\end{eqnarray}
Clearly, only the clover term at $x_0=a$ and $x_0=T-a$ can contribute to the
derivative.  Furthermore the derivative of the magnetic components vanishes,
$\frac{\partial}{\partial\eta}F_{ij}(x)=0$, and we have
$\frac{\partial}{\partial\eta}F_{0k}=-\frac{\partial}{\partial\eta}F_{k0}$ and
$\sigma_{0k}=-\sigma_{k0}$. Thus, for $x \in X = (a,\xvec) \cup (T-a,\xvec)$,
\begin{eqnarray}
  \frac{\partial M\phi(x)}{\partial \eta} 
&=& \frac{i}{2a} \kappa \csw \frac{\partial{}}{\partial \eta}
    \sum_{k=1,2,3} \left\{ \sigma_{k0} \hat F_{k0}(x)
    + \sigma_{0k} \hat F_{0k}(x) \right\} \phi(x) \nonumber\\
&=& \frac{i}{8a} \kappa \csw \sum_{k} \sigma_{k0}
    \left\{ \frac{\partial{}}{\partial \eta} Q_{k0}(x) - \mbox{h.c.} \right\} \phi(x).
\end{eqnarray}
We define for each lattice point $x$ and plane $(\mu,\nu)$ the color matrix
$\xi_{\mu\nu}(x)$ with elements
\beq
  \xi_{\mu\nu,ba}(x) = \psi^\dagger_a(x) \sigma_{\mu\nu} \phi_b(x)
\eeq
and its hermitian part $\chi_{\mu\nu}(x) = \xi_{\mu\nu}(x) +
\xi^\dagger_{\mu\nu}(x)$, where we have explicitly displayed the color indices
of the bermion fields. Using this, we obtain
\begin{eqnarray}
  \left.\frac{\partial \Sb}{\partial \eta}\right|_{\eta=0}
&=& \frac{-\kappa \csw}{4a} \Im \sum_{x\in X} \psi^\dagger(x)
  \sum_k \sigma_{k0} \left( \frac{\partial Q_{k0}(x)}{\partial\eta} - \mbox{h.c.}
  \right) \phi(x) \nonumber\\
&=& \frac{-\kappa \csw}{4a} \Im \Tr \sum_{x\in X} \sum_k \xi_{k0}(x)
  \left( \frac{\partial Q_{k0}(x)}{\partial\eta} - \mbox{h.c.} \right) \nonumber\\
&=& \frac{-\kappa \csw}{4a} \Im \Tr \sum_{x\in X} \sum_k \chi_{k0}(x)
  \frac{\partial Q_{k0}(x)}{\partial\eta}.
\end{eqnarray}
The computation of the derivative is analogous to the gauge action case. For
practical purposes it is advantageous to organize the above sum in terms of
time-like plaquettes at the boundary. Each plaquette contributes to two
different clover leafs. So we can write
\beq
   \left.\frac{\partial \Sb}{\partial \eta}\right|_{\eta=0} 
 = \frac{-\kappa \csw}{4} \frac a L \Im \Tr \sum_\xvec \sum_k
  \left( R^{\rm l}_k(\xvec) + R^{\rm u}_k(\xvec) \right),
\eeq
where we have contributions from the lower boundary
\beqn
  R^{\rm l}_k(\xvec) &=& 
    \frac 1 a U^\dagger(x,0) \Bigl[ i \Omega_0 U(x,k) \Bigr] 
    U(x+a\khat,0) \times\nonumber\\
&&\quad\times
    U^\dagger(x+a\zerohat,k) \chi(x+a\zerohat) \nonumber\\
&& \,\,+\,
    U^\dagger(x+a\zerohat,k) U^\dagger(x,0) \Bigl[ i \Omega_0 U(x,k) \Bigr] 
    \times\nonumber\\
&&\quad\times 
    U(x+a\khat,0) \chi(x+a\zerohat+a\khat)
\eeqn
and the upper boundary
\beqn
  R^{\rm u}_k(\xvec) &=& 
    \frac 1 a U(x-a\zerohat,k) U(x-a\zerohat+a\khat,0)
    \Bigl[ i \Omega_0 U(x,k) \Bigr]^\dagger \times\nonumber\\
&&\quad\times
    U^\dagger(x-a\zerohat,0) \chi(x-a\zerohat) \nonumber\\
&& \,\, +\,
    U(x-a\zerohat+a\khat,0) \Bigl[ i \Omega_0 U(x,k) \Bigr]^\dagger 
    U^\dagger(x-a\zerohat,0) \times\nonumber\\
&&\quad\times
    U(x-a\zerohat,k) \chi(x-a\zerohat+a\khat).
\eeqn
In the implementation of the measurement, bermions do not produce a large
overhead. The computation of $\partial \Sb/\partial\eta$ can be done in a loop
together with $\partial \Sg/\partial\eta$. The effort for a measurement grows
only with $(L/a)^3$, whereas the cost of the same computation for $\Nf=+2$ is
proportional to the lattice volume.

\section{Tests}

Large parts of the simulation program consisted of new code; the gauge field
update, the boson update and the measurement of the boson contribution to the
coupling. In order to check the program carefully for correctness, it is
desirable to separate the test for different parts of the code as much as
possible.

We have done this as follows: in the implementation for unimproved bermions,
the overrelaxation boson update there can be replaced by the heatbath boson
update discussed in section~\ref{sec:BosonUpdate} and setting the improvement
coefficients $\csw=0, \cttilde=1$.  We have confirmed that the modified
program still yields the same results as before.

Then we have run the improved bermion program consisting of the gauge field
update from section~\ref{sec:GaugeUpdate} and the boson update from
section~\ref{sec:BosonUpdate} with various choices of the improvement
parameters and checked a control variable discussed below. This tests the
update algorithm as a whole.

Provided the boson update on a given gauge field generates a sample
distributed according to the action $\Sb$, the measurement routine for
$\partial\Sb/\partial\eta$ can then be checked on the classical background
field by comparing with the tree-level perturbative result.

\subsection{Control variable}

In general, it is advantageous to have a nontrivial observable whose mean is
known exactly for any choice of parameters and bare couplings.  This allows a
consistency check without having to compare results e.g. with the results of
other programs, which may not even be possible. In particular, when comparing
with another Monte Carlo result, one compares two quantities which both are
affected with statistical errors, whereas the comparison with a constant value
only involves one statistical error and is therefore more reliable.

In the bermion model, it is easy to construct such an
observable. We define
\beq
  E = \frac{1}{12 (L/a)^3 [(T/a)-1]} \sum_x | M\phi(x) |^2,
\eeq
where $(L/a)^3[(T/a)-1]$ is the number of sites with fluctuating
boson fields and $12$ is the number of Dirac and color
components.

As a generalization of the property of the one-dimensional Gaussian measure,
\beq
  \langle x^2 \rangle = 1 \;\; \mbox{for} \;\;
  P(x)\propto e^{-\frac 1 2 x^2},
\eeq
we expect the expectation value of $E$, which corresponds to a path
integral with Gaussian measure, to be
\beq
  \langle E \rangle = 1.
\eeq

\subsection{$\partial \Sb/\partial\eta$ with a static gauge field}

By switching off the gauge field update routine, the bermion program can be
transformed into a Monte Carlo simulation for a boson field on a fixed gauge
field background. We have initialized the gauge fields to the classical
background field given by \eqref{eq:BackgroundField} and run the modified
simulation program. The fields then should be distributed like
$\exp(-\Sb[\phi,\phi^\dagger,V])$.

For this distribution, the result for the boson contribution to the coupling,
$\partial \Sb/\partial\eta$, can be calculated exactly using perturbation
theory. We first make this calculation for the fermionic action $\Sf$. The
fermionic contribution to the coupling comes from the term
\beq
  \frac{\partial \Sf}{\partial \eta}
  = \frac i 4 a^5 \csw \sum_{x\mu\nu} \bar\psi(x) \sigma_{\mu\nu} 
    \frac{\partial \hat F_{\mu\nu}(x)}{\partial \eta} \psi(x).
\eeq
Note that with $\psi$, we denote Grassmann fields here.  In the classical
background field, all time-like links are equal to the unity matrix, and all
space-like links are independent of $\xvec$. Therefore the magnetic components
$\partial\hat F/\partial\eta$ vanish. The other components are translational
invariant in the space directions, such that we can write
\beq
  \left\langle \left.\frac{\partial \Sf}{\partial \eta}\right|_{\eta=0}
  \right\rangle_{\rm cl}
  = \frac{\csw L^3}{4} \, [ v(a) + v(L-a) ]
\eeq
with
\beq
  v(x_0) = 2 i a^2 \left\langle \bar\psi(x) \sum_k \sigma_{0k} 
   \left.\frac{\partial \hat F_{0k}(x)}{\partial \eta}\right|_{\eta=0}
   \psi(x) \right\rangle_{\rm cl}.
\eeq
With $\langle\rangle_{\rm cl}$ we denote the fermionic expectation value in
the classical gauge field background.  For $x_0=a$, we have
\beqn
  \left.\frac{\partial \hat F_{0k}(x)}{\partial \eta}\right|_{\eta=0}
&=& \frac{1}{8a^2} \Biggl\{ 
    V(x,k) \frac{\partial V^\dagger(x-a\zerohat,k)}{\partial \eta} \nonumber\\
&&\quad +
    \frac{\partial V^\dagger(x-a\zerohat,k)}{\partial \eta} V(x,k) 
    \Biggr\}_{\eta=0} - \mbox{h.c.} \nonumber\\
&=& -\frac{i}{4a^2} \frac{a}{L} \Omega_0 \,
    \exp\left\{ \frac a T (C_k'-C_k) \right\}\Bigg|_{\eta=0}
    - \mbox{h.c.} \nonumber\\
&=& -\frac{i}{2a^2} \frac{a}{L} \Omega_0 \,
    \cos\left\{\frac{2\pi}{3} \frac{a^2}{LT} \Omega_0 \right\}.
\eeqn
The same result is obtained $x_0=T-a$.  If we now introduce the
diagonal $3\times 3$ matrix
\beq
  r=\Omega_0 \cos\left\{ \frac{2\pi}{3} \frac{a^2}{LT} \Omega_0 \right\},
\eeq
and use $\langle \bar\psi(x) \psi(x)\rangle_{\rm cl} = S(x,x)$, where $S(x,x)$ is
the propagator in the gauge field background, as discussed in
\cite{heplat9606016}, we get
\beq
  v(x_0) = -\Nf \frac a L \sum_\alpha r_{\alpha\alpha} \Tr\left(
    S^\alpha(x,x) \sum_k \sigma_{0k} \right).
\eeq
The results for $v(a)$ and $v(L-a)$ are computed for some parameter set in
\cite{DerivativeOfSWTerm} and listed in table~\ref{tab:v}.  Since the formula
for $v(x_0)$ is linear in $\Nf$, we expect the result for $\partial
\Sb/\partial\eta$ to be of the same value, with the opposite sign.

We have made runs with the parameter set used in table~\ref{tab:v} at
$L/a=4,6$. Within the statistical errors, which had a magnitude of about
3\textperthousand, our numerical simulation results were consistent with these
results.

\begin{table}[htbp]
  \begin{center}
    \begin{tabular}{lll}
      \hline\hline
      $L/a$ & $v(a)$ & $v(L-a)$ \\
      \hline
      4 & 0.05454081493 & 0.02858695070 \\
      6 & 0.01555246822 & 0.00892024138 \\
      8 & 0.00622327486 & 0.00398090210 \\
      \hline\hline
    \end{tabular}
    \caption{\sl Expectation values $v(a)$, $v(L-a)$ on the classical
      background field for $\Nf=+2$, $\kappa=1/8$, $\theta=\pi/5$
      and $\csw=1$.}
    \label{tab:v}
  \end{center}
\end{table}

\section{Tuning and error propagation}
\label{sec:Tuning}

Computing the step scaling function for a given coupling involves a couple of
matching runs at size $L$ and a run at size $2L$. The error of the coupling at
$2L$ is easily obtained as the statistical error of the simulation. But also
the statistical error at $L$ must be taken into account when estimating the
error of the step scaling function. For the calculation of the step scaling
function itself, we also have to consider a mismatch.
Let us assume that $(\beta,\kappa)$ is the pair of bare parameters which
correspond to the renormalized parameters $\gbar^2(L)=u_0, m_1(L)=0$.  Then
the simulations yield results
\beqn
  \gbar^2(L)|_{\beta^*} &=& u_1 \pm \delta u_1 \nonumber\\
  \gbar^2(2L)|_{\beta^*} &=& u_2 \pm \delta u_2,
\eeqn
with $u_1$ being slightly different from $u_0$, and $\beta^*$ having a slight
mismatch compared to the true $\beta$.  We use the expansion
\eqref{eq:SigmaExpansion} to estimate
\beqn
  \Sigma(u_0,a/L) 
&\approx& u_2 - \left\{ \Sigma(u_1,a/L) - \Sigma(u_0,a/L)\right\} \nonumber\\
&=& u_2 - (u_1 + s_0 u_1^2 - u_0 - s_0 u_0^2 + \cdots) \nonumber\\
&=& u_2 - (u_1 - u_0) (1 + 2 s_0 u_0) + \rmO\left((u_1-u_0) u_0^2\right).
\eeqn
For the statistical error, we use the estimate
\beqn
  (\delta \Sigma(u_0,a/L))^2 \approx (\delta u_2)^2 
  + (1 + 2 s_0 u_0)^2 (\delta u_1)^2.
\eeqn

In addition to the renormalized coupling, in the tuning procedure the current
mass $m_1(L/a)$ must be tuned to zero. In order to minimize the amount of
computer time going into the tuning runs, it is helpful to know which
precision is needed in the computation of the current mass. In perturbation
theory one can estimate how a mismatch in the mass propagates into the value
of the step scaling function \cite{SommerSSFUnpublished}. One defines the
derivative of $\Sigma$ with respect to $z=m_1 L$,
\beqn
  \frac \partial {\partial z} \left.\left( 
  \gbar^2(2L)\big\vert_{\gbar^2(L) = u,\ m_1(L)L = z} 
  \right) \right|_{z=0} 
  = \Sigma'_1(a/L)\ u^2 + \cdots.
\eeqn
By approximating this derivative by its continuum value and using the function
$c_{1,1}(z)$ computed in \cite{heplat9508012}, one gets
\beqn
  \Sigma'_1(a/L) \approx \Sigma'_1(0) = \left. 
  -\frac{\Nf}{4\pi} \frac{\partial}{\partial z} c_{1,1}(z)
  \right|_{z=0} = 0.00957\ \Nf.
\eeqn
We conclude that tuning the current mass to a residue value $0.001$ for
example on an $L/a=8$ lattice leads to an error in the step scaling function
smaller than $0.0002 \, u^2$, which can be neglected compared to our
statistical errors. Table~\ref{tab:res_improved_L} contains the results of our
tuning runs and shows that the measured masses are in a safe range.  We have
also found that the best tuning runs for a given pair of $\gbar$ and $m_1$
allow for an interpolation to $m_1=0$. In those cases, we have listed the
value zero with the error given by the fit routine in the table.

\section{Performance}

For the feasibility of our updating strategy, a reasonable acceptance rate in
the gauge field update is essential. In particular, there is no tunable
parameter which could be used to optimize this component of the update. In the
range of bare couplings we have used in our simulations, the acceptance rate
does not vary much. It is roughly $76\%$ at $\gbar^2=0.9793$, and with $70\%$
slightly worse at $\gbar^2=1.5145$.

In the boson sector of the algorithm, the stopping criterion of the solver can
be optimized for performance. In the solver implementation we use, the
iteration scheme is finished as soon as the norm of the residue vector is
smaller than $\epsilon$ times the norm of the solution vector of the
preconditioned system. A solution of the system of equations to machine
precision amounts to setting $\epsilon^2\approx 10^{-14}$.

Since the cost of the whole update process is dominated by the gauge field
update, we could not obtain a great advantage by increasing $\epsilon$. In
test runs on a $4^4$ lattice, the performance maximum is pretty flat in
$\epsilon$, and the decrease in cost only about $10\%$. On larger lattices,
one expects that the accuracy of the solver has to be scaled
\cite{condmat9811025}, so that we expect the benefit of tuning this parameter
to even diminish on the large lattice.  As determining autocorrelation times
accurately requires large statistics, a detailed performance study on large
lattices would be very expensive. Therefore, we have dispensed with this and
instead set the solver precision to a safe value of $\epsilon^2=10^{-10}$,
which gave an acceptance rate of more than $99\%$.

For a comparison of the efficiency of our algorithm with the one for
unimproved Wilson bermions and with simulations with dynamical fermions, we
use the measure $\Mcost$ as defined in \cite{heplat0009027}. It is machine
dependent, with our reference computer being an 8-node APE100, and is defined
as
\beqn
  \Mcost 
&=& \mbox{(update time in seconds on machine M)}\nonumber\\
&& \times (\mbox{error of $\gbar^{-2}$})^2 \times (4a/T) \, (4a/L)^3.
\eeqn

According to this definition, if we want to compute $\gbar^{-2}$ with an
absolute precision of 0.01 on a $4^4$ lattice, we need $10000 \cdot \Mcost$
seconds update time on a Q1 machine. The focus on the absolute error of
$\gbar^{-2}$ is motivated by its relevance in the recursive scheme used to
compute the $\Lambda$ parameter. In 1-loop approximation, the expansion
\eqref{eq:SigmaExpansion} can be written as
\beq
  \gbar^{-2}(2L) \approx \gbar^{-2}(L) - s_0.
\eeq

This means, under an application of the step scaling function, the error of
$\gbar^{-2}$ is conserved. In a recursive computation of the coupling at
scales $L=2^{-k} \Lmax$, the error of $\gbar^{-2}$ in each step is equally
important.

\begin{figure}[htbp]
  \begin{center}
    \epsfig{file=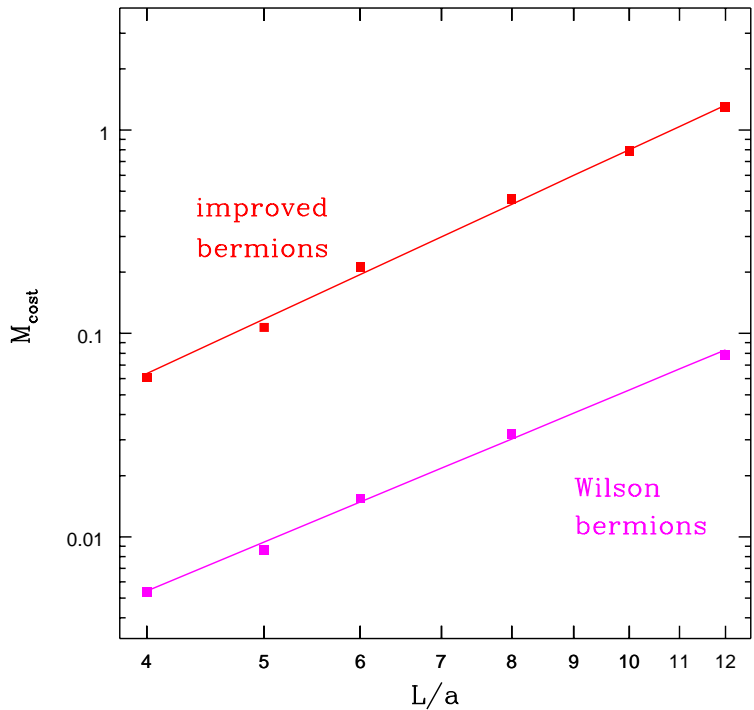,width=11cm,height=11cm}
    \vspace{-0.6cm}
    \caption{\sl Costs of improved and unimproved bermions at $\gbar^2=0.9793$.}
    \label{fig:perf}
  \end{center}
\end{figure}

The cost of our algorithm for the simulated lattice sizes at $\gbar^2=0.9793$
is shown in table~\ref{tab:perf} and figure \ref{fig:perf} in comparison with
the algorithm for unimproved bermions.
\begin{table}[htbp]
  \begin{center}
    \begin{tabular}{rll}
      \hline\hline
      $L/a$ & $M_{\rm cost}$ & $M_{\rm cost}$ \\
            & improved &  unimproved \\
      \hline
      4  & 0.061(2)  & 0.00535(7)              \\
      5  & 0.107(3)  & 0.00866(13)              \\
      6  & 0.212(6)  & 0.0155(2)                \\
      8  & 0.457(11) & 0.0319(4)                \\
      10 & 0.790(17) &                           \\
      12 & 1.30(3)   & 0.0788(12)                \\
      \hline\hline
    \end{tabular}
    \caption{\sl Cost of the measurement of the coupling $\gbar^2=0.9793$ 
      for improved and unimproved bermions. The last two entries 
      for improved bermions are at $\gbar^2 \approx 1.11$.}
    \label{tab:perf}
  \end{center}
\end{table}
When we fit these data with a linear ansatz, excluding the $L/a=4$ lattices,
we find that the cost both for the improved and unimproved theory scales with
about $a^{-2.5}$. The overhead of improvement however turns out to be about a
factor $12$. Considering that unimproved bermions are only a factor $3$ more
expensive than the quenched approximation, this is quite large. In
section~\ref{sec:BermionResults} we will argue that this overhead is
nevertheless more than compensated by a substantially improved extrapolation
to the continuum limit.

Since our motivation for studying the $\Nf=-2$ model was the lower cost in
comparison with dynamical fermions, it is also interesting to compare with
data from \cite{heplat0009027}. We find that on a $L/a=12$ lattice, our
implementation of improved bermions is about a factor $10$ cheaper than
simulations of dynamical fermions.  Also, our algorithm scales slightly
better.

\section{Results}
\label{sec:BermionResults}

Our simulation program was implemented in the TAO language for the APE100
parallel computer.  To give an impression of the cost, the run at
$\gbar^2=\Sigma(0.9793, 1/16)$ needed roughly 20 days on a QH2 machine
with 256 nodes. During this time, 20000 measurements of the couplings have
been accumulated in each of 16 replica. We have also ported the program to the
slightly changed APEmille environment.  Without taking advantage of many
improvements like the extended register file, the performance gain is not more
than a factor~3. About half of the statistics for the measurement of
$\sigma(1.5145, 1/16)$ has been collected on an APEmille crate with 128 nodes.

The measurement of the coupling is quite cheap and its autocorrelation time
only up to about 6 in units of updates (while other observables like the
average plaquette are correlated over 10 updates). Since our main interest is
in the coupling, we have measured it after each update. In contrast, the
measurement of the correlation functions $\fA$ and $\fP$ is quite costly.
Also, the mass $m_1$ fluctuates little and is not required with a great
precision. Therefore, we have only measured these correlation functions after
every 100th or 150th update. For the estimation of statistical uncertainties,
we have employed the methods described in appendix~\ref{app:ErrorAnalysis}.

Our results for the tuning runs are listed in table~\ref{tab:res_improved_L}.
The corresponding results at the same bare parameters are summarized in
table~\ref{tab:res_improved_2L}.

\begin{table}[p]
  \begin{center}
    \begin{tabular}{lllll}
      \hline\hline
      $L/a$ & $\beta$ & $\kappa$ & $\gbar^2(L)$ & $m_1(L/a)$ \\
      \hline
      4 & 10.3488 & 0.131024 &  0.9793(19) &  0.00000(31) \\
      5 & 10.5617 & 0.130797 &  0.9795(21) &  0.00055(13) \\
      6 & 10.7302 & 0.130686 &  0.9793(11) &  0.00000(5)  \\
      8 & 11.0026 & 0.130489 &  0.9793(14) &  0.00000(6)  \\
      \hline
      4 & 8.3378  & 0.132959 &  1.5145(23) &  0.00000(28) \\
      5 & 8.5453  & 0.132637 &  1.5145(17) &  0.00000(7)  \\
      6 & 8.70830 & 0.132433 &  1.5145(33) &  0.00000(4)  \\
      8 & 8.99    & 0.13209  &  1.5145(33) &  0.00066(8)  \\
      \hline\hline
    \end{tabular}
    \caption{\sl Parameters and results for the coupling and the mass at $L$.}
    \label{tab:res_improved_L}
  \end{center}
\end{table}

\begin{table}[htbp]
  \begin{center}
    \begin{tabular}{lllll}
      \hline\hline
      $L/a$ & $\gbar^2(2L)$ & $m_1(2L/a)$ \\
      \hline
      4 &  1.1090(28) &  -0.00300(10) \\
      5 &  1.1079(29) &  -0.00086(5)  \\
      6 &  1.1053(30) &  -0.00094(4)  \\
      8 &  1.1093(40) &  -0.00025(3)  \\
      \hline
      4 &  1.8734(74) &  -0.00266(12) \\
      5 &  1.8648(82) &  -0.00094(7)  \\
      6 &  1.8488(86) &  -0.00070(5)  \\
      8 &  1.8741(100)  & $\,\!$ 0.00002(5)  \\
      \hline\hline
    \end{tabular}
    \caption{\sl Results for the coupling and the PCAC mass at $2L$ at
      the bare parameters given by the listed values of $\gbar^2(L)$.}
    \label{tab:res_improved_2L}
  \end{center}
\end{table}

\begin{table}[htbp]
  \begin{center}
    \begin{tabular}{llll}\hline\hline
      $u$ & $\sigma(u)$ & 
        $\hat\sigma^{\rm 2-loop}(u)$ & 
        $\hat\sigma^{\rm 3-loop}(u)$ \\ 
     \hline
     0.9793 & 1.1086(72)  & 1.10435 & 1.10546 \\ 
     1.5145 & 1.8713(167) & 1.85122 & 1.85938 \\
     \hline\hline
     \end{tabular}
     \caption{\sl Extrapolated simulation results and perturbation theory of the
       massive step scaling function for $\Nf=-2$.} 
    \label{tab:r2}
  \end{center}
\end{table}

\begin{figure}[htbp]
  \begin{center}
    \epsfig{file=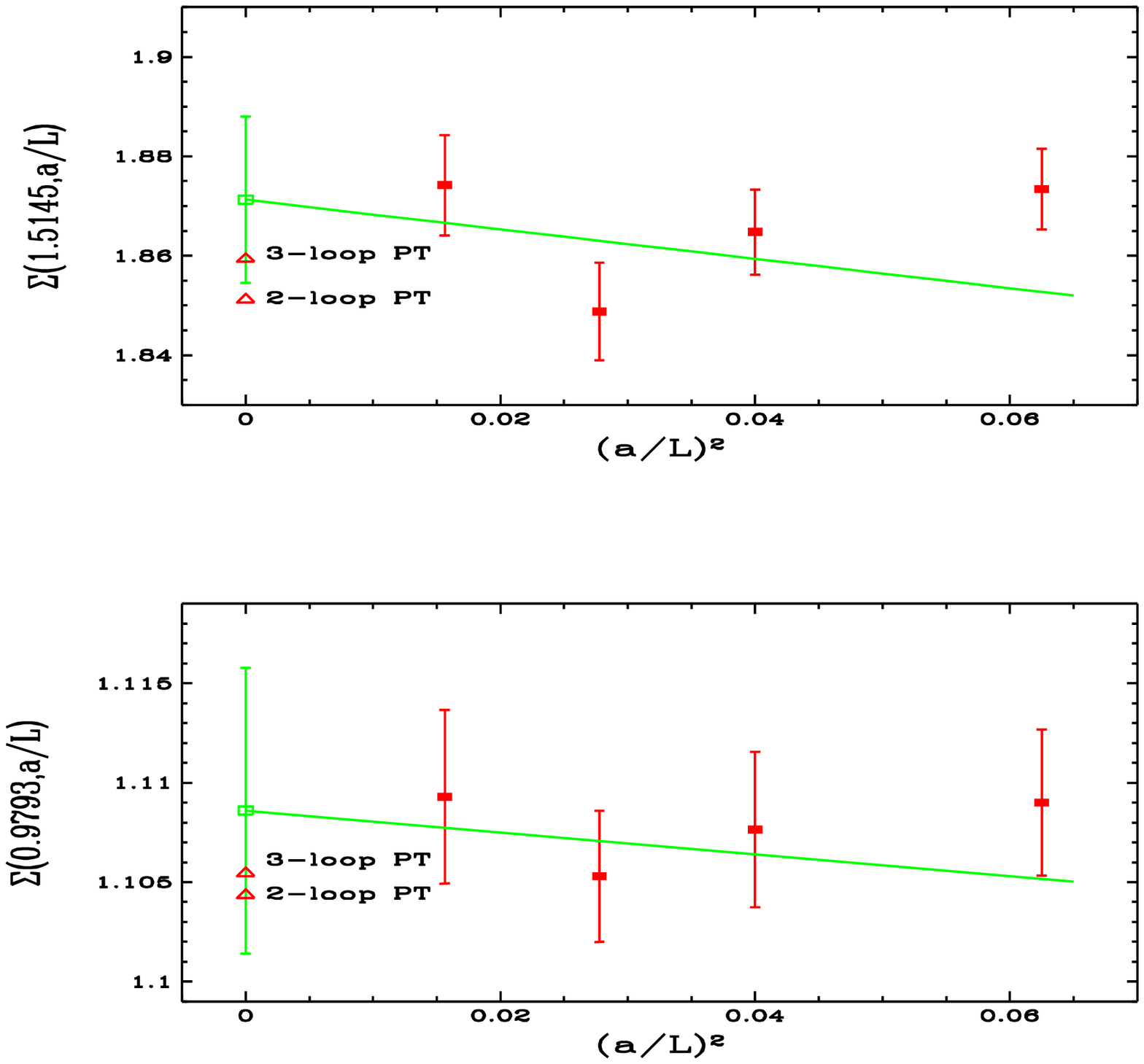,height=19.8cm,width=11cm}
    \vspace{-1.8cm}
    \caption{\sl Step scaling function for improved bermions for the couplings 
      $u=0.9793$ and $u=1.5145$ with fits linear in $(a/L)^2$. Also shown
      is the extrapolated continuum value and the 2- and 3-loop
      values.}
    \label{fig:bermimp}
  \end{center}
\end{figure}

By applying the techniques for the propagation of the coupling mismatch and
error propagation described in section~\ref{sec:Tuning}, we obtain
$\Sigma(u,a/L)$ for $L/a=4,5,6,8$ as shown in figure~\ref{fig:bermimp}.
In the figure linear in $(a/L)^2$, all points lie on a straight line with no
visible appearance of $a/L$ effects. So we fit with the ansatz
\beqn
  \Sigma(u,a/L) = \sigma(u) + \rho(u) (a/L)^2.
\eeqn
In both cases, the slope $\rho(u)$ turns out to be compatible with zero within
the errors. In the computation of the step scaling function in the quenched
approximation, it has been observed that the Schr\"odinger functional coupling
follows perturbation theory to quite low energies. Therefore, we also display
the 2-loop and 3-loop perturbative values in the diagram, which we obtain from
\cite{heplat9911018} and \eqref{eq:SigmaExpansion}. For a direct comparison,
these numbers are summarized in table~\ref{tab:r2}. It can be seen that our
simulation results are consistent with perturbation theory within the
errorbars.  It can also be seen that the 3-loop contribution is of the same
order of magnitude as our statistical error (and much larger than in the
$\Nf=+2$ case).

\begin{figure}[bh]
  \begin{center}
    \epsfig{file=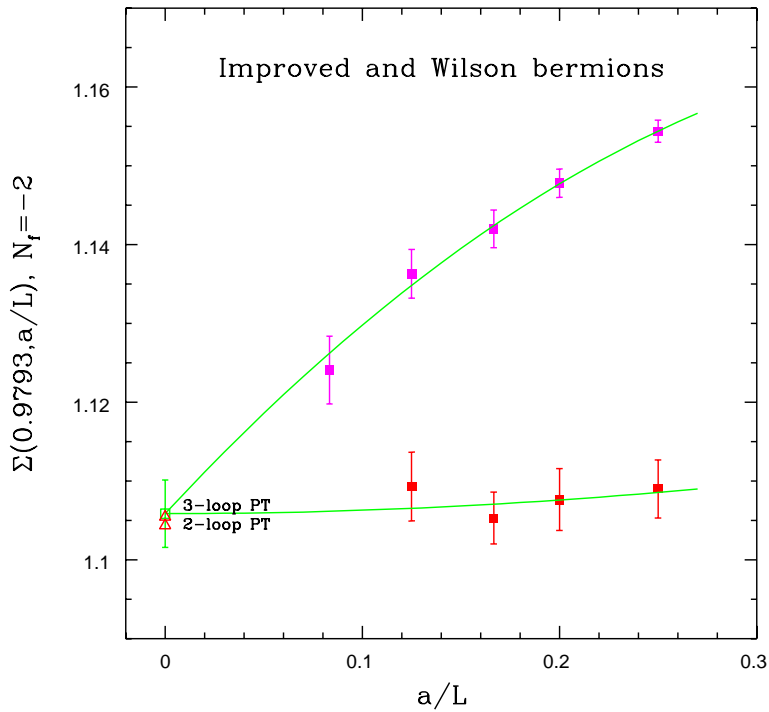,height=11cm,width=11cm}
    \vspace{-0.6cm}
    \caption{\sl Results for the step scaling function at $u=0.9793$, together
      with a quadratic fit under the constraint of universality.}
    \label{fig:bermcombined}
  \end{center}
\end{figure}

Now we want to compare our data with the results for unimproved Wilson
bermions from \cite{heplat9907007}. That study covers the step scaling
function at $u=0.9793$ with resolutions $L/a=4,5,6,8,12$.
Figure~\ref{fig:bermcombined} shows both data sets in a common diagram against
$L/a$. While the unimproved data points clearly show cutoff effects of a few
percent, the improved ones are even consistent with a constant within the
statistical errors. The diagram shows a joint fit of the improved data with a
quadratic term and of the unimproved data with a linear and quadratic term. In
this combined fit, the constraint has been imposed that both curves have the
same continuum limit. The continuum limit from this procedure comes out as
$\sigma_{\rm combined}(0.9793)=1.1059(43)$ which is minimally different from
the value in table~\ref{tab:r2}. This is a strong indication that universality
holds.

An interesting question is whether improvement is profitable. Fitting the
unimproved data alone with a linear plus quadratic function yields a continuum
limit of $\sigma_{\rm unimproved}(0.9793) = 1.103(12)$, i.e. the error
estimate is by a factor $2.7$ higher than for the fit in
figure~\ref{fig:bermimp}. On the other hand, the computational cost was only a
factor $1.7$ less. From these numbers, the conclusion is that the
implementation of $\rmO(a)$ improvement is a cost-effective way of enhancing
the precision of the continuum limit. Surely our observations also amend our
trust in the extrapolation to the continuum limit.

We should note that this is a quantitative comparison against the Wilson
action without any improvement. Experience with the Schr\"odinger functional
boundary conditions in the pure gauge theory indicates that a large part of
the lattice artefacts is caused by the time-like boundaries, which are absent
in the usually simulated torus setups. These cutoff effects can be very
effectively eliminated by setting $\ct$ to its perturbative value, with a
minimal implementation overhead.

So the more interesting effect is the one by the clover term. For the
simulation of dynamical fermions, the balance will bend more towards
improvement than for bermions. One point to consider is the critical slowing
down towards the continuum limit. For bermions, a trustworthy extrapolation of
the unimproved data points needed simulations up to $L/a=24$, whereas we were
satisfied with $L/a=16$ for the improved case. The Hybrid Monte Carlo like
algorithms for dynamical fermions scale worse than our bermion algorithm, and
therefore going to larger lattices is more costly. The other point is that our
algorithm for improved bermions is very expensive compared to the unimproved
program. In a Hybrid Monte Carlo program for $\Nf=+2$, improvement amounts to
the modification of the Dirac operator. Hence the inclusion of the clover term
implies a much lower overhead, for example about 20\% as reported in
\cite{heplat9603008}.

\begin{figure}[htb]
  \begin{center}
    \epsfig{file=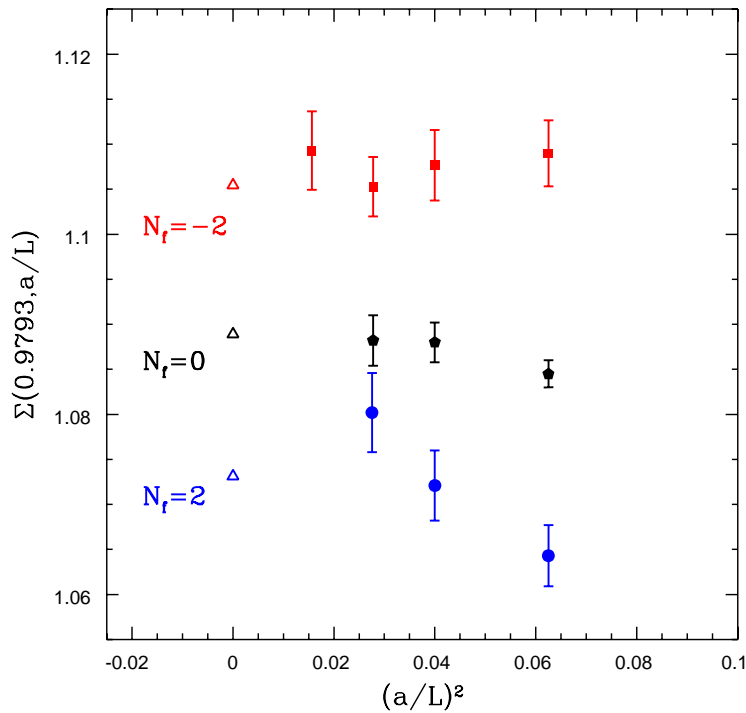,width=11cm,height=11cm}
    \vspace{-0.6cm}
    \caption{\sl Step scaling function for $\Nf=-2,0,+2$ in comparison, in a
      diagram linear in $(a/L)^2$. The 3-loop perturbative values are
      displayed as triangles.}
    \label{fig:sigmaall}
  \end{center}
\end{figure}

Finally, we also want to compare our data with quenched results and data for
$\Nf=+2$ as published in \cite{heplat0105003}. While we use the 2-loop value
for the improvement coefficient $\ct$, it was only known to 1-loop at the time
when the data for $\Nf=0,+2$ was produced.

The cases $\Nf=0$ and also $\Nf=-2$, as discussed before, look well behaved.
One can fit both of them linearly in $(a/L)^2$. When the data point $L/a=4$ is
left out, both data sets are also consistent with a constant, which is an
indication that higher orders in $a$ do not play a significant role.

The $\Nf=+2$ case looks more disturbing, and actually this was one of the main
points that led to the consideration of the bermion model.  Simulation runs at
different couplings have however produced data sets in which the $L/a=5,6$
points differ less \cite{heplat0105003}. One may conclude that in this figure,
the point at $L/a=6$ has a large statistical fluctuation. Nevertheless, this
plot shows that one must be careful with using expectations from the
improvement programme for the extrapolation of data. In particular,
simulations closer to the continuum limit are evidently needed.

\section{Results with perturbative corrections}
\label{sec:PerturbativeCorrections}

\begin{figure}[htbp]
  \begin{center}
    \epsfig{file=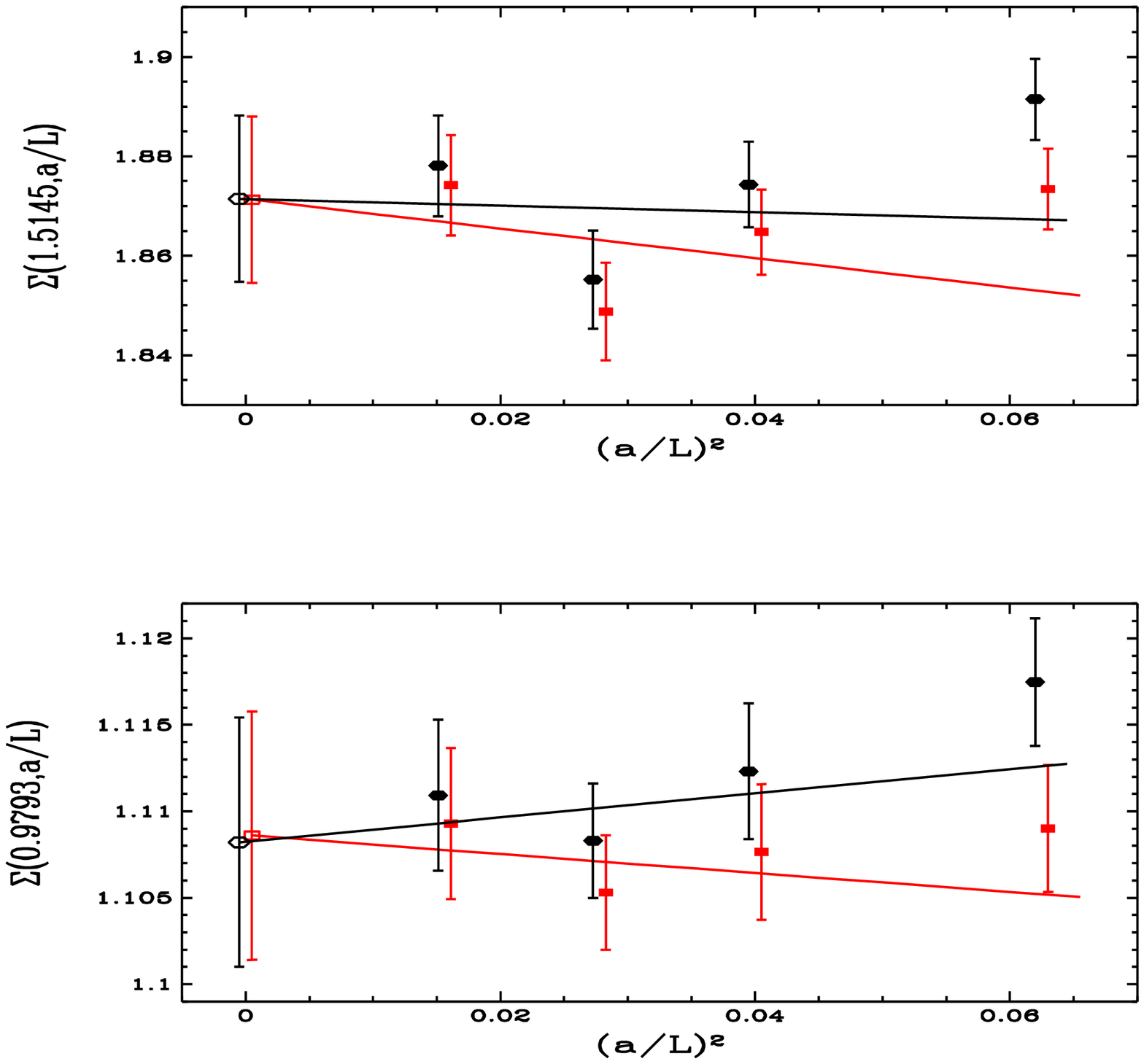,height=19.8cm,width=11cm}
    \vspace{-1.8cm}
    \caption{\sl Step scaling function for improved bermions for the couplings 
      $u=0.9793$ and $u=1.5145$ with fits linear in $(a/L)^2$. The 
      rectangles represent the data points obtained from our simulations,
      whereas the circles represent the data $\Sigma^{(2)}$
      corrected by perturbation theory.}
    \label{fig:bermcorrected}
  \end{center}
\end{figure}

The size of remaining cutoff effects in the step scaling function can be
computed in perturbation theory. In order to get an estimate, one expands the
relative deviation from the continuum limit in a series
\beqn
  \delta(u,a/L) 
&=& \! \frac{\Sigma(u,a/L)-\sigma(u)}{\sigma(u)} \nonumber\\
&=& \! ( \delta_{10} + \delta_{11}\Nf ) u
    + ( \delta_{20}+\delta_{21}\Nf+\delta_{22}\Nf^2 ) u^2 
    + \rmO(u^3).
  \label{eq:delta}
\eeqn
One can hope to smoothen the approach to the continuum limit by canceling the
perturbative contribution to the remaining cutoff effects from the
$\Sigma(u,a/L)$ data from Monte Carlo simulations.  Here we also study the
corrected values
\beq
  \Sigma^{(2)}(u,a/L) 
  = \frac{\Sigma(u,a/L)}{1+\delta_1(a/L)u + \delta_2(a/L)u^2}
\eeq
with $\delta$ coefficients computed to 2-loop order \cite{heplat0106025}.
Figure~\ref{fig:bermcorrected} shows the data for $\Sigma^{(2)}(u,a/L)$ in
comparison with the uncorrected data $\Sigma(u,a/L)$. These points are fitted
in the same way, i.e. linear in $(a/L)^2$ and without the point $L/a=4$. The
perturbative cutoff effects are quite small and do not change the slope of our
extrapolations in a systematic way. The $L/a=4$ points are even pushed away
from the extrapolated value.  As the perturbative corrections are within the
statistical fluctuations of our data, no general statement about the success
of this approach can be made. However we note that the extrapolated value
changes only minimally.

\section{Lattice artefacts in the current mass}

\begin{figure}[htbp]
  \begin{center}
    \epsfig{file=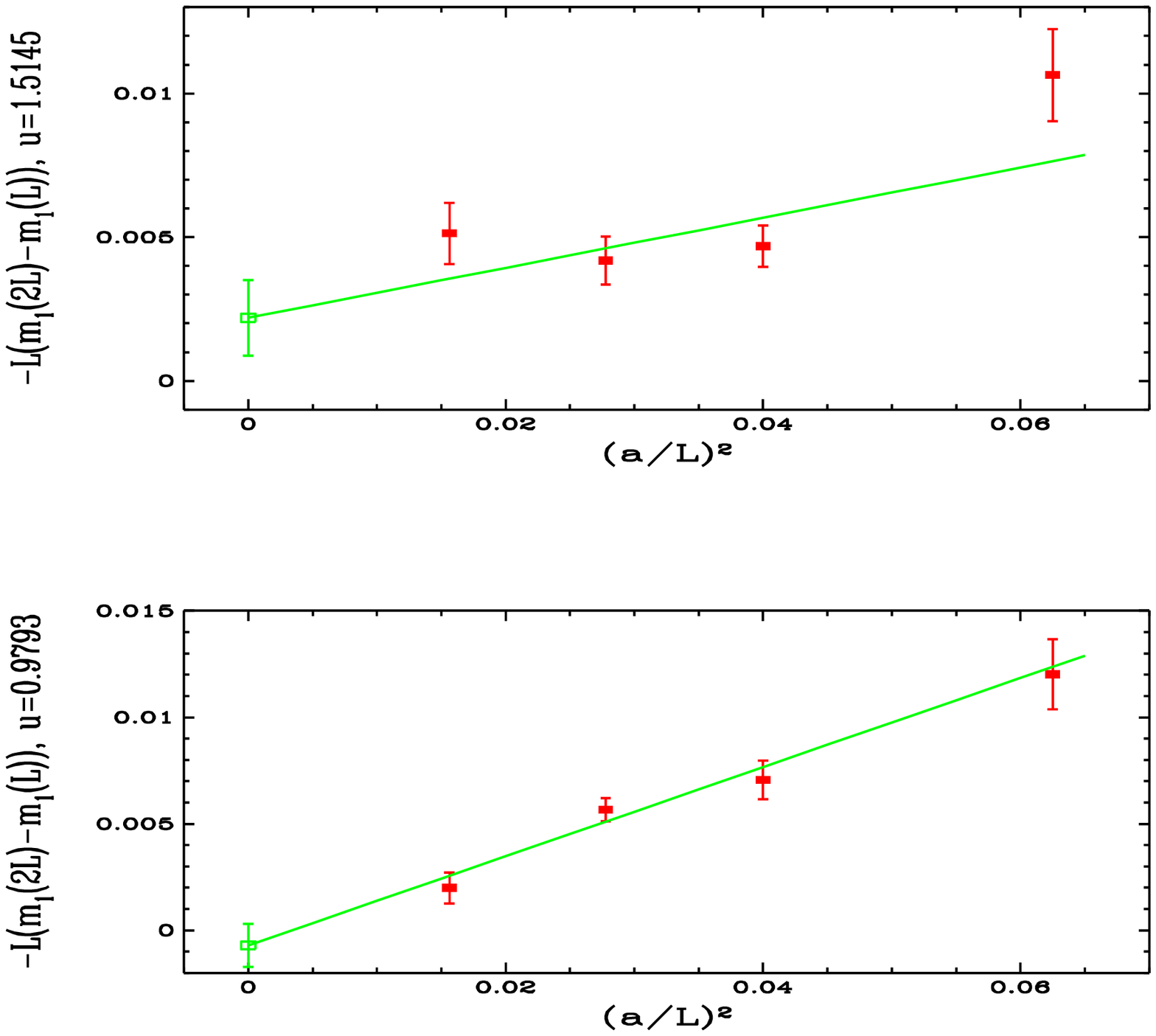,height=19.8cm,width=11cm}
    \vspace{-1.2cm}
    \caption{\sl Lattice artefacts in the PCAC mass $m_1(2L/a)$ at
      $\gbar^2=0.9793$ and $\gbar^2=1.5145$.}
    \label{fig:masses}
  \end{center}
\end{figure}

Another test of the improvement programme and in particular of the PCAC
relation are the lattice artefacts in the fermionic observables.  Our
definition of the step scaling function is such that the current mass
$m_1(L/a)$ is tuned to zero on the small lattices. We then expect that
$m_1(2L/a)$ measured with the same bare parameters vanishes in the continuum
limit and has cutoff effects of order $\rmO(a^2)$. In order to take into
account a small mismatch in $m_1(L/a)$, we examine the quantity
$m_1(2L/a)-m_1(L/a)$ which is plotted against $(a/L)^2$ in
figure~\ref{fig:masses}.
The linear fits in the data points assert the expected scaling
behavior. Only for $\gbar^2=1.5145$ we can see a small deviation which may
stem from a statistical fluctuation and is not a sign of pathological
behavior.

\begin{figure}[h]
  \begin{center}
    \epsfig{file=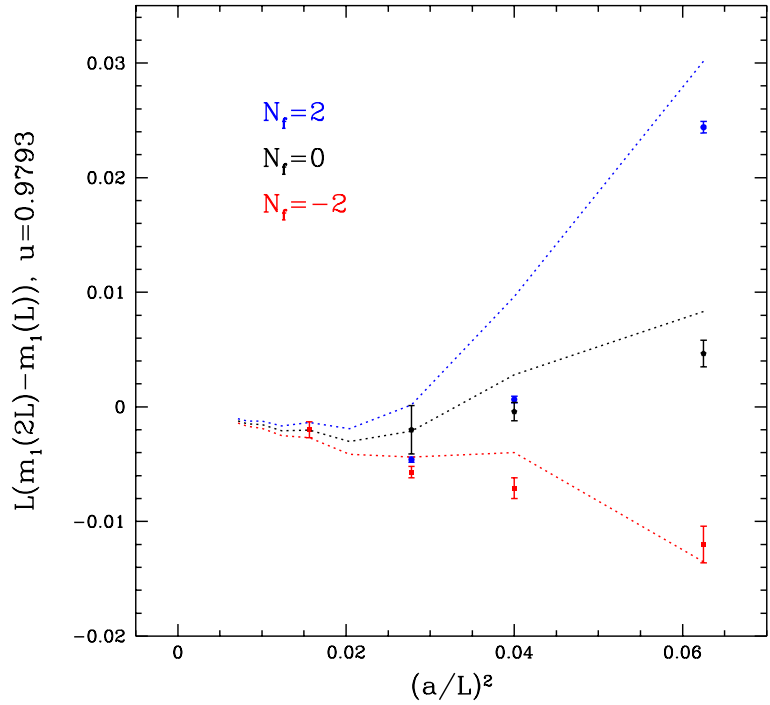,height=11cm,width=11cm}
    \vspace{-0.6cm}
    \caption{\sl Lattice artefacts in the PCAC mass at $u=0.9793$ for
    different flavor numbers. The dashed lines represent (from bottom to top)
    $\Nf=-2,0,+2$ perturbative results. Non-perturbative data are represented
    as rectangles, pentagons and circles.}
    \label{fig:massdiff}
  \end{center}
\end{figure}

Also here it is interesting to see this observable in comparison with the
cases $\Nf=0,+2$. Figure~\ref{fig:massdiff} shows the data points for
the coupling $\gbar^2=0.9793$ in a common diagram. This plot also shows the
perturbative data, which is available for integer $L/a$, connected
with dotted lines. As $m_1$ in \eqref{eq:m1} is defined differently for
even and odd lattice sizes, the perturbative points do not lie on a
smooth curve.

Apparently the points for $\Nf=0,+2$ do not follow
an $\rmO(a^2)$ behavior as well. They still behave inoffensively for
$a\to 0$ and are similar to perturbative expectations. One may infer
that higher orders in $a$ provide the dominant contributions to the
cutoff effects here.

\chapter{Decoupling of heavy flavors}
\label{chap:Decoupling}

Up to now we have only discussed massless schemes, i.e. schemes in which the
renormalization conditions do not depend on the quark masses. This approach
has the advantage that the renormalization group equations assume a simple
form and renormalization constants do not have a dependency on multiple
scales.

For unbroken, non-abelian gauge theories, Appelquist and Carazzone
\cite{AppelQuistCarazzone} have proven a decoupling theorem. It states that
heavy fields coupled to the massless gauge fields do not contribute to the
infrared behavior of the theory, apart from renormalization effects.

The $\MSbar$ scheme - which is certainly the most popular one for high-loop
perturbation theory calculations in QCD - is a massless scheme. While this
approach is in general quite practical, it fails in full QCD with six
non-degenerate flavors, when the coupling is needed at different scales in the
range of the quark masses. For example, the running coupling at an energy of
500~MeV obtained by integrating the renormalization group equation contains
contributions from all six quark flavors, although three of them are so heavy
that they influence the physics on this scale only in a negligible way. Since
the $\beta$-function in this case provides an ``unphysical'' running coupling,
this has to be compensated by the expansion coefficients of the perturbation
series.  Contributions from the heavy quarks must hence be present in all
orders of this series in the form of logarithms of large quark masses. The
reason for the lack of decoupling is that the $\beta$-function and the running
coupling $\alphas$ are not physical observables.  Therefore, the decoupling
theorem cannot be applied directly to them.

This problem can be circumvented by using different effective theories at
different scales. At energies far above the top quark mass, one starts with a
$\beta$-function with six active flavors. When the energy is lowered under the
top quark mass, one considers an effective $\beta$-function with five active
flavors and a different $\Lambda$ parameter.  This method is continued at the
flavor thresholds at lower energies. In this way, the decoupling is
implemented by hand. In general, at the matching scales where one switches
from one effective theory to the other, the running coupling is discontinuous.
Consistency of the pairs of theories below and above the quark masses implies
a set of matching conditions for the running coupling and relations between
the $\Lambda$ parameters. They have been systematically worked out up to
2-loop in \cite{BernreutherWetzel} by connecting the theories above and below
the threshold to the MOM scheme, where the decoupling is explicit. An
introduction into this topic can be found in \cite{hepph9305305}.

In our discussion of the Schr\"odinger functional, we have restricted
ourselves to a mass-in\-de\-pen\-dent step scaling function.  Since the up and
down quarks are in good approximation massless, this is an appropriate
approach for the low energy regime. At some point one may however want to do
research on the running coupling with massive quarks included. As in the
$\beta$-function in the $\MSbar$ scheme, there is no decoupling in the
mass-independent step scaling function. One could work around this difficulty
by using step scaling functions with different numbers of flavors in different
energy regimes. This leaves the problem of connecting the different regimes.

By using a mass dependent step scaling function, one can move through the
flavor thresholds. For example, above the charm quark threshold one simulates
an action with $\Nf=4$ quarks. When the scale is made smaller than the charm
quark mass, its influence on the step scaling function gradually vanishes.  At
some scale the effect of the charm quark becomes negligible in relation to the
precision aimed at, and one can switch to an action with only $\Nf=3$
simulated quarks.
 
\begin{figure}[htb]
  \begin{center}
    \epsfig{file=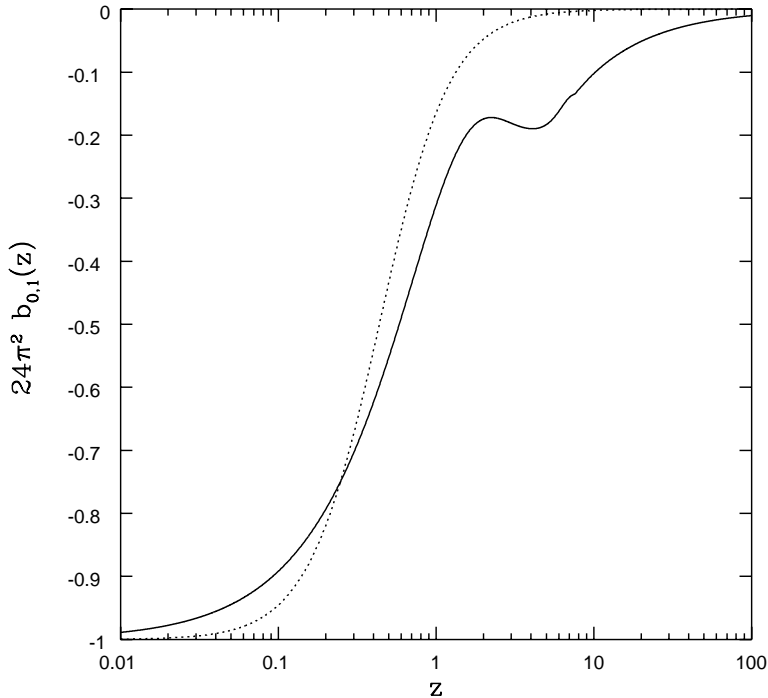,height=11cm,width=11cm}
    \vspace{-0.6cm}
    \caption{\sl Contribution to the 1-loop coefficient of the
      $\beta$-function per flavor. The solid curve shows the behavior in the
      Schr\"odinger functional scheme for $\theta=\pi/5$, whereas the dotted
      curve shows the MOM scheme.}
    \label{fig:b01}
  \end{center}
\end{figure}

Unfortunately, heavy quarks on the lattice can cause sizable lattice
artefacts and make extrapolations to the continuum limit impossible since one
is restricted to lattice sizes that can be simulated on the computer. With the
method described above, one still has to be able to simulate the theory with
the heavy quark included down to the matching point, where the change to the
theory without the heavy quark can be made. The question then is whether one
can find a mass cutoff at which the heavy quark has already decoupled, and at
the same time lattice artefacts are still small.

Figure~\ref{fig:b01} illustrates how the $\beta$-function changes at 1-loop
level when the mass\footnote{In the Schr\"odinger functional scheme
  parametrized by $z=\mR L$, in the MOM scheme by $z=\mR/\mu$.} is increased.
The dashed line shows the behavior in the MOM scheme \cite{GeorgiPolitzer}. As
chiral symmetry holds, the $\beta$-function is an even function of $z$, such
that the asymptotic behavior is proportional to $z^2$ for small masses and
$1/z^2$ for large masses. Therefore, the decoupling is quite abrupt.

In the Schr\"odinger functional scheme, as represented by the solid curve, the
transition region is broader\footnote{The precise shape of the curve depends
  e.g. on the parameter $\theta$. Therefore, the ``dip'' in the right half of
  the plot should not be overrated.}. The temporal fermionic boundary conditions here
violate chiral symmetry even in the continuum limit, hence terms odd in $z$
are possible. Nevertheless, in \cite{heplat9508012} it is argued on the ground of a
1-loop estimation that when a quark with $z\ge z_{\rm cut}=2$ is omitted from
the $\beta$-function, an error in $\gbar^{-2}$ of only 0.003 is made, which is
well below both the typical statistical errors in the ALPHA programme and
experimental measurements. A computation of the lattice artefacts to 1-loop
indicates that lattice artefacts even for several flavors up to $z=2$ are
quite small and follow $(a/L)^2$ expectations when a Symanzik improved action
is used. Putting these observations together, one can conclude that it is
feasible to simulate the range of masses up to $z_{\rm cut}=2$ in the massive
theory and switch to the theory with the heavy quark omitted for higher $z$.

In this chapter, we are going to corroborate whether this statement holds
non-perturbatively. As in the previous chapter, we use the bermion model as a
relatively cheap testing ground. By computing the step scaling function for
increasing $z$, we test the transition from the massive $\Nf=-2$ theory to the
massless $\Nf=0$ theory.

\section{Theory}

We now introduce some of the concepts used in this chapter. We mostly keep the
notation used in \cite{heplat9508012}. Results from there are quoted without
further notice.

As in the massless case, the scale dependence of the coupling is given by the
Callan-Symanzik $\beta$-function,
\beq
  L \frac{\partial \gbar}{\partial L} = -\beta(\gbar),
\eeq
where the derivative is understood to be taken with the renormalized
parameters $\gbar(L_0)$ and $\mbar(L_0)$ fixed at some scale $L_0$.  The
coefficients in the expansion of the $\beta$-function
\beq
  \beta(\gbar) \stackrel{\gbar\to 0}{=} 
  -\gbar^3 \left( b_0(z) + b_1(z) \gbar^2 + \cdots \right)
\eeq
are dependent on the mass parameter
\beq
  z = \mbar(L) L.
\eeq
The step scaling function is generalized in a similar way. On the smaller
lattice, the bare parameters are tuned such that the renormalized coupling and
the parameter $z$ take their prescribed values.  The step scaling function
then is the value of the renormalized coupling on a lattice with twice the
size and the same bare parameters,
\beq
  \sigma(u,z) = \gbar^2(2L) \Big|_{\gbar(L)=u,\mbar(L)=z/L}.
\eeq
In a recursive computation of the running coupling, the quark mass runs, and
therefore in each iteration step the parameter $z$ has to be adjusted
depending on the scale.  This can be done by complementing the step scaling
function for the coupling with a step scaling function $\sigma_{\rm p}$ for
the mass, generalizing the definition in \cite{heplat9810063},
\beq
  \sigma_{\rm p}(u,z) = \lim_{a\to 0}
  \frac{\ZP(g_0,2L/a)}{\ZP(g_0,L/a)}\Bigg|_{\gbar(L)=u,\mbar(L)=z/L}.
\eeq

In order to find the mass dependence of quantities in perturbation theory, it
is convenient to relate their coefficients to a massless scheme like $\MSbar$.
In the expansion
\beq
  g^2_{\MSbar} = \gbar^2 + \frac{c_1(z)}{4\pi} \gbar^4 + \rmO(\gbar^6),
\eeq
the coefficient $c_1(z)$ has a pure gauge theory contribution and a
mass-dependent contribution proportional to the number of flavors,
\beq
  c_1(z) = c_{1,0} + \Nf c_{1,1}(z).
\eeq
For large $z$, the quark contribution diverges as
\beq
  c_{1,1}(z) = \frac{1}{3\pi} \log(z) + \rmO(1/z),
\eeq
with corrections depending on the phase $\theta$ in the fermionic boundary
conditions. Denoting $\hat g^2=\gbar^2(L)|_{z=0}$, this logarithmic divergence
shows up for example in the mass dependence of the coupling at fixed scale,
\beqn
  \gbar^2 - \hat g^2
&=& -\frac{\Nf}{4\pi} 
    \left\{ c_{1,1}(z) - c_{1,1}(0) \right\} \hat g^4 
    + \rmO(\hat g^6) \nonumber\\
&\stackrel{z\to\infty}{=}& -\frac{\Nf}{12\pi^2} \log(z) \, \hat g^4
    + \rmO(\hat g^6).
\eeqn

The massive step scaling function in 1-loop can be computed by switching the
coupling to the $\MSbar$ scheme at the scale $L$ and going back to the
Schr\"odinger functional scheme at the scale $2L$ and the mass parameter $2z$.
We obtain
\beq
  \sigma(u,z) = u + s_0(z) u^2 + \rmO(u^3)
\eeq
with
\beq
  s_0(z) = \frac{\Nf}{4\pi} \left\{ c_{1,1}(z) - c_{1,1}(2z) \right\}
           + 2\log 2 \, b_0(0).
  \label{eq:s0ofz}
\eeq
For large $z$, the mass dependent contribution to the step scaling
function converges to 
\beq
  \sigma(u,z) - \sigma(u,0) \stackrel{z\to\infty}{=}
  \frac{\Nf}{12\pi^2} \log 2 \; u^2 + \rmO(u^3),
\eeq
which is just the difference between the theories with $\Nf$ and zero quarks.

\section{Renormalized mass}

The connection between the renormalized and the PCAC mass is given by the relation
\eqref{eq:RenormalizedAndUnrenormalizedMass}. For our matching runs, we have
always approximated $z\approx m_1 L$. In the following, we will discuss the
neglected quantities appearing there.

In perturbation theory, the improvement coefficient $\bA-\bP$ is
\cite{heplat0009021}
\beq
  (\bA - \bP)(g_0) = -0.00093(8) g_0^2 + \rmO(g_0^4).
\eeq
In the quenched approximation, this value was computed non-perturbatively. It
turns out that only for $\beta\ll 7$ one obtains values that are significantly
larger in magnitude. As this improvement coefficient is multiplied with the
subtracted mass $\mq$, the overall effect on the mass is very tiny compared to
our statistical error.

The renormalization factor $\ZA$ of the axial vector current is scale
independent. To 1-loop in perturbation theory, it is given by
\cite{heplat9611015}
\beqn
  \ZA(g_0) &=& 1 + \ZA^{(1)} g_0^2 + \rmO(g_0^4) \nonumber\\
  \ZA^{(1)} &=& -0.116458(2).
\eeqn
In the quenched approximation, the non-perturbative determination of $\ZA$
deviates about 2 \% from this perturbative formula at the bare couplings used
here.

The axial density renormalization factor $\ZP$ can also be expanded in the
bare coupling. The coefficients are functions of $L/a$ and diverge as powers
of $\log(L/a)$ in the continuum limit \cite{heplat9808013},
\beqn
  \ZP(g_0,L/a) &=& 1 + \ZP^{(1)}(L/a) g_0^2 + \rmO(g_0^4) \nonumber\\
  \ZP^{(1)}(L/a) &=& \frac 4 3 z_{\rm p}(\theta,\rho) 
    - d_0 \log(L/a) + \rmO(a/L).
\eeqn
Here we have followed the notation of \cite{heplat9808013} where the calculation was
done for some choices of $\theta$ and $\rho=T/L$, and without background field.

By neglecting the renormalization factors we make a systematic error in $z$ of
\beq
  \delta z / z = \delta\mbar / \mbar 
  = \left( \ZP^{(1)}-\ZA^{(1)} \right) g_0^2 + \rmO(g_0^4).
\eeq
Numerically, the scale independent part of $\ZP^{(1)}$ cancels $\ZA^{(1)}$
almost exactly. The logarithmic term causes $\delta z/z$ to vary between
roughly $0.051$ on the $L/a=4$ lattices and $0.071$ on the $L/a=8$ lattices,
for the bare couplings we have used. On the one hand, this leads to a
simultaneous shift of all data points for a fixed $z$. On the other hand, it
leads to different errors in the step scaling function with fixed $z$ but
different $L/a$. The latter error can be estimated as follows. When we neglect
the mass dependence of lattice artefacts, an error in $z$ propagates into the
step scaling function as
\beq
  \delta\sigma(u,z) \approx \delta z \, s_0'(z) \, u^2.
\eeq
The derivative of $s_0(z)$ with respect to $z$ can be computed numerically
using data from \cite{heplat9508012}. The largest effect appears for $z=2$,
where
\beq
  s_0'(2) = 0.00349 \, \Nf.
\eeq
The resulting error in $\sigma(u,z)$ then is less than 0.0008, which is an
order of magnitude below our statistical error. Although the use of several
assumptions from perturbation theory with vanishing background field in this
estimate is quite unsatisfactory, we believe it to be not too unrealistic.

\section{Results}

\begin{figure}[htb]
  \begin{center}
    \epsfig{file=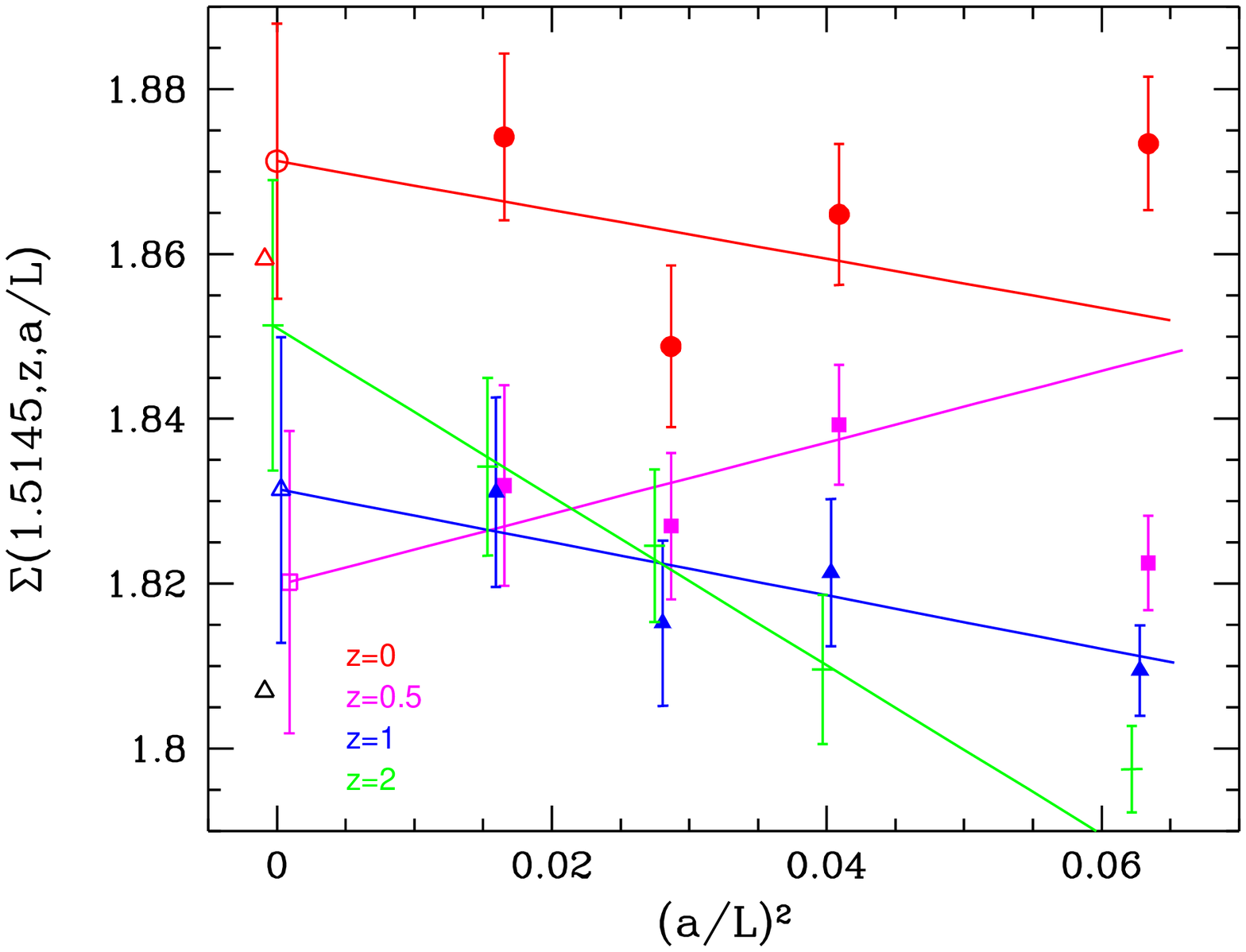,height=11cm,width=11cm}
    \vspace{-0.6cm}
    \caption{\sl Step scaling function at $u=1.5145$ for $z=0,0.5,1,2$
      with fits linear in $(a/L)^2$. Please note that this figure is meant to
      be viewed in color.}
    \label{fig:decoupling}
  \end{center}
\end{figure}

In table~\ref{tab:res_decoupling}, we list our results of the step scaling
function for the lattice sizes $L/a=4,5,6,8$ and for values of the mass
$z=0.5,1,2$. In addition to these results, we use for $z=0$ the results at
$\gbar^2=1.5145$ from chapter~\ref{chap:Bermions}, as listed in
table~\ref{tab:res_improved_2L}. Figure~\ref{fig:decoupling} shows a plot with
all results.

Although the error bars are roughly of the same size as the variation of the
step scaling function for different $L/a$, when $z$ is kept constant, one sees
a slight trend that lattice artefacts become larger for increasing mass. Since
for $L/a=4$, the value $z=2$ is already equivalent with $\mbar a=1/2$, this is
to be expected. Nevertheless, the lattice artefacts do not pose a problem of
principle for the extrapolation of our data with an ansatz linear in
$(a/L)^2$.

\begin{table}[p]
  \begin{center}
    \begin{tabular}{llllll}\hline\hline
    $z_1$ & $L/a$ & $\beta$ & $\kappa$ & $\gbar^2(L)$ & $\gbar^2(2L)$ \\
    \hline
    0.5 & 4 & 8.274636 & 0.129184  &  1.5145(15)  &  1.8225(53)  \\
    0.5 & 5 & 8.474702 & 0.129606  &  1.5145(16)  &  1.8393(69)  \\
    0.5 & 6 & 8.651298 & 0.129873  &  1.5145(24)  &  1.8270(82)  \\
    0.5 & 8 & 8.920407 & 0.130185  &  1.5145(35)  &  1.8320(111) \\
    \hline
    1   & 4 & 8.227619 & 0.125497  &  1.5145(13)  &  1.8095(52)  \\
    1   & 5 & 8.426126 & 0.126580  &  1.5145(16)  &  1.8158(65)  \\
    1   & 6 & 8.599670 & 0.127310  &  1.5145(30)  &  1.8152(91)  \\
    1   & 8 & 8.855909 & 0.128237  &  1.5145(41)  &  1.8311(99)  \\
    \hline
    2   & 4 & 8.166523 & 0.118802  &  1.5145(13)  &  1.7975(49)  \\
    2   & 5 & 8.357652 & 0.120885  &  1.5145(17)  &  1.8096(87)  \\
    2   & 6 & 8.521765 & 0.122377  &  1.5145(20)  &  1.8246(88)  \\
    2   & 8 & 8.793347 & 0.124347  &  1.5145(33)  &  1.8342(97)  \\
    \hline
    \end{tabular}
    \caption{\sl Tuning results for $\beta$ and $\kappa$ for 
      the given values of $z$ and $L/a$, together with the measured coupling
      at $L$ and $2L$.}
    \label{tab:res_decoupling}
  \end{center}
\end{table}

\begin{table}[htbp]
  \begin{center}
    \begin{tabular}{llll}\hline\hline
      $z$ & \mbox{without $L/a=4$} & \mbox{with $L/a=4$} \\
     \hline
     0   & 1.871(17) & 1.859(11) \\
     0.5 & 1.820(18) & 1.840(11) \\
     1.0 & 1.831(19) & 1.833(11) \\
     2.0 & 1.851(18) & 1.844(11) \\
     \hline\hline
     \end{tabular}
     \caption{\sl Extrapolated simulation results of the
       massive step scaling function $\sigma(1.5145,z)$ for $\Nf=-2$.} 
    \label{tab:sigma_decoupling}
  \end{center}
\end{table}

\begin{table}[htbp]
  \begin{center}
    \begin{tabular}{llll}\hline\hline
       & $n=1$ & $n=2$ & $n=3$ \\
     \hline
     $\hat\sigma^{n \rm -loop}(1.5145,0)\vert_{\Nf=-2}$
       & 1.811555 & 1.851225 & 1.859385 \\
     $\hat\sigma^{n \rm -loop}(1.5145,0)\vert_{\Nf=0}$
       & 1.773940 & 1.803887 & 1.806922 \\
     \hline
     $\sigma^{n \rm -loop}(1.5145,0)\vert_{\Nf=-2}$
       & 1.762845 & 1.828158 & 1.849306 \\
     $\sigma^{n \rm -loop}(1.5145,0)\vert_{\Nf=0}$
       & 1.735997 & 1.788089 & 1.801806 \\
     \hline\hline
     \end{tabular}
     \caption{\sl Perturbation theory for the step scaling function
       for $\Nf=-2$ and $\Nf=0$ to 1-loop, 2-loop and 3-loop.}
    \label{tab:sigmapert}
  \end{center}
\end{table}

We have fitted our data with the $L/a=4$ point dropped. This extrapolation is
shown in figure~\ref{fig:decoupling}. To see how much the $L/a=4$ point
influences the extrapolation, we have also computed the fit with this point,
and listed the results in table~\ref{tab:sigma_decoupling}. All these numbers
are also to be compared with the perturbative values
\beqn
  \hat\sigma^{\rm 3-loop}(1.5145,0)
    \vert_{\Nf=-2} &=& 1.859385 \nonumber\\
  \hat\sigma^{\rm 3-loop}(1.5145,0)
    \vert_{\Nf=0} &=& 1.806922.
\eeqn

In the figure, these 3-loop results are shown as red and black triangles,
respectively.  From these numbers, it is clear that our accuracy is not
sufficient for resolving the step scaling functions with $z=0.5,1,2$. In fact,
when we leave out the $L/a=4$ point, the extrapolated values appear in the
``wrong'' order, but with a statistical error that is roughly of the same
magnitude as the difference between the values.

In \cite{heplat9508012}, the lattice artefacts parametrized as in
\eqref{eq:delta} were calculated to 1-loop for the massive step scaling
function, both for an improved and unimproved action. The mass definition used
there is the pole mass, which is to 1-loop equivalent to our mass definition
with the background field switched off. At least for $z=0$, the lattice
artefacts do not differ much between different mass definitions, so that the
qualitative behavior is probably independent of it
\cite{SommerSSFUnpublished}.

The lattice artefacts we see in our simulations are of the same magnitude as
those computed perturbatively. The perturbative lattice artefacts also behave
like $(a/L)^2$, which is consistent with our observation that an ansatz linear
in $(a/L)^2$ works. It would be desirable to have the $\delta_1(a/L)$ values or
even $\delta_2(a/L)$ for $z\neq 0$ with our mass definition. Then one could
use the method described in section~\ref{sec:PerturbativeCorrections} to
cancel the perturbative contributions to the lattice artefacts. Since the
lattice artefacts cause the step scaling function at finite $a/L$ to decouple
faster than in the continuum, this method should reduce the slope of the fit,
which here depends on the mass parameter.

We would also like to note here that the size of lattice artefacts depends on
the number of flavors. With $\Nf=-2$ used here, the quark contribution has the
same sign as the pure gauge theory contribution. This situation changes for
$\Nf=2$ and $\Nf=4$. There, the perturbative lattice artefacts, for instance
with $z=2$, are \emph{smaller} than in the pure gauge theory.

A quantitative comparison of our extrapolated data with perturbation theory is
difficult. In table~\ref{tab:sigmapert}, the perturbative results for the step
scaling in 1-loop to 3-loop are listed, according to the definitions
\eqref{eq:SigmaPerturbative} and \eqref{eq:SigmahatPerturbative}.

Clearly, the 2-loop and higher contributions are so significant that a
comparison of non-perturbative results to the 1-loop results is impossible.
On the other hand, the massive $\beta$-function has not been computed beyond
1-loop. An alternative to studying absolute values of $\sigma(u,z)$ might be
to consider only the mass dependent contribution $\sigma(u,z)-\sigma(u,0)$.
But also for this quantity, higher order terms are a problem. We expect
\beq
  \sigma(u,z)\vert_{\Nf=-2} - \sigma(u,0)\vert_{\Nf=-2}
    \stackrel{z\to\infty}{=}
    \sigma(u,0)\vert_{\Nf=0} - \sigma(u,0)\vert_{\Nf=-2}.
\eeq
But to 1-loop,
\beqn
  \hat\sigma^{\rm 1-loop}(1.5145,0)\vert_{\Nf=0}
  - \hat\sigma^{\rm 1-loop}(1.5145,0)\vert_{\Nf=-2}
  = -0.038 \nonumber\\
  \sigma^{\rm 1-loop}(1.5145,0)\vert_{\Nf=0}
  - \sigma^{\rm 1-loop}(1.5145,0)\vert_{\Nf=-2}
  = -0.027.
\eeqn
This indicates that $\rmO(u^3)$ corrections to this difference are of the same
order of magnitude as the $z$-dependence we want to investigate.

\chapter{Summary and Outlook}

In this thesis, we have investigated aspects of the step scaling function
in the Schr\"odinger functional scheme. 

In chapter~\ref{chap:Performance} we have compared the efficiency of various
algorithms in pure $\SUthree$ gauge theory. The emphasis in this study is on
the coupling $\gbar$, which is directly relevant for the computation of the
step scaling function. The favored method for this kind of theories is the
Hybrid Overrelaxation (HOR) algorithm. However, this algorithm is not easily
generalizable to theories with an action that is quadratic in the individual
link variables. The other algorithms in this study are based on molecular
dynamics. While the global Hybrid Monte Carlo (HMC) is the standard algorithm
for the simulation of dynamical fermions, the Local Hybrid Monte Carlo (LHMC)
is a local version of it. The important results of this study are a factor of
3 in the cost of LHMC, and even a factor of 26 for HMC compared to the HOR
algorithm, for the lattice size $L/a=8$.  These numbers may for instance serve
as a guideline for the choice of suitable algorithms for actions with a pure
gauge theory like term plus a correction. The bermion model studied in this
work is an example for such a theory.

Chapter~\ref{chap:Bermions} is devoted to the step scaling function for two
flavors of massless bermions with $\rmO(a)$ improvement. The motivation for
considering this model is the reduced computational cost in comparison with
dynamical fermions. We have studied in detail the approach to the continuum
limit and lattice artefacts. We have found that lattice artefacts follow the
expected behavior and an extrapolation linearly in $(a/L)^2$ works well and
yields results compatible with renormalized perturbation theory. A comparison
with unimproved Wilson bermions illustrates the success of $\rmO(a)$ Symanzik
improvement and confirms that the unimproved and improved theory converge to a
universal continuum limit. Unfortunately, algorithmic difficulties arising
from the inclusion of the clover term cause a large overhead of our simulation
program compared to the unimproved one, so that the efficiency advantage over
fermions is only a factor 10. Consequently, we cannot reach lattice sizes much
larger than with dynamical fermions. Nevertheless, our results put the methods
used in the $\Nf=2$ case on a firmer ground and increase our trust that a
solid determination of the $\Lambda$ parameter for two massless fermions can
soon be finished. The running coupling for $\Nf=2$ will also serve as the
basis for quark mass renormalization and phenomenological investigations.

In chapter~\ref{chap:Decoupling} we have investigated the massive step scaling
function in the bermion model. Since heavy quarks are expected to decouple at
low energies, this quantity can be used to move through a flavor threshold,
below which one can switch to an effective theory without the heavy quark. Our
aim was to corroborate statements from previous perturbative studies, which
are essential for the feasibility of the recursive evaluation of the running
coupling in QCD with several massive quarks. We have found lattice artefacts
to be of tractable magnitude in our range of simulated masses. The
extrapolation of the massive step scaling function showed that the decoupling
effect is not totally different from our expectation. Resolving different
values of the mass correctly would require to increase our statistical
precision significantly in comparison to the mass dependence of the step
scaling function.

\begin{appendix}

\chapter{Notation}
\label{app:Notation}

As a convention, we use letters $\mu,\nu,\ldots$ from the middle of the greek
alphabet to denote Lorentz indices running from 0 to 3. When we are referring
to spatial components only, we use latin letters $k,l,\ldots$. Dirac spinors
carry indices $A,B,\ldots$. Color vectors in the fundamental representation of
$\SUN$ have indices $\alpha,\beta$ from the beginning of the greek alphabet.
Repeated indices are always summed over.

We distinguish space-like 3-vectors $\xvec=(x_1,x_2,x_3)$ from 4-vectors
$x=(x_0,x_1,x_2,x_3)$ by a bold font.

\section*{SU($2$) matrices}

Any $2 \times 2$ matrix can be written in the form
\beq
  w = w_0 + i w_k \sigma_k.
\eeq  
where $\sigma_k$ are the Pauli matrices
\beq
  \sigma_1 = \left( \begin{array}{cc}
    0 & 1 \\
    1 & 0 \end{array} \right) \quad
  \sigma_2 = \left( \begin{array}{cc}
    0 & -i \\
    i & 0 \end{array} \right) \quad
  \sigma_3 = \left( \begin{array}{cc}
    1 & 0 \\
    0 & -1 \end{array} \right).
\eeq
The set of complex coefficients are called the quaternionic
representation. In this representation,
\beqn
  \det w &=& w_\mu w_\mu \nonumber\\
  \Tr \, w &=& 2 w_0.
\eeqn

For $\SUtwo$ matrices $w$, these coefficients are real
and fulfill $w_\mu w_\mu=1$.

Another possible parametrization of $\SUtwo$ is the exponential
mapping with group generators. This will not be used here, but
only in the \SUN case where $N\ge 3$.

\section*{SU($N$) matrices}

Every $\SUN$ matrix $U$ can be parametrized by
\beq
  U = \exp\left\{ \sum_{a=1}^{N^2-1} i T_a \omega_a \right\}.
\eeq
The coefficients $\omega_a$ are real-valued. $T_a \in \sun$ are 
called generators of the Lie algebra associated with the group. 
Their representation is as $N \times N$ matrices. They fulfill
\beq
  X^\dagger = -X \,\,\mbox{and}\,\, \Tr X = 0.
\eeq
For $\SUthree$, we write $T_j=\lambda_j$, where $\lambda_j$ are the
Gell-Mann matrices
\beqn
&&  \lambda_1 = \left( \begin{array}{ccc}
  0 & 1 & 0  \\
  1 & 0 & 0  \\
  0 & 0 & 0  \end{array} \right) \quad
  \lambda_2 = \left( \begin{array}{ccc}
  0 & -i & 0 \\
  i & 0 & 0  \\
  0 & 0 & 0  \end{array} \right) \quad
  \lambda_3 = \left( \begin{array}{ccc}
  1 & 0  & 0 \\
  0 & -1 & 0 \\
  0 & 0  & 0 \end{array} \right) \\
&&  \lambda_4 = \left( \begin{array}{ccc}
  0 & 0 & 1  \\
  0 & 0 & 0  \\
  1 & 0 & 0  \end{array} \right) \quad
  \lambda_5 = \left( \begin{array}{ccc}
  0 & 0 & -i \\
  0 & 0 & 0  \\
  i & 0 & 0  \end{array} \right) \quad
  \lambda_6 = \left( \begin{array}{ccc}
  0 & 0 & 0  \\
  0 & 0 & 1  \\
  0 & 1 & 0  \end{array} \right) \\
&&  \lambda_7 = \left( \begin{array}{ccc}
  0 & 0 & 0  \\
  0 & 0 & -i \\
  0 & i & 0  \end{array} \right) \quad
  \lambda_8 = \left( \begin{array}{ccc}
  1/\sqrt 3 & 0 & 0 \\
  0 & 1/\sqrt 3 & 0 \\
  0 & 0 & -2/\sqrt 3 \end{array} \right) 
\eeqn

\section*{Haar measure}

As integration measure for $\SUN$, we use the Haar measure $dU$.
It has the (for a gauge theory natural) property of being invariant
under left and right multiplication with group elements,
\beq
  \int dU f(U) = \int dU f(gUh^{-1}), \quad g,h \in \SUN
\eeq
and it is normalized,
\beq
  \int dU = 1.
\eeq
In the quaternionic representation of $\SUtwo$, the Haar measure
takes the form
\beq
  dw = \frac{1}{2\pi^2} \delta(w^2-1) d^4w.
  \label{eq:HaarMeasureQuaternionic}
\eeq

\section*{Dirac matrices}

We choose Dirac matrices in the chiral representation, where
\beqn
  \gamma_{1,2,3} &=& \left( \begin{array}{cc}
    0 & -i\sigma_{1,2,3} \\
    i\sigma_{1,2,3} & 0 \end{array} \right) \nonumber\\
  \gamma_0 &=& \left( \begin{array}{cc}
    0 & -1 \\
    -1 & 0 \end{array} \right) \\
  \gamma_5 &=& \left( \begin{array}{cc}
    1 & 0 \\
    0 & -1 \end{array} \right).
\eeqn
These matrices fulfill
\beq
  \gamma_\mu^\dagger=\gamma_\mu, \quad
  \gamma_\mu^2 = 1, \quad
  \{ \gamma_\mu, \gamma_\nu \} = 2 \delta_{\mu,\nu}.
\eeq
Also, in this representation, $\gamma_5=\gamma_0\gamma_1\gamma_2\gamma_3$ has
the property $\gamma_5^\dagger=\gamma_5$ and $\gamma_5^2=1$.
The matrices
\beq
  \sigma_{\mu\nu} = \frac i 2 [\gamma_\mu,\gamma_\nu] 
\eeq
are hermitian and are explicitly given by
\beqn
  \sigma_{ok} &=& \left( \begin{array}{cc}
    \sigma_k & 0 \\
    0 & -\sigma_k \end{array} \right) \nonumber\\
  \sigma_{ij} &=& \epsilon_{ijk} \left( \begin{array}{cc}
    \sigma_k & 0 \\
    0 & -\sigma_k \end{array} \right),
\eeqn
where $\epsilon_{ijk}$ is the totally antisymmetric tensor.

\section*{Lattice derivatives}

On the lattice, derivatives have to be discretized. We define forward and
backward derivatives by
\beqn
  \partial_\mu f(x) = \frac{f(x+a\muhat)-f(x)}{a} \nonumber\\
  \partial^*_\mu f(x) = \frac{f(x)-f(x-a\muhat)}{a},
\eeqn
where $\muhat$ denotes the unit vector in direction $\mu$.

\chapter{Random numbers}
\label{app:RandomNumbers}

\section*{Generator}

All updating algorithms used in this work require a large number of random
numbers. In practice, it is only feasible to use \emph{pseudo} random numbers,
which are produced by deterministic process. Here we use the generator by
L\"uscher \cite{heplat9309020,RandomNumbers} which is based on one proposed by
Marsaglia and Zaman \cite{MarsagliaZaman}.  It is known to have a large
periodicity time and its underlying dynamical system is chaotic, so it is
assumed to generate numbers which do not introduce a bias into simulation
results.

\section*{Gaussian distribution}

Random numbers with a distribution proportional to $e^{-y^2}$
can be generated in pairs. This means, we generate the combined
distribution
\beq
  P(y_1, y_2) = \exp(-y_1^2-y_2^2),
\eeq
which gives two independent random numbers with the desired distribution. The
variables $y_1$ and $y_2$ can be interpreted as coordinates in a
two-dimensional plane, i.e. they can be transformed into polar coordinates
\beqn 
  y_1 &=& \rho \cos\theta \nonumber\\
  y_2 &=& \rho \sin\theta.
\eeqn
The transformed variables have a distribution $\tilde P(\rho,\theta)$ with
\beqn
  P(y_1, y_2) dy_1 dy_2 = \tilde P(\rho,\theta) d\rho d\theta \nonumber\\
  \det \frac{\partial(y_1,y_2)}{\partial(\rho,\theta)} = \rho.
\eeqn
The distribution for the polar coordinates is thus given by
\beqn
  \tilde P(\rho,\theta) 
  &=& \rho e^{-\rho^2} \nonumber\\
  &\propto& \frac{d}{d\rho} (1-e^{-\rho^2}).
\eeqn
$\theta$ can simply be computed by multiplying a uniformly in $[0,1[$
distributed number by $2\pi$. $\rho$ can be obtained by drawing $u$ from a
uniform distribution in $[0,1[$ and computing\footnote{The usage of $1-u$
  avoids logarithms of zero, since $u$ takes values smaller than $1$.}
\beq
  \rho = (-\log(1-u))^{\frac 1 2}.
\eeq

\chapter{Error analysis}
\label{app:ErrorAnalysis}

An important aspect of the extraction of results from Monte Carlo runs is the
estimation of errors. As the set of data from a finite simulation run is a
sample from a statistical ensemble, there is of course a naive statistical
error proportional to $\sqrt{1/N}$.

Furthermore, any Monte Carlo run needs some time before it reaches the thermal
equilibrium, where field configurations are distributed according to the
ensemble given by the Boltzmann factor. Including the thermalization phase
into the average of an observable leads to an initialization bias. The only
practical way to avoid this is to throw away a sufficiently high number of
measurements at the beginning of the run.

Another source of errors is the correlation between successive measurements.
In general, an updating algorithm does not generate a chain of configurations
which are really independent of each other. One can express the real
statistical error of an observable $\calO$ as the product of the ``naive
error'' and a factor $2\tauint$, where $\tauint$ is an integrated
autocorrelation time as described below. The autocorrelation time may differ
between different observables.

In the evaluation of autocorrelation times, it is essential to distinguish
between primary quantities and secondary quantities.  Primary quantities, such
as the inverse coupling or the average plaquette, are directly extracted from
a field configuration, i.e. they are expectation values of functions of the
field variables. Secondary quantities are functions of two or more, in general
correlated, primary quantities. In this work, an example for a secondary
quantity is the current mass which is calculated as a quotient of fermionic
correlation functions $\fA$ and $\fP$.

\section*{Primary quantities}

Let us a sequence of measurements $a^i, i=1 \ldots N$ of an observable $\hat A$
which has the exact mean value $A=\langle A \rangle$. We assume that these
stem from a large enough ensemble in the thermal equilibrium, such that they
have a Gaussian distribution following the central limit theorem. The natural
estimator for $A$ is the sample average,
\beq
  A \approx \bar a := \frac{1}{N} \sum_{i=1}^N a^i.
\eeq
This is the best estimator in the sense that the expectation value
of $\bar a$ over an infinite ensemble of Monte Carlo runs is the
correct value,
\beq
  \langle \bar a - A \rangle_{\rm MC} = 0.
  \label{eq:ExpectationOfMeanPrimary}
\eeq

If all configurations in the sample are uncorrelated, the
variance of the normal distribution is
\beqn
  \sigma^2 
&=& \frac{ \langle ( \hat A - A )^2 \rangle }{N} \nonumber\\
&=& \frac{ \langle (\bar a - A)^2 \rangle_{\rm MC}}{N}.
\label{eq:SigmaNaiv}
\eeqn
In the realistic case, where configurations are correlated,
this underestimates the error. We introduce the autocorrelation
function
\beq
  \Gamma(i-j) = \Bigl\langle (a^i - A ) (a^j - A) \Bigr\rangle_{\rm MC}
\eeq
between the i'th and the j'th measurement. $\Gamma$ depends only
on the distance between the measurements. Its value at zero is
\beq
  \Gamma(0) = \langle ( \hat A - A )^2 \rangle.
\eeq
An estimator for $\Gamma$ is 
\beq
  \Gamma(t) \approx \frac{1}{N-t} \sum_{i=t+1}^N
    (a_i - \bar a) (a_{i-t} - \bar a).
  \label{eq:Gamma}
\eeq
In general, we neglect the error of $\Gamma(t)$.  The integrated
autocorrelation time is defined as\footnote{This definition is one
common convention, where $\tauint$ becomes $1/2$ when there are no
autocorrelations.}
\beq
  \tauint = \frac 1 2 
  \sum_{i=-\infty}^{\infty} \frac{\Gamma(i)}{\Gamma(0)}.
\eeq

We can now express the ``true'' variance of the estimator $\bar a$
by the autocorrelation function,
\beqn
  \sigma^2 
  &=& \frac{1}{N^2} \left\langle 
      \left(\sum_{i=1}^N (a_i - A)^2 \right) \right\rangle \nonumber\\
  &=& \frac{1}{N^2} \sum_{i=1}^N \sum_{j=1}^N \Gamma(i-j) \nonumber\\
  &=& \frac{1}{N^2} \sum_{t=-(N-1)}^{N-1} 
      \! ( N - |t| ) \, \Gamma(t) \nonumber\\
  &\approx& \Gamma(0) \, \frac{2\tauint}{N}
  \quad \mbox{for $N \gg t$}.
  \label{eq:SigmaOfGammaAndTauint}
\eeqn
A comparison with \eqref{eq:SigmaNaiv} shows that the error behaves as if the
effective number of measurements was $N/(2\tauint)$ instead of $N$.
Intuitively speaking, generating a completely decorrelated configuration from
a given one requires $2\tauint$ update steps.

A popular method called \emph{binning} takes this statement verbatim: in the
binning method, one averages measurements over bins of length $\Nbin$ and
looks at the naively computed error of the binned data set. When $\Nbin$ is
increased, subsequent bins should gradually become uncorrelated after it
reaches the value $2\tauint$. One expects the error in dependence of the bin
length to converge to a plateau. In practice, the autocorrelation function has
statistical errors which are enhanced for increasing $\Nbin$ (which implies a
decreasing number of bins) and the plateau is washed out. The determination of
$\tauint$ and in particular its error is therefore difficult.

Here, we use a different method explained in \cite{Sokal}. The main point here
is not to sum up $\Gamma$ over the whole possible range of times, but only
over a window $[-M,M]$,
\beq
  \hat\tauint = \frac 1 2 \sum_{t=-M}^{M} \frac{\Gamma(t)}{\Gamma(0)}.
\eeq
Making this window larger increases the variance of the estimated $\tauint$,
making it smaller increases its bias. As a compromise, Madras and Sokal
\cite{MadrasSokal} propose to choose the window to be the smallest one which
fulfills the condition $M \ge c \hat\tauint$, with a suitably chosen $c$. As
$\hat\tauint$ depends on the choice of $M$, this plays the role of a
self-consistency condition. We usually choose $c=6$. With this method, the
error of $\tauint$ can be estimated as
\beq
  \frac{\delta\tauint}{\tauint} = \sqrt{\frac{2(2M+1)}{N}}.
\eeq

\section*{Secondary quantities}

Let us consider a set of primary quantities $\hat A_\alpha$ with exact mean
values $A_\alpha$. For each of them, there are measurement data $a_\alpha^i,
i=1 \ldots N$, which have sample means $\bar a_\alpha$. Now we want to study a
secondary quantity which is given as a function $f=f(\{\hat A_{\alpha}\})$. A
natural estimator for the value $F=f(\{A_{\alpha}\})$ is $f(\{\bar
a_\alpha\})$.  For primary quantities, we have seen in
\eqref{eq:ExpectationOfMeanPrimary} that the expectation value of the
estimator is identical with the exact mean and the variance goes with $1/N$,
\beqn
  \langle \bar a_\alpha - A_\alpha \rangle_{\rm MC} &=& 0 \nonumber\\
  \langle (\bar a_\alpha - A_\alpha)^2 \rangle_{\rm MC} &=& \rmO(1/N).
  \label{eq:ExpectationOfVarPrimary}
\eeqn
For secondary quantities, this is a bit different. To see this, we follow
\cite{heplat0009027} and expand $f(\{\bar a_{\alpha}\})$ around the set of
$A_\alpha$. We denote the derivative of $f$ with respects to its $\alpha$'th
argument as $f_\alpha$. Then the Taylor expansion to first order reads
\beq
  f(\{\bar a_\alpha\}) = F 
  + \sum_{\alpha} (\bar a_\alpha - A_\alpha) f_\alpha(\{A_\alpha\}) 
  + \rmO\!\left(\sum_\alpha(\bar a_\alpha-A_\alpha)^2\right).
\eeq
Thus, the estimate has an unavoidable bias
\beq
  \Bigl\langle f(\{\bar a_\alpha\}) - F \Bigr\rangle_{\rm MC} = \rmO(1/N).
\eeq
This can be reduced by making $N$ large enough. The variance is
\beqn
  \sigma^2 
  &=& \Bigl\langle ( f(\{\bar a_\alpha\}) - F )^2 \Bigr\rangle_{\rm MC} \nonumber\\
  &=& \left\langle \left[ \sum_\alpha (\bar a_\alpha - A_\alpha) 
      f_\alpha(\{A_\alpha\}) \right]^2 \right\rangle_{\rm MC}
      + \rmO(1/N^2).
\eeqn
This can be written in a more intuitive form by defining projected observables
\parbox{1cm}{\hspace{1cm}}
\parbox{4.5cm}{\beqn
  A_{\rm H} &=& \sum_\alpha A_\alpha f_\alpha(\{A_\alpha\}) \nonumber\\
  \bar a_{\rm H} &=& \sum_\alpha \bar a_\alpha f_\alpha(\{A_\alpha\}) \nonumber
\eeqn}
\parbox{1cm}{\hspace{1cm}}
\parbox{4.5cm}{\beqn
  A_{\rm\bar h} &=& \sum_\alpha A_\alpha f_\alpha(\{\bar a_\alpha\}) \nonumber\\
  \bar a_{\rm\bar h} &=& \sum_\alpha \bar a_\alpha f_\alpha(\{\bar a_\alpha\}), \nonumber
\eeqn}
\hfill
\parbox{1cm}{\beqn
\eeqn}

\noindent such that
\beqn
  \sigma^2 
&=& \Bigl\langle ( A_{\rm H} - \bar a_{\rm H} )^2 
    \Bigr\rangle_{\rm MC} + \rmO(1/N^2) \nonumber\\
&\approx& \Bigl\langle ( A_{\rm\bar h} - \bar a_{\rm\bar h})^2 
          \Bigr\rangle_{\rm MC}.
\eeqn

This is formally the same situation as in the case of primary observables.
Consequently, the same methods can be applied here.  In particular, given the
analytical form of the derivatives of $f$ with respect to all its arguments,
we can compute the projected data $a^i_{\rm\bar h}$ and use them to determine the
autocorrelation function $\Gamma(t)$ according to \eqref{eq:Gamma}. This
allows to compute the autocorrelation time $\tauint$ and the error $\sigma^2$.

As is argued in \cite{heplat0009027}, the advantage of this ``summation
method'' compared to jackknife is that the error of $\tauint$ decays much
faster with the number of data points.  Therefore, error estimates can be be
accurate, which is particularly important for performance studies, where
$\tauint$ directly enters the definition of the cost of an algorithm.

In this thesis, this aspect is less important. In the algorithms used here,
the Schr\"odinger functional coupling can be measured as a primary quantity
(in contrast to e.g. the PHMC algorithm, where the coupling is obtained by a
reweighting procedure).  Nevertheless, we have used the method described here
e.g. for the determination of $\csw$ in section~\ref{sec:csw}. The variable
$\Delta M$ used there is computed from 12 primary observables, which means
that 12 derivatives have to be implemented for the computation of the
projected observables. This shows the increased complexity compared to the
more ``universal'' jackknife method.

\end{appendix}

\bibliographystyle{unsrt}
\bibliography{diss}
\addcontentsline{toc}{chapter}{Bibliography}

\chapter*{Acknowledgements}

I want to thank all people who have made this thesis possible, in
particular:
\begin{itemize}
\item my supervisor Ulli Wolff for giving me the opportunity to work 
  in the framework of the ALPHA collaboration, his willingness to answer 
  any question and his scientific thoroughness and honesty.
\item Rainer Sommer for the idea and several discussions about of the 
  mass decoupling project.
\item Juri Rolf for a good collaboration on the bermion project. His
  careful and critical reading of this thesis was invaluable.
\item Burkhard Bunk for sharing his experience and Francesco Knechtli for 
  many discussions.
\item the COM group at Humboldt University and the theory group in Zeuthen
  for a pleasant stay in Berlin.
\item the Graduiertenkolleg 271 for financial support.
\item DESY/NIC for financial support in the last months of my work and
  for support with using the APE computers.
\end{itemize}

\selectlanguage{german}

\chapter*{Lebenslauf}

\begin{tabular}{ll}


1978 - 1982       & Besuch der Grundschule in Gelsenkirchen-Buer \\

1982 - 1991       & Besuch des Annette-von-Droste-H\"ulshoff-Gymnasiums \\
                  & in Gelsenkirchen-Buer \\

6/1991            & Abitur \\

7/1991  - 6/1992  & Wehrdienst \\

10/1992 - 01/1998 & Studium an der Universit\"at M\"unster \\
                  & in der Fachrichtung Physik \\

6/1998  - 6/2001  & Doktoranden-Stipendium im Graduiertenkolleg 271 \\
                  & \glqq{}Strukturuntersuchungen, Pr\"azisionstests und Erwei- \\
                  & terungen des Standardmodells der Elementarteilchen- \\
                  & physik\grqq{} an der Humboldt-Universit\"at zu Berlin \\

9/2001  - 12/2001 & Wissenschaftlicher Mitarbeiter am Deutschen \\
                  & Elektronensynchrotron (DESY), Zeuthen\\

\end{tabular}

\section*{Publikationsliste}

\noindent Bernd Gehrmann, Ulli Wolff. Efficiencies and optimization of HMC algorithms
in pure gauge theory. {\em Nucl. Phys. Proc. Suppl.}, 83:801-803, 2000.

\vspace{2mm}
\noindent ALPHA collaboration. First results on
the running coupling in QCD with two massless flavors. {\em Phys. Lett.},
B515:49-56, 2001.

\vspace{2mm}
\noindent Bernd Gehrmann, Juri Rolf, Stefan Kurth, Ulli Wolff. Schr\"odinger
functional at negative flavor number. {\em Nucl. Phys.}, B612:3-24, 2001.

\vspace{2mm}
\noindent Bernd Gehrmann, Juri Rolf, Stefan Kurth, Ulli Wolff. Schr\"odinger
functional at $\Nf=-2$. {\em Nucl. Phys. Proc. Suppl.}, 106:793-795, 2002.

\chapter*{Selbst\"andigkeitserkl\"arung}

Hiermit erkl\"are ich, die vorliegende Arbeit selbst\"andig ohne fremde 
Hilfe verfa{\ss}t zu haben und nur die angegebene Literatur und Hilfsmittel 
verwendet zu haben.\\

\vspace{5cm}
\noindent \dcauthorname  \dcauthorsurname \\
\dcdatesubmitted

\end{document}